\documentclass[10pt]{amsart}       
\usepackage{graphicx}
\usepackage{amsmath}
\usepackage{amsfonts}
\usepackage{amsthm}
\usepackage{amssymb}
\usepackage{subfigure}
\usepackage{verbatim}
\usepackage{color}
\newtheorem{thm}{Theorem}[section]
\newtheorem{prop}[thm]{Proposition}
\newtheorem{defn}[thm]{Definition}
\newtheorem{lemma}[thm]{Lemma}

\newtheorem{cor}[thm]{Corollary}

\newcommand{\bbm}{\begin{bmatrix}}
\def\ebm{\end{bmatrix}}
\def\e{\varepsilon}
\def\BB{\mathbf{B}}
\def\BBi{\overline{\mathbf{B}}}
\def\EE{\mathbf{E}}
\def\GG{\mathbf{G}}
\def\hh{\mathbf{h}}
\def\HH{\mathbf{H}}
\def\HHt{\mathbf{\widetilde{H}}}
\def\jj{\mathbf{j}}
\def\JJ{\mathbf{J}}
\def\JJz{\underline{\mathbf{J}}}
\def\JJi{\overline{\mathbf{J}}}

\def\jjz{\underline{\mathbf{j}}}
\def\jji{\overline{\mathbf{j}}}
\def\KK{\mathbf{K}}
\def\KKz{\underline{\mathbf{K}}}
\def\KKi{\overline{\mathbf{K}}}

\def\mi{\overline{m}}

\def\MMi{\overline{\mathbf{M}}}
\def\MMz{\underline{\mathbf{M}}}
\def\mmi{\overline{\mathbf{m}}}
\def\mmz{\underline{\mathbf{m}}}

\def\Qi{\overline{Q}}
\def\QQz{\underline{\mathbf{Q}}}
\def\QQi{\overline{\mathbf{Q}}}

\def\eq{\begin{equation}}
\def\endeq{\end{equation}}
\def\eqarr{\begin{eqnarray}}
\def\endeqarr{\end{eqnarray}}
\def\eqnn{\begin{equation*}}
\def\endeqnn{\end{equation*}}
\def\ds{\begin{displaystyle}}
\def\endds{\end{displaystyle}}
\def\sech{\text{sech}}

\begin{document}

\title{Exact Solutions of Semiclassical Non-characteristic Cauchy Problems for the Sine-Gordon Equation}
\author{Robert Buckingham}
   \address{\noindent Department of Mathematics, University of Michigan, Ann Arbor, MI 48109}
   \email{robbiejb@umich.edu}
\author{Peter D. Miller}
   \email{millerpd@umich.edu}

\begin{abstract}
  The use of the sine-Gordon equation as a model of magnetic flux
  propagation in Josephson junctions motivates studying the
  initial-value problem for this equation in the semiclassical limit
  in which the dispersion parameter $\e$ tends to zero.  Assuming
  natural initial data having the profile of a moving $-2\pi$ kink at
  time zero, we analytically calculate the scattering data of this
  completely integrable Cauchy problem for all $\e>0$ sufficiently
  small, and further we invert the scattering transform to calculate
  the solution for a sequence of arbitrarily small $\e$. This sequence
  of exact solutions is analogous to that of the well-known
  $N$-soliton (or higher-order soliton) solutions of the focusing
  nonlinear Schr\"odinger equation.  Plots of exact solutions for
  small $\e$ reveal certain features that emerge in the semiclassical
  limit.  For example, in the limit $\epsilon\rightarrow 0$ one
  observes the appearance of nonlinear caustics, i.e.\@ curves in
  space-time that are independent of $\e$ but vary with the initial
  data and that separate regions in which the solution is expected to
  have different numbers of nonlinear phases.

  In the appendices we give a self contained account of the Cauchy
  problem from the perspectives of both inverse scattering and
  classical analysis (Picard iteration).  Specifically, Appendix
  \ref{appendix} contains a complete formulation of the
  inverse-scattering method for generic $L^1$-Sobolev initial data,
  and Appendix \ref{app-well-posed} establishes the well-posedness for
  $L^p$-Sobolev initial data (which in particular completely justifies
  the inverse-scattering analysis in Appendix \ref{appendix}).
\end{abstract}

\maketitle

\section{Introduction}
The sine-Gordon equation 
\eq
\label{sine-Gordon}
\e^2u_{tt}-\e^2u_{xx} + \sin(u) = 0
\endeq
describes a broad array of physical and mathematical phenomena.  The
partial differential equation \eqref{sine-Gordon} may be regarded as
the continuum limit of a chain of pendula subject to an external
(gravity) force and coupled to their nearest neighbors via Hooke's
law.  In nonlinear optics, the sine-Gordon equation is a special case
of the Maxwell-Bloch equations and describes self-induced transparency
in the sharp-line limit \cite{Maimistov:2001}.  In biology, the
sine-Gordon equation models transcription and denaturation in DNA
molecules \cite{Salerno:1991}.  B\"acklund showed a correspondence
between solutions of the sine-Gordon equation and surfaces of constant
negative curvature \cite{Fordy:1994-book}.

In solid-state physics, the sine-Gordon equation models idealized
magnetic flux propagation along the insulating barrier between two
superconductors in a \emph{Josephson junction}.  Here the length
$\ell_0$ of the transmission line corresponds to a length of $1$ in
terms of the dimensionless coordinate $x$ measuring distance along the
junction.  Let $L$ be the inductance per unit length and $C$ be the
capacitance per unit length.  Then $v:=(LC)^{-1/2}$ is the typical
velocity parameter, and the macroscopic time scale $t$ measures one
dimensionless unit when $\ell_0/v$ seconds have passed.  The parameter
$\e$ is the ratio of the Josephson length $\ell_J$ to the transmission
line length $\ell_0$.  The Josephson length $\ell_J$ is in turn
proportional to $\Phi_0^{1/2}$, where $\Phi_0:=
h/(2\e)\approx2.064\times10^{-15}$ V sec is the quantum unit of
magnetic flux.  Laboratory experiments by Scott, Chu, and Reible
\cite{Scott:1976} analyzed flux propagation in Josephson junctions of
length $\ell_0=35$ cm for which $\ell_J$ was approximately $10^{-4}$
to $10^{-3}$ m.  Therefore, in these experiments,
$\e:=\ell_J/\ell_0\approx0.0005$.  The period of a signal input to the
transmission line in these experiments was typically on the order of
$10^{-9}$ seconds, which is approximately one dimensionless time unit
on the $t$-scale.  Together with $\e$ being small, this motivates the
study of the \emph{semiclassical} (or zero-dispersion) limit as
$\e\downarrow 0$.  For analytical convenience we choose to study the
Cauchy initial-value problem on the real line $x\in\mathbb{R}$.
Formulating a \emph{semiclassical Cauchy problem} means fixing
functions $f$ and $g$ independent of $\e$, and then, for all $\e>0$
sufficiently small, posing the Cauchy problem for \eqref{sine-Gordon}
with initial data of the form $u(x,0;\e)=f(x)$, $\e u_t(x,0;\e)=g(x)$.
See Appendix \ref{app-well-posed} for an account of the well-posedness
theory of this Cauchy problem for $\e>0$ fixed.  Solving the
semiclassical Cauchy problem means obtaining the one-parameter family
of solutions $u(x,t;\e)$.  We are usually most interested in the
asymptotic behavior of the solution as $\e\downarrow 0$.

In this paper, we consider the sine-Gordon equation \eqref{sine-Gordon} 
for all $\e>0$ sufficiently small with the initial condition
\eq
\label{init-cond}
\sin \left(\frac{f}{2}\right) := \sech(x), \quad \cos\left(\frac{f}{2}\right):=\tanh(x), \quad g := 2\mu\,\sech(x), \quad x\in\mathbb{R}
\endeq
where $\mu\in\mathbb{R}$ is a parameter.  We refer to the solution of the 
Cauchy problem as $u(x,t;\e,\mu)$.  The 
\emph{topological charge} (or winding number) of solutions satisfying 
\eqref{init-cond} is a constant of motion given by
\eq
w[u]:= \frac{1}{2\pi}\int_{-\infty}^{+\infty} u_x \, dx =-1.
\endeq  
From one point of view, the initial data \eqref{init-cond} are natural to 
study, because $u(x,t;\sqrt{\mu^2+1},\mu)$ is 
an exact mathematical antikink solution of the sine-Gordon equation 
explicitly given by
\eq
\begin{split}
\cos(u(x,t;\sqrt{\mu^2+1},\mu)) & = 1-2\,\sech^2\left(\frac{1}{\e w}(x-vt)\right), \\
\sin(u(x,t;\sqrt{\mu^2+1},\mu)) & = 2\,\sech\left(\frac{1}{\e w}(x-vt)\right)\tanh\left(\frac{1}{\e w}(x-vt)\right)
\end{split}
\endeq
with velocity $v$ and width parameter $w$ given by
\eq
v = \frac{1}{\mu^2+1-\mu\sqrt{\mu^2+1}}-1, \quad w = -\frac{1}{2}\sqrt{1-v^2}.
\endeq
From these formulae we see that $u$ is a traveling wave with velocity
$v$ bounded by 1 (the light speed), demonstrating the hyperbolicity of
the sine-Gordon equation.  This solution admits a natural relativistic
interpretation since the relationship between $v$ and $w$ corresponds
to Lorentz contraction in special relativity.

For $\e\neq\sqrt{\mu^2+1}$, the initial data \eqref{init-cond} no
longer corresponds to simply one soliton, but in general excites a
nonlinear superposition of kinks, antikinks, breathers, and radiation.
It is interesting to observe that the initial data \eqref{init-cond}
satisfies the advection equation $u_t + vu_x = 0$ with constant
velocity $v=\mu/\e$.  In this sense, we may consider the initial data
as being in uniform motion to the right with velocity $v$.  Note,
however, that if $\mu\neq 0$, then for $\e>0$ sufficiently small, the
velocity $v$ of the initial data exceeds the constraint $|v|\le 1$
imposed by the hyperbolic nature of the sine-Gordon equation
\eqref{sine-Gordon}.  In this situation, one might expect the
sine-Gordon equation to regularize the superluminal velocity of the
initial data for $t>0$ by some kind of catastrophic effect that
destroys the profile of the initial data.  In fact, we will show (see
figures \ref{reflectionless-plots-mu1} and
\ref{reflectionless-plots-N16}) that the regularization of the
velocity takes place via the emission of a large number (inversely
proportional to $\e$) of kink-antikink pairs.

The family of solutions corresponding to the initial data
\eqref{init-cond} may be viewed as an analogue for the sine-Gordon
equation of the $N$-soliton (or higher-order soliton) solution to the
cubic focusing nonlinear Schr\"odinger (NLS) equation
\begin{equation}
iq_t+\frac{1}{2}q_{xx}+|q|^2q=0\,.
\label{eq:unscaledNLS}
\end{equation}
Satsuma and Yajima \cite{Satsuma:1974} found that with initial data
$q(x,0;N)$=$N\,\sech(x)$ the scattering data relevant for the focusing
NLS equation can be found in closed form for any $N\in\mathbb{R}$.
Furthermore, if $N\in\mathbb{Z}$ then the scattering data are
reflectionless and so the solution $q(x,t;N)$ can be found
more-or-less explicitly.  In \cite{Miller:1998} it was noted that,
with $\e=c/N$ and $\tau=Nt/c$, the function
$\phi(x,\tau;\e)=cq(x,t;N)/N$ satisfies the initial-value
problem \eq
\label{NLS-N-soliton}
i\e\phi_\tau + \frac{\e^2}{2}\phi_{xx} + |\phi|^2\phi = 0, \quad \phi(x,0;\e) = c\,\sech(x).
\endeq
The functions $\phi(x,\tau;\e)$ therefore solve a semiclassical Cauchy
problem since $\phi(x,0;\e)$ is independent of $\e$.
Numerical reconstruction of the inverse-scattering solution for
$\e=\e_N=c/N$, $N=1,2,3,\dots$ in \cite{Miller:1998} revealed
a spatio-temporal pattern for $\phi$ emerging as $\e\downarrow 0$
consisting of a fixed macrostructure with \emph{nonlinear caustics}
(phase transition boundaries or breaking curves) separating regions
of the space-time plane consisting of oscillations of different local
genus (number of nonlinear phases).  At least two caustic curves
appear in the dynamics (a \emph{primary caustic} $t=t_1(x)$ and a
\emph{secondary caustic} $t=t_2(x)>t_1(x)$).  The semiclassical
asymptotics for times $t$ up to and just beyond the primary caustic
were obtained in \cite{Kamvissis:2003-book} and these results were
extended to times $t$ just beyond the secondary caustic (requiring a
substantial modification of the method that captures the primary
caustic) in \cite{Lyng:2007}.  In a related result, Tovbis and
Venakides \cite{Tovbis:2000} generalized the calculation of Satsuma
and Yajima by computing the scattering data associated with the
semiclassically-scaled focusing NLS equation \eqref{NLS-N-soliton}
explicitly for all sufficiently small $\e>0$ when the initial data is
given in the form \eq
\label{Tovbis-Venakides}
\phi(x,0;\mu) = A(x)e^{iS(x)/\e}, \quad S'(x)=-\mu\tanh(x), \,\, A=c\,\sech(x)\,.
\endeq
Subsequently, the Cauchy problem for \eqref{NLS-N-soliton} with
this initial data has been studied by Tovbis, Venakides, and Zhou
\cite{Tovbis:2004, Tovbis:2006}.  In this paper, we present a
calculation of the scattering data for
(\ref{sine-Gordon})--(\ref{init-cond}) analogous to the work in
\cite{Satsuma:1974} and \cite{Tovbis:2000}, and we also present an
explicit computation of $u(x,t;\e,\mu)$ as $\e\downarrow0$ analogous
to \cite{Miller:1998}.  The asymptotic analysis of the semiclassical
Cauchy problem for sine-Gordon corresponding to the work in
\cite{Kamvissis:2003-book, Lyng:2007, Tovbis:2004, Tovbis:2006} will
be carried out in a later work.

The sine-Gordon equation \eqref{sine-Gordon} is an integrable system,
possessing a Lax pair (see \eqref{eigenvalue-eqn2} and
\eqref{time-evolution-eqn}) and admitting all the benefits thereof,
including the existence of inverse-scattering transforms for solving
Cauchy problems in various coordinate systems.  We consider the Cauchy
problem in laboratory coordinates and we use the Riemann-Hilbert
formulation of inverse scattering.  For the sine-Gordon equation in
characteristic coordinates, the inverse-scattering method was first
given in \cite{Ablowitz:1973} and \cite{Zakharov:1974}.  The
inverse-scattering method corresponding to the (noncharacteristic)
Cauchy problem for the sine-Gordon equation in laboratory coordinates
was worked out by Kaup \cite{Kaup:1975}.  An account of the
Riemann-Hilbert method for carrying out the inverse step in laboratory
coordinates can be found in the text of Faddeev and Takhtajan
\cite{Faddeev:1987-book}, and further developments to the theory were
made by Zhou \cite{Zhou:1995} and Cheng et al.\@ \cite{Cheng:1999}.
In our paper, we add to this literature by giving in Appendix
\ref{appendix} a complete description of the Riemann-Hilbert
formulation of the solution of the Cauchy problem in laboratory
coordinates assuming that at each instant of time the solution $u$ has
$L^1$-Sobolev regularity.  That the sine-Gordon equation
\eqref{sine-Gordon} preserves this degree of regularity if it is
present at $t=0$ is established by independent arguments in Appendix
\ref{app-well-posed}.

Briefly, the inverse-scattering method proceeds as follows.  Cauchy
data for the sine-Gordon equation characterize a set of scattering
data, which consist of the \emph{reflection coefficient}
$\rho:\mathbb{R}\rightarrow\mathbb{C}$, the \emph{eigenvalues}
$\{z_n\}$, and the \emph{modified proportionality constants}
$\{c_n\}$.  The scattering data are used to formulate a
Riemann-Hilbert problem with an explicit, elementary dependence on $x$
and $t$.  While it is not in general possible to solve a
Riemann-Hilbert problem in closed form, for \emph{reflectionless}
Cauchy data (i.e.\@ for which $\rho(z)\equiv0$) the
Riemann-Hilbert problem can be reduced to the solution of a system of
linear algebraic equations.

In Section \ref{forward-scattering}, we explicitly calculate the
scattering data corresponding to viewing the initial data
\eqref{init-cond} as a kind of potential in the linear scattering
problem \eqref{eigenvalue-eqn2} associated with the Cauchy problem for
the sine-Gordon equation \eqref{sine-Gordon}.  Our analysis will be
valid for all $\mu \in \mathbb{R}$ and $\e>0$ sufficiently small.
Furthermore, we show that if $\e$ lies in the sequence
\begin{equation}
\e=\e_N(\mu):=\frac{\sqrt{\mu^2+1}}{2N+1}\,, \quad\quad
N\in\mathbb{Z}^+:=\{0,1,2,\dots\}
\label{eq:epsilonsequence}
\end{equation}
(note that this sequence converges to zero as $N\rightarrow\infty$),
then the scattering data are reflectionless ultimately implying via
inverse-scattering theory that $u$ can be computed explicitly (that
is, $u$ can be expressed by a finite number of arithmetic operations).
The inverse step is carried out for $\e$ in the sequence
\eqref{eq:epsilonsequence} corresponding to reflectionless initial
data in Section \ref{inverse-scattering}, where $\cos(u)$ and
$\sin(u)$ are extracted by considering an appropriate limit of the
solution to the Riemann-Hilbert problem.  As $\e\downarrow0$ through
this sequence, a pattern emerges in which $u$ consists of modulated
wave trains of wave number and frequency inversely proportional to
$\e$ with one or more nonlinear phases.  The spatio-temporal scale of
the modulation is fixed as $\e\downarrow 0$.  Regions of space-time
containing waves with different numbers of nonlinear phases are
separated by nonlinear caustics that are independent of $\e$ for fixed
$\mu$.  See figures \ref{reflectionless-plots-mu0},
\ref{reflectionless-plots-mu1}, and \ref{reflectionless-plots-N16} for
plots of $\cos(u)$ exhibiting these features for various values of
$\e$ and $\mu$.  At a qualitative level, these features resemble those
observed for solutions of the semiclassical Cauchy problem for the
focusing NLS equation.  Section \ref{conclusion} discusses the
limitations inherent in an approach to the semiclassical limit based
upon calculations of complexity and sensitivity increasing with
$N\sim\e^{-1}$, and explores possible extensions.

\

\noindent {\it Remark 1}.  Under the scalings $x=\e X$ and $t=\e T$
and the choice $\e=\e_N(\mu)$ (see
\eqref{eq:epsilonsequence}), equations (\ref{sine-Gordon}) and
(\ref{init-cond}) become \eq
\label{scaled-SG}
U_{TT} - {U}_{XX} + \sin(U) = 0,
\endeq
\eq
\label{scaled-ICs}
\begin{split}
\sin\left(\frac{1}{2}U(X,0;N,\mu)\right)=
\sech\left(\e_N(\mu)X\right), \quad 
&\cos\left(\frac{1}{2}U(X,0;N,\mu)\right)=\tanh\left(\e_N(\mu)X\right), \\ 
U_{T}(X,0;N,\mu)=2\mu\,\sech&\left(\e_N(\mu)X\right),
\end{split}
\endeq
where $U(X,T;N,\mu)=u(x,t;\e,\mu)$.  This is a fixed-dispersion Cauchy
problem with a sequence of different initial conditions depending on
$N$, just as in the problem for the NLS equation studied by Satsuma
and Yajima.  The initial conditions all have topological charge
$w[U]=-1$ but $U(X,0;N,\mu)$ and $U_T(X,0;N,\mu)$ become more dilated
in $X$ (slowly-varying) as $N$ increases.  Therefore, an alternate way
of viewing our result is that we can find exact solutions to the
fixed-dispersion initial-value problem
(\ref{scaled-SG})--(\ref{scaled-ICs}) for $N\in\mathbb{Z}^+$.
As an example of an explicit solution of \eqref{scaled-SG} obtained in this way,
when $N=1$ and $\mu=0$ we have
\begin{equation}
\label{exact-cosu-example}
\cos(U(X,T;1,0)) = 1-
\frac{2n(X,T)^2}{d(X,T)^2}\,,
\end{equation}
where
\begin{equation}
\begin{split}
n(X,T)&:=11+
\cos\left(\frac{4\sqrt{2}}{3}T\right) + 
8\cosh\left(\frac{2}{3}X\right) + 
4 \cos\left(\frac{2\sqrt{2}}{3}T\right)\left[2\cosh\left(\frac{2}{3}X\right) + 
\cosh\left(\frac{4}{3}X\right)\right]\\
d(X,T)&:=
4\left[2+3\cos\left(\frac{2\sqrt{2}}{3}T\right)\right]
\cosh\left(\frac{1}{3}X\right)+\left[9+\cos\left(\frac{4\sqrt{2}}{3}T\right)\right]\cosh(X)+2\cosh\left(\frac{5}{3}X\right)\,.
\end{split}
\end{equation}
The focusing NLS equation \eqref{eq:unscaledNLS} admits a scaling
symmetry in which scaling the independent variable $x$ is equivalent
to scaling the dependent variable (amplitude) $q$ and the time $t$.
Thus, the $N$-soliton (or higher-order soliton) solutions of the
focusing NLS equation that were originally obtained by Satsuma and
Yajima \cite{Satsuma:1974} by considering a fixed-width pulse with
variable amplitude can just as easily be viewed as a fixed-amplitude
pulse with variable width.  From the point of view of semiclassical
asymptotics, dilation in $x$ is the more natural interpretation of the
higher-order solitons as the presence of the parameter $\e$ in
\eqref{NLS-N-soliton} amounts to rescaling $x$ and $t$, and thus the
variable width of the pulse is absorbed into the semiclassical
parameter $\e$ as in \cite{Miller:1998,Kamvissis:2003-book}.  Of
course, the sine-Gordon equation does not admit the amplitude/dilation
symmetry enjoyed by the focusing NLS equation, so we are not free to
interpret the family of exact solutions we obtain in this paper in
terms of scaling of amplitude.  It seems that perhaps a more generally
fruitful approach to seeking analogues of the higher-order soliton in
other integrable systems is to consider pulse width dilation as being
more fundamental than amplitude dilation.  As more evidence of the
utility of this approach (beyond the sine-Gordon example), the
modified NLS equation (which includes an additional term in
\eqref{eq:unscaledNLS} that breaks the scaling symmetry needed to
exchange amplitude for width) does not have higher-order solitons in
the sense of Satsuma and Yajima, but it does have exact solutions
corresponding to arbitrarily width-dilated pulses that are useful in
semiclassical analysis \cite{DiFranco:2007}.

\

\noindent {\it Remark 2}.  In characteristic or light-cone coordinates
$\chi$ and $\tau$ defined by $x=\chi+\tau$ and $t=\chi-\tau$, the
sine-Gordon equation (\ref{sine-Gordon}) is
$\e^2v_{\chi\tau}=\sin(v)$, where $v(\chi,\tau;\e)=u(x,t;\e)$.  The
associated $\chi$ evolution equation in the Lax pair is the
Zakharov-Shabat eigenvalue equation, which is the same eigenvalue
equation as for the focusing NLS equation \cite{Zakharov:1972}.  Thus
it is possible to solve a semiclassical characteristic Cauchy problem
with special initial data $v(\chi,0;\e)=c\,\sech(\chi)$ using the
Satsuma-Yajima higher-order soliton solution.  However, in many
applications (as in Josephson junction theory), the correct problem is
the non-characteristic Cauchy problem with two independent initial
conditions: $u(x,0;\e)=f(x)$, $\e u_t(x,0;\e)=g(x)$.  The
Satsuma-Yajima solution to the semiclassical problem posed along a
characteristic $\tau=0$ or $x=t$ will have a very complicated and
unwieldy form and an undesired dependence on $\e$ upon restriction to
$t=0$, and therefore is probably not relevant to the
non-characteristic semiclassical
Cauchy problem we wish to consider.\\

\noindent {\it On notation}.  As will be explained in detail in
Section \ref{forward-scattering} and Appendix \ref{appendix}, we will
use three different gauges for the eigenvalue problem.  Objects
associated with the \emph{infinity gauge} will be denoted by an
overline ($\JJi$).  Objects associated with the \emph{zero gauge} will
be denoted by an underline ($\JJz$).  Finally, objects associated with
the \emph{symmetric gauge} will not have a bar ($\JJ$).  The complex
conjugate of $a$ is denoted by $a^*$.  We use the notation $f(a,b;c)$
to emphasize that $c$ is a parameter.  The dependence on parameters
may be suppressed by writing $f(a,b)$ in place of $f(a,b;c)$.  We also
make frequent use of the standard Pauli matrices defined as \eq
\sigma_1 := \bbm 0 & 1 \\ 1 & 0 \ebm, \quad \sigma_2 := \bbm 0 & -i \\
i & 0 \ebm, \quad \sigma_3 := \bbm 1 & 0 \\ 0 & -1 \ebm.
\endeq
Vectors will be denoted by bold lower-case letters and matrices by bold upper-case letters, with the exception of the Pauli matrices.  The transpose of 
a vector $\mathbf{v}$ is denoted by $\mathbf{v}^\mathsf{T}$, and the 
conjugate-transpose of a matrix $\mathbf{A}$ is denoted by 
$\mathbf{A}^\dagger$.  Finally, $\chi_S(\cdot)$ denotes the characteristic
function (indicator function) of a set $S$, that is $\chi_S(x)=1$ if
$x\in S$ and $\chi_S(x)=0$ otherwise.

\section{Scattering Theory for the Special Initial Data}
\label{forward-scattering}
The quantities $z\pm z^{-1}$ appear throughout the scattering and
inverse-scattering theory of the sine-Gordon equation
\eqref{sine-Gordon}, and so for convenience we define \eq
\label{D-and-E}
D(z):=\frac{1}{4}\left(z+\frac{1}{z}\right), \quad E(z):=\frac{1}{4}\left(z-\frac{1}{z}\right).
\endeq
Fix the initial condition \eqref{init-cond}.  To find the scattering
data it is necessary to solve the following generalized eigenvalue
problem for $\JJi(x;z,t,\e,\mu)$ (see, for example, \cite{Kaup:1975}):
\eq
\label{eigenvalue-eqn}
4i\e\JJi_x = \bbm 4E+z^{-1}(1-\cos(u)) & -z^{-1}\sin(u)-i\e(u_x+u_t) \\ -z^{-1}\sin(u)+i\e(u_x+u_t) & -4E-z^{-1}(1-\cos(u)) \ebm \JJi.
\endeq
This formulation of the eigenvalue problem is useful in the study of
solutions $\JJi$ when $z$ is bounded away from $0$ (see
\cite{Kaup:1975} and Proposition \ref{m-LHP-at-infinity} below), and
for this reason, we say that \eqref{eigenvalue-eqn} is written in the
\emph{infinity gauge}.  The use of alternate gauges proves to be
beneficial.  For example, the gauge transformation \eqref{zero-gauge}
(see Appendix \ref{appendix}) casts the eigenvalue problem
\eqref{eigenvalue-eqn} into an alternate form that is useful in the
analysis of solutions corresponding to bounded $z$, and in particular
near $z=0$ (see \cite{Kaup:1975} and Proposition \ref{m-LHP-at-zero}).
Therefore, we refer to the coordinate system arrived at via the
transformation \eqref{zero-gauge} as the \emph{zero gauge}.  While the
infinity gauge and the zero gauge are useful in the analysis of the
scattering problem required to formulate an inverse-scattering theory,
to calculate the scattering data corresponding to \eqref{init-cond} we
found it to be useful to introduce a gauge transformation that
symmetrizes the appearance of $z$ and $z^{-1}$ in the eigenvalue
problem and at the same time also removes the function $u_x$ from the
coefficients.  It is in this third, \emph{symmetric gauge} that it is
easiest to see the eigenvalue problem is in fact hypergeometric for
the initial data \eqref{init-cond}.  Once it is clear from working in
the symmetric gauge that the eigenvalue problem has exactly three
regular singular points, it is possible to use the theory of Euler
transforms to analyze the asymptotic behavior of the Jost solutions
and thus obtain the scattering data.

\subsection{Transformation to a hypergeometric equation}
The first step in transforming \eqref{eigenvalue-eqn} into a
hypergeometric equation is to introduce an appropriate gauge
transformation.  If $\JJi$ satisfies equation (\ref{eigenvalue-eqn}),
then the invertible transformation 
(having an interpretation as a rotation at each $x$ by an angle $-u/4$) 
\eq
\label{symmetric-gauge}
\JJ(x;z,t,\e,\mu) = \mathbf{A}\JJi := \bbm \displaystyle
\cos\left(\frac{u}{4}\right) & \displaystyle \sin\left(\frac{u}{4}\right) \\ \\
\displaystyle -\sin\left(\frac{u}{4}\right) & \displaystyle
\cos\left(\frac{u}{4}\right) \ebm \JJi
\endeq
yields a solution $\JJ$ of the eigenvalue problem written in the
\emph{symmetric gauge}: 
\eq
\label{symmetric-eigenvalue-eqn}
4i\e\JJ_x = \bbm \displaystyle 4E\cos\left(\frac{u}{2}\right) & \displaystyle
-4D\sin\left(\frac{u}{2}\right)-i\e u_t \\ \\
\displaystyle -4D\sin\left(\frac{u}{2}\right)+i\e u_t & 
\displaystyle -4E\cos\left(\frac{u}{2}\right) \ebm \JJ.
\endeq
Written in this form\footnote{The absence of $u_x$ in the symmetrized
  form \eqref{symmetric-eigenvalue-eqn} of the eigenvalue problem
  provides an alternate framework in which to consider discontinuous
  initial data without the use of delta functions (\emph{cf.\@}
  \cite{Kalbermann:2007}).}, the eigenvalue problem appears similar to
one used by Faddeev and Takhtajan (see \cite{Faddeev:1987-book} part
2, chapter 2, equation 4.1).  The Jost solutions $\JJ^\pm$ in the
symmetric gauge are defined to be the fundamental solution matrices of
the linear problem \eqref{symmetric-eigenvalue-eqn} for \emph{real}
values of $z$, normalized by the conditions \eq
\begin{split}
\label{Psi-symmetric-normalization}
&\JJ^- = \bbm e^{iEx/\e} & 0 \\ 0 & -e^{-iEx/\e}\ebm +o(1) \quad \text{as } x\to -\infty \text{ for } z\in\mathbb{R},\\
&\JJ^+ = \bbm e^{-iEx/\e} & 0 \\ 0 & e^{iEx/\e} \ebm +o(1) \quad \text{as } x\to +\infty \text{ for } z\in\mathbb{R}.
\end{split}
\endeq
We denote the columns in this way: $\JJ^\pm =: [\jj_1^\pm, \jj_2^\pm]$.  
They are related to the Jost solutions for the infinity gauge (see 
\eqref{Psi-infinity-normalizations}) by
\eq
\label{symmetric-Jost-solutions}
\jj_1^- = \mathbf{A}\jji_2^-, \quad \jj_2^- = \mathbf{A}\jji_1^-, \quad \jj_1^+ = \mathbf{A}\jji_1^+, \quad \jj_2^+ = \mathbf{A}\jji_2^+.
\endeq

For the choice of initial condition (\ref{init-cond}), the eigenvalue
equation (\ref{symmetric-eigenvalue-eqn}) takes the form 
\eq
\label{Psi-x-eqn}
2i\e\JJ_x = \bbm 2E\tanh(x) & \left(-2D-i\mu\right)\sech(x) \\ \left(-2D+i\mu\right)\sech(x) & -2E\tanh(x) \ebm \JJ.
\endeq
With the change of independent variable
\eq
y=\tanh(x),
\endeq
the eigenvalue problem (\ref{Psi-x-eqn}) becomes
\eq
\label{Psi-y-eqn}
2i\e(1-y^2)\JJ_y = \bbm 2Ey & \left(-2D-i\mu\right)(1-y^2)^{1/2} \\ \left(-2D+i\mu\right)(1-y^2)^{1/2} & -2Ey \ebm \JJ.
\endeq
Here $-1<y<1$ and the positive square root is chosen.  There are two ways 
to eliminate the square roots in the coefficient matrix.  The first is to 
introduce the linear transformation
\eq
\label{g-for-Psi-plus-1}
\jj_1^+ = \bbm 1 & 0 \\ 0 & (1-y^2)^{1/2} \ebm \mathbf{g},
\endeq
which results in a differential equation satisfied by 
$\mathbf{g}(y;z,\e,\mu)$:
\eq
\label{g-diffeq}
2i\e(1-y^2)\mathbf{g}_y = \bbm 2Ey & \left(-2D-i\mu\right)(1-y^2) \\ -2D+i\mu & \left(-2E+2i\e\right)y \ebm \mathbf{g}.
\endeq
This equation has exactly three regular singular points
$y\in\{-1,1,\infty\}$ and can be written in hypergeometric form.  We
will use \eqref{g-diffeq} to find expressions for $\jj_1^+$ and
$\jj_1^-$.

\

\noindent
{\it Remark}.  If we had taken \eqref{eigenvalue-eqn} instead of 
\eqref{symmetric-eigenvalue-eqn} as our starting point and followed analogous 
steps, namely (i) substitution of the initial data using double-angle formulae, 
(ii) the independent variable transformation $y=\tanh(x)$, and (iii) the use 
of the gauge transformation \eqref{g-for-Psi-plus-1} to reduce the problem to 
rational form, we would have arrived at
\eq
\label{not-clearly-hg}
2i\e(1-y^2)\mathbf{g}_y = \bbm 2E+z^{-1}(1-y^2) & -z^{-1}(1-y^2)y + i\e(1-y^2) \\ -z^{-1}y-i\e & -2E-z^{-1}(1-y^2)+2i\e y \ebm \mathbf{g}
\endeq
instead of \eqref{g-diffeq}.  Let $v=y^{-1}$.  Then near $v=0$,
\eqref{not-clearly-hg} has the leading-order form
$\mathbf{g}_v=O(v^{-3})\cdot \mathbf{g}$, whereas \eqref{g-diffeq} has
the leading-order form $\mathbf{g}_v=O(v^{-2})\cdot \mathbf{g}$.
After some calculation it is possible to see that the method of
Frobenius still applies to \eqref{not-clearly-hg} near $v=0$ even with
the additional growth at $y=\infty$ due to special identities holding
among the entries of the matrix coefficients of the leading-order
terms.  However, the local (Frobenius) analysis of \eqref{g-diffeq} is
more straightforward with only a double pole at $v=0$.  Later we will
also see that it is more difficult to obtain integral representations
for solutions of \eqref{not-clearly-hg} than for \eqref{g-diffeq}.

\

An alternative to the linear transformation \eqref{g-for-Psi-plus-1} is
\eq
\label{h-change-of-variables}
\jj_2^+ = \bbm 1 & 0 \\ 0 & (1-y^2)^{-1/2} \ebm \mathbf{h},
\endeq
which after substitution into \eqref{Psi-y-eqn} yields the
differential equation for $\mathbf{h}(y;z,\e,\mu)$:
\eq
\label{h-diffeq}
2i\e(1-y^2)\mathbf{h}_y = \bbm 2Ey & -2D-i\mu \\ \left(-2D+i\mu\right)(1-y^2) & \left(-2E-2i\e\right)y \ebm \mathbf{h}.
\endeq
This equation also has exactly three regular singular points
$y\in\{-1,1,\infty\}$.  It will be used to find expressions for
$\jj_2^+$ and $\jj_2^-$.

\subsection{Integral representations for Jost solutions}
We now use the theory of Euler transforms \cite{Hille:1997-book} to derive 
integral representations for the four Jost solutions, starting with 
$\jj_1^+$ and $\jj_1^-$.  Define
\eq
\gamma=\gamma(\mu): = \sqrt{\mu^2+1}.
\endeq
\begin{prop}
\label{Psi-plus-1-prop}
Choose the principal branches of the functions
$(s\pm1)^{-1/2+\gamma/2\e}$ with branch cuts on the real $s$-axis from
$\mp1$ to $-\infty$.  Also choose the principal branch of
$(s-y)^{-iE/\e-\gamma/2\e-1}$ with branch cut on the real $s$-axis
from $y$ to $-\infty$.  Take $\Sigma^+$ to be a closed contour in the
$s$-plane passing through the branch point $s=-1$ and encircling $s=1$
once in the counterclockwise direction 
\begin{figure}[h]
        \subfigure[The contour $\Sigma^+$.] {
                \label{sigma-plus-pic}
                \begin{minipage}[b]{0.45\textwidth}
                  \centering
                        \includegraphics[width=2.5in]{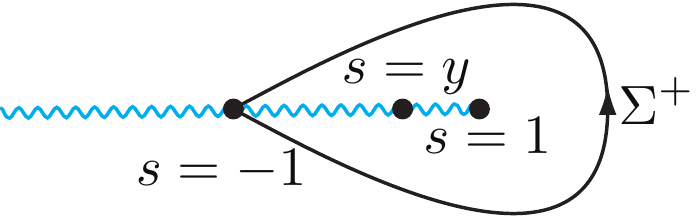}
                \end{minipage} } 
        \subfigure[The contour $\Sigma^-$.]{
                \label{sigma-minus-pic}
                \begin{minipage}[b]{0.45\textwidth}
                  \centering
                        \includegraphics[width=2.5in]{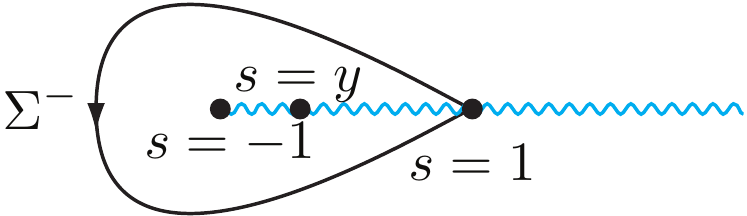}
                \end{minipage} }
  \caption{The integration contours and branch cuts for the Jost functions.}
        \label{sigma-plus-and-minus}
\end{figure}
(see figure \ref{sigma-plus-pic}).  Then, for
$z\in\mathbb{R}\backslash\{0\}$, $\jj_1^+=:[J_{11}^+,
J_{21}^+]^\mathsf{T}$ is given by \eq
\label{form-of-J-11-plus}
J_{11}^+=C_1(1-y^2)^{iE/2\e}\int_{\Sigma^+}(s-1)^{-1/2+\gamma/2\e}(s+1)^{-1/2+\gamma/2\e}(s-y)^{-iE/\e-\gamma/2\e}\,ds 
\endeq
\eq
\label{form-of-Psi-plus-21}
J_{21}^+=C_1(1-y^2)^{iE/2\e+1/2}\left(\frac{z-(\gamma+\mu)i}{z+(\gamma+\mu)i}\right)\int_{\Sigma^+}(s-1)^{-1/2+\gamma/2\e}(s+1)^{-1/2+\gamma/2\e}(s-y)^{-iE/\e-\gamma/2\e-1}\,ds
\endeq
with
\eq
\label{C-one}
C_1=C_1(z,\e,\mu) := \frac{\Gamma\left(\frac{1}{2}+\frac{iE}{\e}\right)\Gamma\left(1+\frac{\gamma}{2\e}-\frac{iE}{\e}\right)}{2^{1+\gamma/2\e}i\pi\Gamma\left(\frac{1}{2}+\frac{\gamma}{2\e}\right)}.
\endeq
\end{prop}
\begin{proof}
  We begin by computing the Frobenius exponents of \eqref{g-diffeq}.
  Assume that, for some number $p^+\in\mathbb{C}$, $\mathbf{g}$ has a
  Frobenius series about $y=1$ of the form 
\eq \mathbf{g}(y) =
  \sum_{n=0}^\infty \mathbf{g}_n^+(1-y)^{p^++n}
\endeq
for some vector-valued coefficients $\mathbf{g}_n^+$.  Substituting
this series into equation (\ref{g-diffeq}) and considering the
leading-order terms immediately gives the (indicial) eigenvalue equation
\eq -4i\e p^+\mathbf{g}_0^+ = \bbm 2E & 0 \\ -2D+i\mu & -2E+2i\e \ebm
\mathbf{g}_0^+.
\endeq
Therefore, the Frobenius exponents at $y=1$ are
\eq
p^+ = \frac{iE}{2\e}, \quad -\frac{iE}{2\e}-\frac{1}{2}.
\endeq
Similarly, substituting a series of the form 
\eq
\mathbf{g}(y) = \sum_{n=0}^\infty \mathbf{g}_n^-(1+y)^{p^-+n}
\endeq
into equation (\ref{g-diffeq}) and considering the leading-order terms shows 
that the Frobenius exponents at $y=-1$ are exactly the same:
\eq
p^- = \frac{iE}{2\e}, \quad -\frac{iE}{2\e}-\frac{1}{2}.
\endeq
We now shift two of the exponents to zero via the substitution
\eq
\mathbf{g}(y) = (1-y^2)^{iE/2\e}\mathbf{f}(y).
\endeq
It follows that $\mathbf{f}(y;z,\e,\mu)$ satisfies the differential
equation 
\eq
\label{f-diffeq}
2i\e(1-y^2)\mathbf{f}_y = \bbm 0 & \left(-2D-i\mu\right)(1-y^2) \\ -2D+i\mu & \left(-4E+2i\e\right)y \ebm \mathbf{f}.
\endeq
We attempt to express $\mathbf{f}=:[f_1,f_2]^\mathsf{T}$ as
\eq
\label{euler-transforms}
f_1(y) = \int_{\Sigma^+}  F_1(s)(s-y)^\alpha ds, \quad f_2(y) = \int_{\Sigma^+}  F_2(s)(s-y)^\beta ds
\endeq
where the Euler transforms $F_{1,2}(s;z,\e,\mu)$ and the constant
exponents $\alpha=\alpha(z,\e,\mu)$ and $\beta=\beta(z,\e,\mu)$ remain
to be chosen.  Substituting the expressions (\ref{euler-transforms})
into the system \eqref{f-diffeq} shows that the first equation $2i\e
f_{1y} = \left(-2D-i\mu\right)f_2$ can be easily solved by choosing
\eq
\label{F2-ito-F1}
F_2(s) = \frac{2i\e\alpha}{\left(2D+i\mu\right)}F_1(s) \quad \text{and} \quad \beta=\alpha-1.
\endeq

\

\noindent {\it Remark}.  The fact that there is such a simple
relationship between $F_1$ and $F_2$ is related to the fact that $f_2$
could be easily eliminated to write a second-order differential
equation for $f_1$ that is essentially the Gauss hypergeometric
equation.  On the other hand, if we had worked from the beginning in
the infinity gauge, the elimination of $f_2$ using the first row of
\eqref{not-clearly-hg} would have resulted in a second-order equation
that is not obviously of hypergeometric form.  This, in turn, would
lead to further complications in the following analysis leading from
\eqref{F1-eqn} to \eqref{F1-answer}.

\

It remains to satisfy the second equation of the system \eqref{f-diffeq}:
\begin{equation}
2i\e(1-y^2)f_{2y} = \left(-2D+i\mu\right)f_1 + \left(-4E+2i\e\right)yf_2.  
\end{equation}
Writing $y=s-(s-y)$ and $1-y^2=(1-s^2)+2s(s-y)-(s-y)^2$ and using
equations (\ref{F2-ito-F1}) gives 
\eqarr \nonumber
& & \hspace{-.8in}\frac{-2i\e^2\alpha(\alpha-1)}{\left(2D+i\mu\right)}\int_{\Sigma^+}  
F_1(s)\left[(1-s^2)(s-y)^{\alpha-2}+2s(s-y)^{\alpha-1}-(s-y)^\alpha\right]\,ds \\
\label{F1-eqn}
& = & \left(iD+\frac{\mu}{2}\right)\int_{\Sigma^+}  F_1(s)(s-y)^\alpha\, ds \\
\nonumber
& & +\left(2iE+\e\right)\frac{2i\e\alpha}{\left(2D+i\mu\right)}\int_{\Sigma^+}  F_1(s)\left[s(s-y)^{\alpha-1}-(s-y)^\alpha\right]\,ds.
\endeqarr
If we now choose $\alpha$ to satisfy the quadratic equation 
\eq
\frac{2i\e^2\alpha(\alpha-1)}{\left(2D+i\mu\right)} = \left(iD+\frac{\mu}{2}\right) - \left(2iE+\e\right)\frac{2i\e\alpha}{\left(2D+i\mu\right)}
\endeq
then the $(s-y)^\alpha$ terms will cancel in equation (\ref{F1-eqn}).
Specifically, we choose 
\eq
\label{alpha}
\alpha = -\frac{iE}{\e}-\frac{\gamma}{2\e}.
\endeq
Using integration by parts to eliminate the $(s-y)^{\alpha-2}$ term yields
\eq
\begin{split}
-\e\int_{\Sigma^+} \frac{d}{ds}\left[F_1(s)(1-s^2)\right](s-y)^{\alpha-1}\,ds 
+ 2\e(\alpha-1)\int_{\Sigma^+}  F_1(s)s(s-y)^{\alpha-1}\,ds \hspace{.6in}\\
= \left(-2iE-\e\right)\int_{\Sigma^+}  F_1(s)s(s-y)^{\alpha-1}\,ds.\hspace{.6in}
\end{split}
\endeq
Setting the integrands equal and using equation (\ref{alpha}) for
$\alpha$ gives the first-order linear differential equation 
\eq
\e(1-s^2)\frac{d}{ds}F_1(s) = (\e-\gamma)sF_1(s)
\endeq
or
\eq
\label{F1-diffeq}
\frac{d}{ds}\log F_1(s) = 
\left(\frac{\gamma}{2\e}-\frac{1}{2}\right)\frac{1}{s-1} +
\left(\frac{\gamma}{2\e}-\frac{1}{2}\right)\frac{1}{s+1} 
\endeq
for $F_1(s)$.  The general solution 
\eq
\label{F1-answer}
F_1(s) = C_1(s-1)^{-1/2+\gamma/2\e}(s+1)^{-1/2+\gamma/2\e}\,,
\endeq
where $C_1$ is an integration constant, gives equations
(\ref{form-of-J-11-plus}) and (\ref{form-of-Psi-plus-21}) up to the
choice of the constant $C_1$.

The constant $C_1$ is to be chosen so that $\jj_1^+$ is normalized as
required in equation (\ref{Psi-symmetric-normalization}).  Consider
equation (\ref{form-of-J-11-plus}) for $J_{11}^+$ as $y\to 1$.  Now
for $s\in\Sigma^+$, 
\begin{multline}
\left\vert(s-1)^{-1/2+\gamma/2\e}(s+1)^{-1/2+\gamma/2\e}
(s-y)^{-iE/\e-\gamma/2\e}\right\vert \\
\begin{aligned}\leq  & \left\vert(s-1)^{-1/2+\gamma/2\e}(s-y)^{-iE/\e-\gamma/2\e}
\right\vert|s+1|^{-1/2+\gamma/2\e}\\
\leq &  K|s+1|^{-1/2+\gamma/2\e}
\end{aligned}
\end{multline}
for some constant $K>0$, as $\Sigma^+$ is bounded away from 
$s=y$ and $s=1$.  Since $\gamma/2\e>0$, the function $|s+1|^{-1/2+\gamma/2\e}$ 
is integrable on 
$\Sigma^+$.  Therefore, by Lebesgue's dominated convergence theorem, 
\eq
\label{J-11-plus-for-large-y}
J_{11}^+ = C_1(1-y^2)^{iE/2\e}\left[\int_{\Sigma^+}(s+1)^{-1/2+\gamma/2\e}(s-1)^{-iE/\e-1/2}\,ds+o(1)\right] \text{ as } y\to 1.
\endeq
The integrand is integrable at $s=1$, so we can collapse $\Sigma^+$ to 
the upper and lower edges of the branch cut $[-1,1]$, yielding
\begin{multline}
\int_{\Sigma^+}(s+1)^{-1/2+\gamma/2\e}(s-1)^{-iE/\e-1/2}\,ds \\
\begin{aligned} = &2i\sin\left(\pi\left(\frac{1}{2}+\frac{iE}{\e}\right)\right)\int_{-1}^1(s+1)^{-1/2+\gamma/2\e}(1-s)^{-1/2-iE/\e}ds\\
= &\frac{2i\pi}{\Gamma\left(\frac{1}{2}+\frac{iE}{\e}\right)\Gamma\left(\frac{1}{2}-\frac{iE}{\e}\right)}\int_{-1}^1(s+1)^{-1/2+\gamma/2\e}(1-s)^{-1/2-iE/\e}\,ds.
\end{aligned}
\end{multline}
In the last step we used the reflection identity
\eq
\label{reflection-identity}
\sin(\pi z) = \frac{\pi}{\Gamma(z)\Gamma(1-z)}.
\endeq
The remaining integral is a beta integral, which may be expressed in terms 
of gamma functions.  Indeed, making the change of variables $s=2w-1$ gives
\eq
\begin{split}
\int_{-1}^1(s+1)^{-1/2+\gamma/2\e}(1-s)^{-1/2-iE/\e}\,ds & = 
2^{\gamma/2\e-iE/\e}\int_0^1 w^{-1/2+\gamma/2\e}(1-w)^{-1/2-iE/\e}\,dw\\
 & = 2^{\gamma/2\e-iE/\e}\frac{\Gamma\left(\frac{1}{2}+\frac{\gamma}{2\e}\right)\Gamma\left(\frac{1}{2}-\frac{iE}{\e}\right)}{\Gamma\left(1+\frac{\gamma}{2\e}-\frac{iE}{\e}\right)}
\end{split}
\endeq
using the identity
\eq
\label{beta-integral-identity}
\int_0^1 w^{a-1}(1-w)^{b-1}\,dw = \frac{\Gamma(a)\Gamma(b)}{\Gamma(a+b)}
\endeq
valid for $\Re(a),\Re(b)>0$.  Also note that, as $y\to 1$,
\eq
1-y^2 = \sech^2(x) = 2^2e^{-2x}(1+o(1)).
\endeq
Therefore, as $x\to +\infty$, 
\eq
\label{J-11-plus-for-large-x}
J_{11}^+ = C_1 e^{-iEx/\e}\left[\frac{2^{1+\gamma/2\e}i\pi\Gamma\left(\frac{1}{2}+\frac{\gamma}{2\e}\right)}{\Gamma\left(\frac{1}{2}+\frac{iE}{\e}\right)\Gamma\left(1+\frac{\gamma}{2\e}-\frac{iE}{\e}\right)} + o(1)\right].
\endeq
Comparing equations (\ref{Psi-symmetric-normalization}) and 
(\ref{J-11-plus-for-large-x}) gives the expression (\ref{C-one}) for the 
constant $C_1$.
\end{proof}

\begin{prop}
\label{Psi-minus-1-prop}
Choose the principal branches of the functions $(\pm
1-s)^{-1/2+\gamma/2\e}$ with branch cuts on the real $s$-axis from
$\pm 1$ to $+\infty$.  Also choose the principal branch of
$(y-s)^{-iE/\e-\gamma/2\e-1}$ with branch cut on the real $s$-axis
from $y$ to $+\infty$.  Take $\Sigma^-$ to be a closed contour in the
$s$-plane passing through the branch point $s=1$ and encircling $s=-1$
once in the counterclockwise direction (see figure
\ref{sigma-minus-pic}).  Then, for $z\in\mathbb{R}\backslash\{0\}$,
$\jj_1^-=:[ J_{11}^-, J_{21}^-]^\mathsf{T}$ is given by 
\eq J_{11}^- =
C_1(1-y^2)^{iE/2\e}\int_{\Sigma^-}(-1-s)^{-1/2+\gamma/2\e}(1-s)^{-1/2+\gamma/2\e}(y-s)^{-iE/\e-\gamma/2\e}\,ds
\endeq
\eq
J_{21}^- = -C_1(1-y^2)^{iE/2\e+1/2}\left(\frac{z-(\gamma+\mu)i}{z+(\gamma+\mu)i}\right)\int_{\Sigma^-}(-1-s)^{-1/2+\gamma/2\e}(1-s)^{-1/2+\gamma/2\e}(y-s)^{-iE/\e-\gamma/2\e-1}\,ds 
\endeq
with $C_1$ given by equation (\ref{C-one}).
\end{prop}
\begin{proof}
The construction follows that of $\jj_1^+$, except with $\Sigma^-$ in place of 
$\Sigma^+$, and choice of the solution 
\eq
F_1(s) = C_1(-1-s)^{-1/2+\gamma/2\e}(1-s)^{-1/2+\gamma/2\e}
\endeq
to equation (\ref{F1-diffeq}).
\end{proof}
\noindent
Now we use the system of differential equations in the form
\eqref{h-diffeq} resulting from the transformation
\eqref{h-change-of-variables} to find expressions for the Jost
solutions $\jj_2^+$ and $\jj_2^-$.
\begin{prop}
\label{Psi-plus-2-prop}
Take $(s\pm1)^{-1/2+\gamma/2\e}$, $(s-y)^{-iE/\e-\gamma/2\e-1}$, and 
$\Sigma^+$ as in Proposition \ref{Psi-plus-1-prop}.  Then, for 
$z\in\mathbb{R}\backslash\{0\}$, 
$\jj_2^+ =: [J_{12}^+, J_{22}^+]^\mathsf{T}$ is given by 
\eq
\label{Psi-plus-12}
J_{12}^+ = -C_2(1-y^2)^{-iE/2\e+1/2}\left(\frac{z+(\gamma+\mu)i}{z-(\gamma+\mu)i}\right)\int_{\Sigma^+}(s-1)^{-1/2+\gamma/2\e}(s+1)^{-1/2+\gamma/2\e}(s-y)^{iE/\e-\gamma/2\e-1}\,ds
\endeq
\eq
\label{Psi-plus-22}
J_{22}^+ = C_2(1-y^2)^{-iE/2\e}\int_{\Sigma^+}(s-1)^{-1/2+\gamma/2\e}(s+1)^{-1/2+\gamma/2\e}(s-y)^{iE/\e-\gamma/2\e}\,ds
\endeq
with
\eq
\label{C-two}
C_2=C_2(z,\e,\mu) := \frac{\Gamma\left(\frac{1}{2}-\frac{iE}{\e}\right)\Gamma\left(1+\frac{\gamma}{2\e}+\frac{iE}{\e}\right)}{2^{1+\gamma/2\e}i\pi\Gamma\left(\frac{1}{2}+\frac{\gamma}{2\e}\right)}.
\endeq
\end{prop}
\begin{proof}
The Frobenius exponents $p^-$ around $y=-1$ and $p^+$ around $y=1$ for 
$\mathbf{h}$ satisfying \eqref{h-diffeq} are 
\eq
p^- = \frac{iE}{2\e}, \hspace{.1in} -\frac{iE}{2\e}+\frac{1}{2}, \quad\quad p^+ = \frac{iE}{2\e}, \hspace{.1in} -\frac{iE}{2\e}+\frac{1}{2}.
\endeq
Therefore, defining $\mathbf{f}(y;z,\e,\mu) = [f_1(y;z,\e,\mu), 
f_2(y;z,\e,\mu)]^\mathsf{T}$ in terms of $\mathbf{h}$ by
\eq
\mathbf{h}(y) = (1-y^2)^{-iE/2\e+1/2}\mathbf{f}(y)
\endeq
has the effect of shifting one exponent to zero near each of the points 
$y=\pm 1$.  By direct 
calculation, $\mathbf{f}$ satisfies
\eq
2i\e(1-y^2)\mathbf{f}_y = \bbm \left(4E+2i\e\right)y & -2D-i\mu \\ \left(-2D+i\mu\right)(1-y^2) & 0 \ebm \mathbf{f}.
\endeq
Assume integral representations of the form
\eq
f_1(y) = \int_{\Sigma^+}  F_1(s)(s-y)^\alpha ds, \quad f_2(y) = \int_{\Sigma^+}  F_2(s)(s-y)^\beta ds.
\endeq
Proceeding as in Proposition \ref{Psi-plus-1-prop}, we obtain
\eq
\begin{split}
F_1(s) = -\left(\frac{z+(\gamma+\mu)i}{z-(\gamma+\mu)i}\right)F_2(s), \quad F_2(s) = C_2(s-1)^{-1/2+\gamma/2\e}(s+1)^{-1/2+\gamma/2\e},\\
\alpha=\beta-1, \quad \beta = \frac{iE}{\e}-\frac{\gamma}{2\e},\hspace{1.6in}
\end{split}
\endeq
where $C_2$ is a constant of integration.

The constant $C_2$ is chosen so $\jj_2^+$ is normalized as required in 
equation (\ref{Psi-symmetric-normalization}).  Starting with equation 
(\ref{Psi-plus-22}) for $J_{22}^+$, for $s\in\Sigma^+$ we have
\eq
\left\vert(s-1)^{-1/2+\gamma/2\e}(s+1)^{-1/2+\gamma/2\e}(s-y)^{iE/\e-\gamma/2\e}\right\vert \leq K|s+1|^{-1/2+\gamma/2\e}
\endeq
for some constant $K>0$.  By dominated convergence,
\eq
J_{22}^+ = C_2(1-y^2)^{-iE/2\e}\left[\int_{\Sigma^+}(s-1)^{-1/2+\gamma/2\e}(s+1)^{iE/\e-1/2}ds+o(1)\right] \text{ as } y\to 1.
\endeq
This is the same expression as equation (\ref{J-11-plus-for-large-y}) for 
$J_{11}^+$ with $C_1$ and $E$ replaced with $C_2$ and $-E$, 
respectively.  Therefore, as $y\to 1$,
\eq
\begin{split}
\label{Psi-plus-22-for-large-x}
J_{22}^+ & = C_2 e^{iEx/\e}\left[2^{1+\gamma/2\e}i\sin\left(\pi\left(\frac{1}{2}-\frac{iE}{\e}\right)\right)\frac{\Gamma\left(\frac{1}{2}+\frac{\gamma}{2\e}\right)\Gamma\left(\frac{1}{2}+\frac{iE}{\e}\right)}{\Gamma\left(1+\frac{\gamma}{2\e}+\frac{iE}{\e}\right)} + o(1)\right] \\
 & = C_2 e^{iEx/\e}\left[\frac{2^{1+\gamma/2\e}i\pi\Gamma\left(\frac{1}{2}+\frac{\gamma}{2\e}\right)}{\Gamma\left(\frac{1}{2}-\frac{iE}{\e}\right)\Gamma\left(1+\frac{\gamma}{2\e}+\frac{iE}{\e}\right)} + o(1)\right] 
\end{split}
\endeq
using the identity (\ref{reflection-identity}), which gives equation (\ref{C-two}).
\end{proof}
\begin{prop}
Define $(\pm1-s)^{-1/2+\gamma/2\e}$, $(y-s)^{-iE/\e-\gamma/2\e-1}$, and 
$\Sigma^-$ as in Proposition \ref{Psi-minus-1-prop}.  Then, for 
$z\in\mathbb{R}\backslash\{0\}$, 
$\jj_2^- =: [ J_{12}^- , J_{22}^- ]^\mathsf{T}$ is given by 
\eq
\label{Psi-minus-21}
J_{12}^- = -C_2(1-y^2)^{-iE/2\e+1/2}\left(\frac{z+(\gamma+\mu)i}{z-(\gamma+\mu)i}\right)\int_{\Sigma^-}(-1-s)^{-1/2+\gamma/2\e}(1-s)^{-1/2+\gamma/2\e}(y-s)^{iE/\e-\gamma/2\e-1}\,ds
\endeq
\eq
\label{Psi-minus-22}
J_{22}^- = C_2(1-y^2)^{-iE/2\e}\int_{\Sigma^-}(-1-s)^{-1/2+\gamma/2\e}(1-s)^{-1/2+\gamma/2\e}(y-s)^{iE/\e-\gamma/2\e}\,ds
\endeq
with $C_2$ given by equation (\ref{C-two}).
\end{prop}
\begin{proof}
The integral representation for $\jj_2^-$ is derived in the same way as the 
representation for $\jj_2^+$ in Proposition \ref{Psi-plus-2-prop}, with 
$\Sigma^-$ replacing $\Sigma^+$ and the choice 
$F_2(s) = C_2(-1-s)^{-1/2+\gamma/2\e}(1-s)^{-1/2+\gamma/2\e}$.
\end{proof}

\subsection{The scattering data}
\label{scattering-data}
We now use the integral formulae from Section \ref{scattering-data} to
calculate the scattering matrix $\mathbf{S}$, the eigenvalues
$\{z_n\}$, and (in certain special cases of interest) the
proportionality constants $\{\eta_n\}$.
\begin{prop}
The coefficient $S_{22}(z)=S_{22}(z;\e,\mu)$ is given by
\eq
\label{S22-coefficient}
S_{22}(z) =
\frac{(z-(\gamma+\mu)i)}{(z+(\gamma+\mu)i)}\cdot\frac{\left[\Gamma\left(\frac{1}{2}-\frac{iE}{\e}\right)\right]^2}{\Gamma\left(1-\frac{\gamma}{2\e}-\frac{iE}{\e}\right)\Gamma\left(\frac{\gamma}{2\e}-\frac{iE}{\e}\right)},
\quad z\in\mathbb{R}.
\endeq
\end{prop}
\begin{proof}
Take $z\in\mathbb{R}\backslash\{0\}$.  By equations (\ref{phi-plus-ito-phi-minus}) and (\ref{symmetric-Jost-solutions}),
\eq
\label{symmetric-Psi-plus-2-ito-Psi-minus}
\jj_2^+ = S_{22}\jj_1^- + S_{12}\jj_2^-.
\endeq
To determine $S_{22}$ we use 
\eq
\label{S22-from-Psi-plus-21}
S_{22} = \frac{\det\bbm\jj_2^+,\jj_2^-\ebm}{\det\bbm\jj_1^-,\jj_2^-\ebm} = -\lim_{x\to-\infty}\det\bbm\jj_2^+,\jj_2^-\ebm = \lim_{x\to-\infty}\left(J_{12}^+e^{-iEx/\e}\right).
\endeq
We now analyze $J_{12}^+$ as $x\to-\infty$.  Consider the integral 
\eq
I_1(y;z,\e,\mu):=\int_{\Sigma^+}(s-1)^{-1/2+\gamma/2\e}(s+1)^{-1/2+\gamma/2\e}(s-y)^{iE/\e-\gamma/2\e-1}\,ds.
\endeq
Making the substitution $s=-1+(y+1)v$ gives
\eq
\begin{split}
I_1 & = (y+1)^{iE/\e-1/2}\int_{\widehat{\Sigma}^+}(-2+(y+1)v)^{-1/2+\gamma/2\e}v^{-1/2+\gamma/2\e}(v-1)^{iE/\e-\gamma/2\e-1}\,dv \\
  & =: (y+1)^{iE/\e-1/2}I_2.
\end{split}
\endeq
Here we take $\widehat{\Sigma}^+$ to be the counterclockwise-oriented
contour starting at $v=0$, following the semicircle in the lower
half-plane of unit radius centered at $v=1$ to $v=2$, proceeding
along the real axis to $v=2/(y+1)$, coming back along the real axis to
$v=2$ along the top side of the branch cut, and then returning to
$v=0$ along the semicircle in the upper half-plane of unit radius
centered at $v=1$.  
\begin{figure}
        \subfigure[The contour $\widehat{\Sigma}^+$.] {
                \label{sigma_hat_plus}
                \begin{minipage}[b]{0.45\textwidth}
                  \centering
                        \includegraphics[]{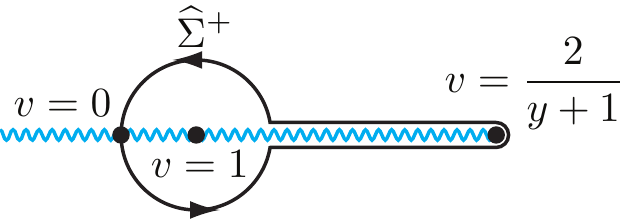}
                        \vspace{.2in}
                \end{minipage} } 
        \subfigure[The contours $\Sigma^{1+}$ and $\Sigma^{2+}$.]{
                \label{sigma_1_and_2_plus}
                \begin{minipage}[b]{0.45\textwidth}
                  \centering
                        \includegraphics[]{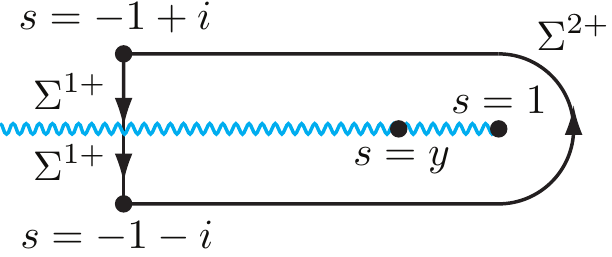}
                \end{minipage} }
              \caption{The integration contours used to calculate
                $S_{22}(z)$ and $S_{12}(z)$.}
        \label{S22-and-S12-contours}
      \end{figure}
See figure \ref{sigma_hat_plus}.  Now for
$v\in\widehat{\Sigma}^+$, 
\eq |-2+(y+1)v| =
|y+1|\cdot\left|v-\frac{2}{y+1}\right|\leq |y+1|\cdot\frac{2}{|y+1|} =
2.
\endeq
The inequality above follows because the contour $\widehat{\Sigma}^+$
lies inside the circle of radius $2/(y+1)$ centered at $v=2/(y+1)$.
Therefore, for $v\in\widehat{\Sigma}^+$ with $|v-1|=1$, 
\eq
\left\vert(-2+(y+1)v)^{-1/2+\gamma/2\e}v^{-1/2+\gamma/2\e}(v-1)^{iE/\e-\gamma/2\e-1}\right\vert
\leq K|v|^{-1/2+\gamma/2\e}
\endeq
for some constant $K>0$.  The right-hand side is 
integrable on the part of $\widehat{\Sigma}^+$ with $|v-1|=1$.  For the 
part of $I_2$ involving integration over the upper and lower edges of the 
branch cut $[2,2/(y+1)]$, we have $|-2+(y+1)v|\leq|2y|\leq 2$ and 
$\chi_{[2,2/(y+1)]}(v)\leq\chi_{[2,+\infty)}(v)$.  Therefore, since 
$-1+\gamma/\e>0$ for $\e>0$ sufficiently small,
\eq
|-2+(y+1)v|^{-1/2+\gamma/2\e}\chi_{[2,2/(y+1)]}(v) \leq 2^{-1/2+\gamma/2\e}\chi_{[2,+\infty]}(v).
\endeq
So, for $v\in[2,2/(y+1)]$,
\eq
\left\vert(-2+(y+1)v)^{-1/2+\gamma/2\e}v^{-1/2+\gamma/2\e}(v-1)^{iE/\e-\gamma/2\e-1}\right\vert \leq 2^{-1/2+\gamma/2\e}v^{-1/2+\gamma/2\e}(v-1)^{-\gamma/2\e-1},
\endeq
which is integrable on $[2,+\infty)$.  Define $\Sigma^+_\text{LHP}$ to be a 
contour going from $v=0$ to $v=2$ in the lower half-plane bounded away from 
$v=1$ and then from $v=2$ to $v=+\infty$ along the real axis.  Also define 
$\Sigma^+_\text{UHP}$ to be a contour going from $v=+\infty$ to $v=2$ along 
the real axis, and then from $v=2$ to $v=0$ in the upper half-plane, bounded 
away from $v=1$.  By dominated convergence, we may pass to the limit 
$y\to-1$ in the integrand for $I_2$:
\eq
\begin{split}
I_2(y;z,\e,\mu) = 2^{-1/2+\gamma/2\e}&\bigg[\int_{\Sigma^+_\text{LHP}}e^{i\pi(1/2-\gamma/2\e)}v^{-1/2+\gamma/2\e}(v-1)^{iE/\e-\gamma/2\e-1}\,dv \\
&+ \int_{\Sigma^+_\text{UHP}}e^{-i\pi(1/2-\gamma/2\e)}v^{-1/2+\gamma/2\e}(v-1)^{iE/\e-\gamma/2\e-1}\,dv + o(1)\bigg]
\end{split}
\endeq
as $y\to-1$.  Deforming both contours so they lie on the negative real 
axis gives
\eq
\begin{split}
I_2&=2^{1/2+\gamma/2\e}i\sin\left(\pi\left(\frac{\gamma}{2\e}+\frac{iE}{\e}\right)\right)\left[\int_{-\infty}^0(-v)^{-1/2+\gamma/2\e}
(-v+1)^{iE/\e-\gamma/2\e-1}\,dt+o(1)\right]\\
&=\frac{2^{1/2+\gamma/2\e}i\pi}{\Gamma\left(\frac{\gamma}{2\e}+\frac{iE}{\e}\right)\Gamma\left(1-\frac{\gamma}{2\e}-\frac{iE}{\e}\right)}\left[\int_{-\infty}^0(-v)^{-1/2+\gamma/2\e}(-v+1)^{iE/\e-\gamma/2\e-1}\,dt+o(1)\right]
\end{split}
\endeq
by the identity (\ref{reflection-identity}).
Using the change of variables $v=w/(w-1)$ and the identity 
(\ref{beta-integral-identity}),
\eq
\begin{split}
\int_{-\infty}^0(-v)^{-1/2+\gamma/2\e}(-v+1)^{iE/\e-\gamma/2\e-1}\,dv & = 
\int_0^1w^{-1/2+\gamma/2\e}(1-w)^{-1/2-iE/\e}\,dw \\
& = \frac{\Gamma\left(\frac{1}{2}+\frac{\gamma}{2\e}\right)\Gamma\left(\frac{1}{2}-\frac{iE}{\e}\right)}{\Gamma\left(1+\frac{\gamma}{2\e}-\frac{iE}{\e}\right)}\,.
\end{split}
\endeq
Also, as $y\to-1$,
\eq
\label{one-plus-y-small}
(1+y)^{iE/2\e} = 2^{iE/2\e}e^{iEx/\e}(1+o(1)).
\endeq
Using equations (\ref{S22-from-Psi-plus-21}), (\ref{Psi-plus-12}), (\ref{C-two}), and (\ref{one-plus-y-small}) and the factorial identity
\eq
\Gamma(1+z) = z\Gamma(z)
\endeq
therefore gives equation (\ref{S22-coefficient}).  
\end{proof}
\begin{prop}
\label{eigenvalues-prop}
The eigenvalues in the upper half of the complex $z$-plane are 
\begin{enumerate}
\item[1.] (Antikink) $z=(\sqrt{\mu^2+1}-\mu)i$,
\item[2.] (Kink-antikink pairs) $z=e^b i$ with $b\in\mathbb{R}$ satisfying
$\cosh(b) = \sqrt{\mu^2+1}-2n\e$ for $n\in\mathbb{Z}^+$ such that 
$1\leq n \leq (\sqrt{\mu^2+1}-1)/2\e$,
\item[3.] (Breathers) $z=e^{i\theta}$ with $\sin(\theta) = 
\sqrt{\mu^2+1}-2n\e$ for $n\in\mathbb{Z}^+$ such that 
$(\sqrt{\mu^2+1}-1)/2\e < n \leq \sqrt{\mu^2+1}/2\e$.
\end{enumerate}
\end{prop}
\noindent
\begin{proof}
  By general scattering theory, $S_{22}(z)$ has an analytic extension
  from the real line into the upper half-plane (see Theorem
  \ref{half-plane-lemma}), and, by definition, the eigenvalues are the
  zeros of this analytic continuation in the open upper half-plane.
  For the special case of the initial data \eqref{init-cond}, the
  analyticity of $S_{22}(z)$ can be seen from the explicit formula
  \eqref{S22-coefficient}.  Indeed, $\Gamma(z)$ has no zeros and simple poles
  at $z\in-\mathbb{Z}^+$.  It follows that
  $\Gamma(\frac{1}{2}-\frac{iE}{\e})$ and
  $\Gamma(\frac{\gamma}{2\e}-\frac{iE}{\e})$ have no poles for $z$ in
  the upper half-plane.  Therefore, the zeros of $S_{22}(z)$ are exactly
  $(\sqrt{\mu^2+1}-\mu)i$ (case 1) and the poles of
  $\Gamma(1-\frac{\gamma}{2\e}-\frac{iE}{\e})$ (cases 2 and 3).
\end{proof}

\begin{prop}
The coefficient $S_{12}(z)=S_{12}(z;t,\e,\mu)$ at $t=0$ is given by
\label{S12}
\eq
\label{S12-coefficient}
S_{12}(z) = -\frac{\Gamma(\frac{1}{2}-\frac{iE}{\e})\Gamma(\frac{1}{2}+\frac{iE}{\e})}{\Gamma(\frac{1}{2}-\frac{\gamma}{2\e})\Gamma(\frac{1}{2}+\frac{\gamma}{2\e})}, \quad z\in\mathbb{R}.
\endeq
\end{prop}
\begin{proof}
Assume $z\in\mathbb{R}\backslash\{0\}$.  Using equations 
(\ref{symmetric-Psi-plus-2-ito-Psi-minus}) and (\ref{Psi-plus-22}),
\eq
\begin{split}
S_{12}(z) = & \frac{\det\bbm\jj_1^-,\jj_2^+\ebm}{\det\bbm\jj_1^-,\jj_2^-\ebm} = -\lim_{x\to-\infty}\det\bbm\jj_1^-,\jj_2^+\ebm = -\lim_{x\to-\infty}(J_{22}^+e^{iEx/\e}) \\
     = & -\lim_{y\to-1}2^{-iE/\e}C_2\int_{\Sigma^+}(s-1)^{-1/2+\gamma/2\e}(s+1)^{-1/2+\gamma/2\e}(s-y)^{iE/\e-\gamma/2\e}ds.
\end{split}
\endeq
To analyze $J_{22}^+$ in the limit $x\rightarrow -\infty$, we begin by
deforming $\Sigma^+$ to the contour $\Sigma^{1+}\cup\Sigma^{2+}$,
where $\Sigma^{1+}$ is the contour running in a vertical line from
$-1+i$ to $-1-i$, and $\Sigma^{2+}$ is a horseshoe-shaped contour
running from $-1-i$ to 2 to $-1+i$, staying bounded away from $s=1$,
$y$, and $-1$.  See figure \ref{sigma_1_and_2_plus}.  For
$s\in\Sigma^{1+}$, 
\eq |s-y|^{-\gamma/2\e} =
|(s+1)-(y+1)|^{-\gamma/2\e} = (|s+1|^2+(y+1)^2)^{-\gamma/4\e} \leq
|s+1|^{-\gamma/2\e}
\endeq
because $\gamma/\e>0$.  
Then
\begin{equation}
\begin{split}
\left|(s-1)^{-\frac{1}{2}+\frac{\gamma}{2\e}}
(s+1)^{-\frac{1}{2}+\frac{\gamma}{2\e}}
(s-y)^{\frac{iE}{\e}-\frac{\gamma}{2\e}}\right| &
\leq  |s-1|^{-\frac{1}{2}+\frac{\gamma}{2\e}}
|s+1|^{-\frac{1}{2}+\frac{\gamma}{2\e}}
\left|(s-y)^{\frac{iE}{\e}-\frac{\gamma}{2\e}}\right| \\
    & =  |s-1|^{-\frac{1}{2}+\frac{\gamma}{2\e}}|s+1|^{-\frac{1}{2}+\frac{\gamma}{2\e}}|s-y|^{-\frac{\gamma}{2\e}}e^{-E\text{Arg}(s-y)/\e} \\
&\leq  K|s+1|^{-\frac{1}{2}+\frac{\gamma}{2\e}}
\end{split}
\end{equation}
for some constant $K>0$, and again, since $\gamma/\e>0$, the integrand 
is integrable on $\Sigma^{1+}$.  
Also, the integrand is bounded and therefore integrable on $\Sigma^{2+}$.
Thus, by dominated convergence,
\eq
S_{12}(z) = 2^{-iE/\e}C_2
\int_{\Sigma^+}(s-1)^{-1/2+\gamma/2\e}(s+1)^{-1/2+iE/\e}\,ds.
\endeq
Next, deform $\Sigma^+$ to the contour running from $-1$ to $1$ on the
real axis along the lower edge of the branch cut for
$(s-1)^{-1/2+\gamma/2\e}$ and then from $1$ to $-1$ on
the real axis along the upper edge of the branch cut.  Using the
change of variables $s=2w-1$ and equation
(\ref{beta-integral-identity}), 
\begin{multline}
\begin{aligned}
\int_{\Sigma^+}(s-1)^{-\frac{1}{2}+\frac{\gamma}{2\e}}
(s+1)^{-\frac{1}{2}+\frac{iE}{\e}}\,ds
 = & \int_{-1}^{1}e^{-i\pi(-\frac{1}{2}+\frac{\gamma}{2\e})}
(1-s)^{-\frac{1}{2}+\frac{\gamma}{2\e}}(1+s)^{-\frac{1}{2}+\frac{iE}{\e}}\,ds \\
  & \quad\quad{}+ \int_{1}^{-1}e^{i\pi(-\frac{1}{2}+\frac{\gamma}{2\e})}
(1-s)^{-\frac{1}{2}+\frac{\gamma}{2\e}}(1+s)^{-\frac{1}{2}+\frac{iE}{\e}}\,ds \\
=& 2i\sin\left(\pi\left(\frac{1}{2}-\frac{\gamma}{2\e}\right)\right)
\int_{-1}^1(1-s)^{-\frac{1}{2}+\frac{\gamma}{2\e}}(1+s)^{-\frac{1}{2}+\frac{iE}{\e}}\,ds
 \\
 = & 2^{1+\frac{\gamma}{2\e}+\frac{iE}{\e}}i
\sin\left(\pi\left(\frac{1}{2}-\frac{\gamma}{2\e}\right)\right)
\int_0^1(1-w)^{-\frac{1}{2}+\frac{\gamma}{2\e}}w^{-\frac{1}{2}+\frac{iE}{\e}}\,dw \\
= & 
2^{1+\frac{\gamma}{2\e}+\frac{iE}{\e}}i
\sin\left(\pi\left(\frac{1}{2}-\frac{\gamma}{2\e}\right)\right)
\frac{\Gamma\left(\frac{1}{2}+\frac{\gamma}{2\e}\right)
\Gamma\left(\frac{1}{2}+\frac{iE}{\e}\right)}
{\Gamma\left(1+\frac{\gamma}{2\e}+\frac{iE}{\e}\right)}.
\end{aligned}
\end{multline}
Therefore,
\eq
S_{12}(z) = -\frac{1}{\pi}\Gamma\left(\frac{1}{2}-\frac{iE}{\e}\right)\Gamma\left(\frac{1}{2}+\frac{iE}{\e}\right)\sin\left(\pi\left(\frac{1}{2}-\frac{\gamma}{2\e}\right)\right),
\endeq
which completes the proof after the use of the reflection identity 
(\ref{reflection-identity}).
\end{proof}
Proposition \ref{S12} gives immediately
\begin{prop}
  $S_{12}(z) \equiv 0$ for $\e = \e_N(\mu)$ (see
  \eqref{eq:epsilonsequence}), where $N\in\mathbb{Z}^+$.
\end{prop}
\noindent
The significance of this result, combined with Proposition
\ref{S22-and-S21-ito-S11-and-S12} relating $S_{21}(z)$ to $S_{12}(z)$,
is that $\e = \gamma, \gamma/3, \gamma/5, \dots$ gives a sequence of
values of $\e$ tending to zero for which the reflection coefficient
$\rho(z):=S_{21}(z)/S_{22}(z)$ is identically zero, and thus the
scattering data are reflectionless and the corresponding solution of
the Cauchy problem can be constructed from discrete spectral data
only.  The inverse-scattering transform may then be carried out more
or less explicitly, a calculation we will perform in Section
\ref{inverse-scattering}.

Together, the formulae \eqref{S22-coefficient} and
\eqref{S12-coefficient} show that $\rho(z)=S_{21}(z)/S_{22}(z)$
admits, in this special case of the initial data \eqref{init-cond}, a
meromorphic continuation into the upper half-plane (generally
$S_{21}(z)$ admits no continuation of any kind from the real axis
$z\in\mathbb{R}$).  The meromorphic continuation of $\rho(z)$ to the
upper half-plane that is available in this case will have poles not
only at the zeros of $S_{22}(z)$ (these are, by definition, the
eigenvalues), but also at the poles of $S_{21}(z)$.  These latter
poles are those of $\Gamma\left(\frac{1}{2}+\frac{iE}{\e}\right)$
(again using Proposition \ref{S22-and-S21-ito-S11-and-S12} to relate
$S_{21}$ to $S_{12}$); in the upper half-plane these are:
\begin{itemize}
\item (Imaginary axis) $z=e^b i$ with $b\in\mathbb{R}$ satisfying $\cosh(b)=(2n+1)\e$ for $n\in\mathbb{Z}^+$ such that $n\geq(1-\e)/2\e$.
\item (Unit circle) $z=e^{i\theta}$ with $\sin(\theta)=(2n+1)\e$ for $n\in\mathbb{Z}^+$ such that $n<(1-\e)/2\e$.
\end{itemize}
These ``phantom poles'' (poles of $S_{21}(z)$ in the upper half-plane)
are not (necessarily) eigenvalues.  However, when the reflection
coefficient is nonzero, they will affect deformations of the
Riemann-Hilbert problem that are used in asymptotic analysis.  See the
discussion at the end of Appendix \ref{appendix}.

Next we calculate the proportionality constants $\{\eta_n\}$, defined by 
$\jji_1^-(x;z) = \eta_n \jji_2^+(x;z)$ when $z$ is an eigenvalue, for the 
reflectionless cases $\e=\e_N(\mu)$.

\begin{prop}
\label{proportionality-t-zero}
Let $\e=\e_N(\mu)$ (see \eqref{eq:epsilonsequence}) where
$N\in\mathbb{Z}^+$.  Let $z$ be an eigenvalue in the closed first
quadrant and set $n=(\gamma+2iE(z))/2\e$ (note $n\in\mathbb{N}$).
Then the corresponding proportionality constant is $\eta_n = (-1)^{n-1}$.
\end{prop}
\begin{proof}
  From equation (\ref{sigma-n-def}), $\mathbf{A}\jji_1^-(x;z) = \eta_n
  \mathbf{A}\jji_2^+(x;z)$, where $z$ is the above eigenvalue indexed
  by $n$, and $\eta_n$ is its associated proportionality constant.
  Thus \eq \jj_2^-(x;z) = \eta_n \jj_2^+(x;z).
\endeq
The second entry gives in particular $J_{22}^-(x;z) = \eta_n
J_{22}^+(x;z)$.  We evaluate equations (\ref{Psi-plus-22}) and
(\ref{Psi-minus-22}) at $\e=\e_N(\mu)$.  Note that
\begin{equation}
\frac{iE(z)}{\e} - \frac{\gamma}{2\e} = n-2N-1 \quad \text{and} \quad -\frac{1}{2}+\frac{\gamma}{2\e} = N.
\end{equation}
With these substitutions, equation (\ref{Psi-plus-22}) takes the form
\eq
J_{22}^+\left(x;z,\e_N(\mu),\mu\right) = 
C_2(1-y^2)^{-(n-2N-1)/2}\int_{\Sigma^+}(s+1)^N(s-1)^N(s-y)^{n-2N-1}\,ds.
\endeq
Since $N$ is a nonnegative integer, we may deform the contour $\Sigma^+$ away 
from $s=-1$ and $s=1$ to a small circle $\Sigma^y$ around $s=y$.  Thus
\eq
\label{J22plus}
J_{22}^+\left(x;z,\e_N(\mu),\mu\right) = C_2(1-y^2)^{-(n-2N-1)/2}\int_{\Sigma^y}(s+1)^N(s-1)^N(s-y)^{n-2N-1}\,ds.
\endeq
Likewise, equation (\ref{Psi-minus-22}) becomes
\begin{equation}
\begin{split}
J_{22}^-\left(x;z,\e_N(\mu),\mu\right) & =  
C_2(1-y^2)^{-(n-2N-1)/2}\int_{\Sigma^-}(-1-s)^N(1-s)^N(y-s)^{n-2N-1}\,ds \\
 & = (-1)^{n-1}C_2(1-y^2)^{-(n-2N-1)/2}\int_{\Sigma^y}(s+1)^N(s-1)^N(s-y)^{n-2N-1}
\,ds \\
 & = (-1)^{n-1}J_{22}^+\left(x;z,\e_N(\mu),\mu\right),
\end{split}
\end{equation}
and so $\eta_n = (-1)^{n-1}$ by comparison with equation 
\eqref{J22plus}.
\end{proof}
The results for the scattering data
are summarized in Theorem 
\ref{scattering-data-theorem}.
\begin{thm}
\label{scattering-data-theorem}
The scattering data for the sine-Gordon equation \eqref{sine-Gordon}
at $t=0$ with initial condition \eqref{init-cond} are as follows.
\begin{equation}
S_{22}(z)=
\frac{(z-(\gamma+\mu)i)}{(z+(\gamma+\mu)i)}\cdot
\frac{\left[\Gamma\left(\frac{1}{2}-\frac{iE}{\e}\right)\right]^2}
{\Gamma\left(1-\frac{\gamma}{2\e}-\frac{iE}{\e}\right)
\Gamma\left(\frac{\gamma}{2\e}-\frac{iE}{\e}\right)}\,,
\end{equation}
\begin{equation}
  S_{12}(z) = 
-\frac{\Gamma(\frac{1}{2}-\frac{iE}{\e})
\Gamma(\frac{1}{2}+\frac{iE}{\e})}{\Gamma(\frac{1}{2}-\frac{\gamma}{2\e})
\Gamma(\frac{1}{2}+\frac{\gamma}{2\e})}\,,
\end{equation}
and
\begin{equation}
  S_{11}(z) = S_{22}(-z)\,, \quad \quad S_{21}(z) = -S_{12}(-z)\,.
\end{equation}
The eigenvalues in the upper half-plane are 
\begin{enumerate}
\item[1.] (Antikink) $z=(\sqrt{\mu^2+1}-\mu)i$,
\item[2.] (Kink-antikink pairs) $z=e^b i$ on the imaginary axis with 
$b\in\mathbb{R}$ satisfying
$\cosh(b) = \sqrt{\mu^2+1}-2n\e$ for each $n\in\mathbb{Z}^+$ such that $1\leq n \leq (\sqrt{\mu^2+1}-1)/2\e$,
\item[3.] (Breathers) $z=e^{i\theta}$ on the unit circle with $\sin(\theta) = \sqrt{\mu^2+1}-2n\e$ for each $n\in\mathbb{Z}^+$ satisfying $(\sqrt{\mu^2+1}-1)/2\e < n \leq \sqrt{\mu^2+1}/2\e$.
\end{enumerate}
The eigenvalues are generically (with respect to $\mu\in\mathbb{R}$
and $\e>0$) all simple.  The scattering data are reflectionless for
$\e=\e_N(\mu)$ where $N\in\mathbb{Z}^+$.  In the reflectionless cases,
the proportionality constants are $\eta_n =
(-1)^{n-1}$,
where $n=(\gamma+2iE(z))/2\e\in\mathbb{N}$
for any eigenvalue $z=z_n$ in the upper half-plane, and the modified
proportionality constants are $c_n^0=\eta_n/S_{22}'(z_n)$ where
\begin{equation}
\label{S22-prime}
S_{22}'(z_n) = \begin{cases} \displaystyle
\frac{(N!)^2}{2i(\gamma+\mu)(2N)!}\,, & n=0 \\ 
\displaystyle
(-1)^n\frac{i}{4\gamma}(2N+1)
\frac{[(N-n)!]^2(n-1)!}{(2N-n)!}
\frac{z_n-(\gamma+\mu)i}{z_n+(\gamma+\mu)i}
\left(1+\frac{1}{z_n^2}\right)\,, & 
n>0\,. 
\end{cases}
\end{equation}
\end{thm}
\noindent{\it Remark.} More generally, for reflectionless potentials,
\eq S_{22}'(z_n) = \prod_{k\neq n}(z_n-z_k)\prod_k(z_n-z_k^*)^{-1},
\label{eq:S22-prime-general}
\endeq
where the product runs over all eigenvalues (presumed simple) in the
upper half-plane.  From the point of view of the numerical
inverse-scattering method used in this paper, the specialized formulae
\eqref{S22-prime} (which are adapted to the initial data
\eqref{init-cond}) are especially useful because many of the factors in
\eqref{eq:S22-prime-general} involve differences of nearly-equal
numbers which lead to loss of accuracy in finite precision arithmetic,
while the products in \eqref{eq:S22-prime-general} have been converted
into products of integers in \eqref{S22-prime} that can be evaluated
with exact arithmetic.  On the other hand, the more general formula
\eqref{eq:S22-prime-general} leads to a representation of the modified
proportionality constants $\{c_n^0\}$ as residues of a meromorphic
function, and such a representation is useful in the context of
deformations introduced to study the asymptotic ($N\rightarrow\infty$)
behavior of the meromorphic Riemann-Hilbert problem of reflectionless
inverse scattering \cite{Kamvissis:2003-book}.

According to Theorem~\ref{scattering-data-theorem}, the eigenvalues
for the sine-Gordon problem with initial data \eqref{init-cond} lie on
the imaginary axis and the unit circle.  Those on the positive
imaginary axis come in pairs symmetric with respect to reflection
through the unit circle, except for a single distinguished eigenvalue
at $(\sqrt{\mu^2+1}-\mu)i$.  This eigenvalue contributes the net
topological charge $-1$ of the solution $u$.  The eigenvalues on the
unit circle have imaginary parts that are equally spaced.  
\begin{figure}[h]
        \subfigure[$\mu=1$, $N=4$] {
                \label{eigenvalues-N4-mu1}
                \begin{minipage}[b]{0.31\textwidth}
                  \centering
                        \includegraphics[width=2in,height=1.9in]{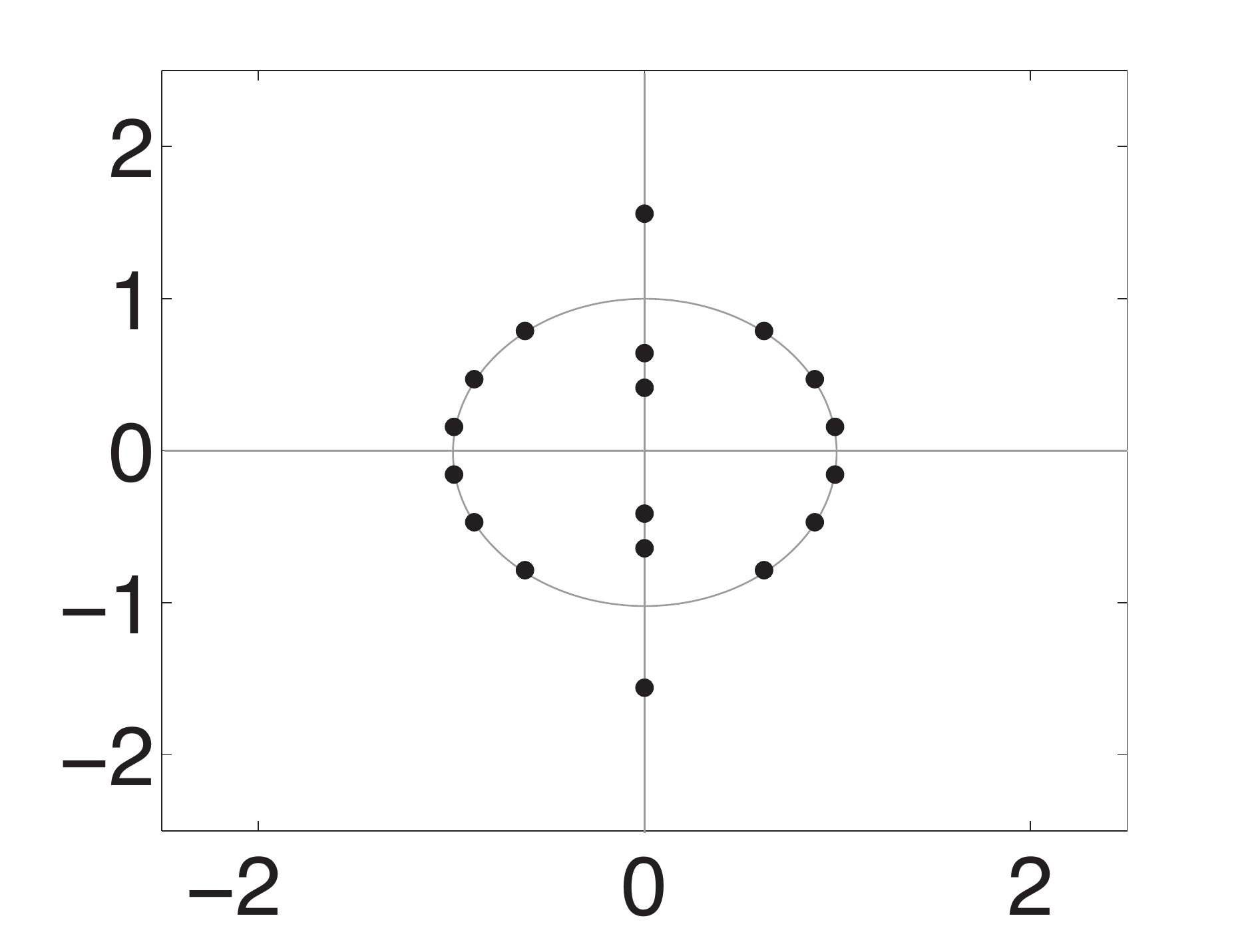}
                      \end{minipage} } 
        \subfigure[$\mu=1$, $N=8$]{
                \label{eigenvalues-N8-mu1}
                \begin{minipage}[b]{0.31\textwidth}
                  \centering
                        \includegraphics[width=2in,height=1.9in]{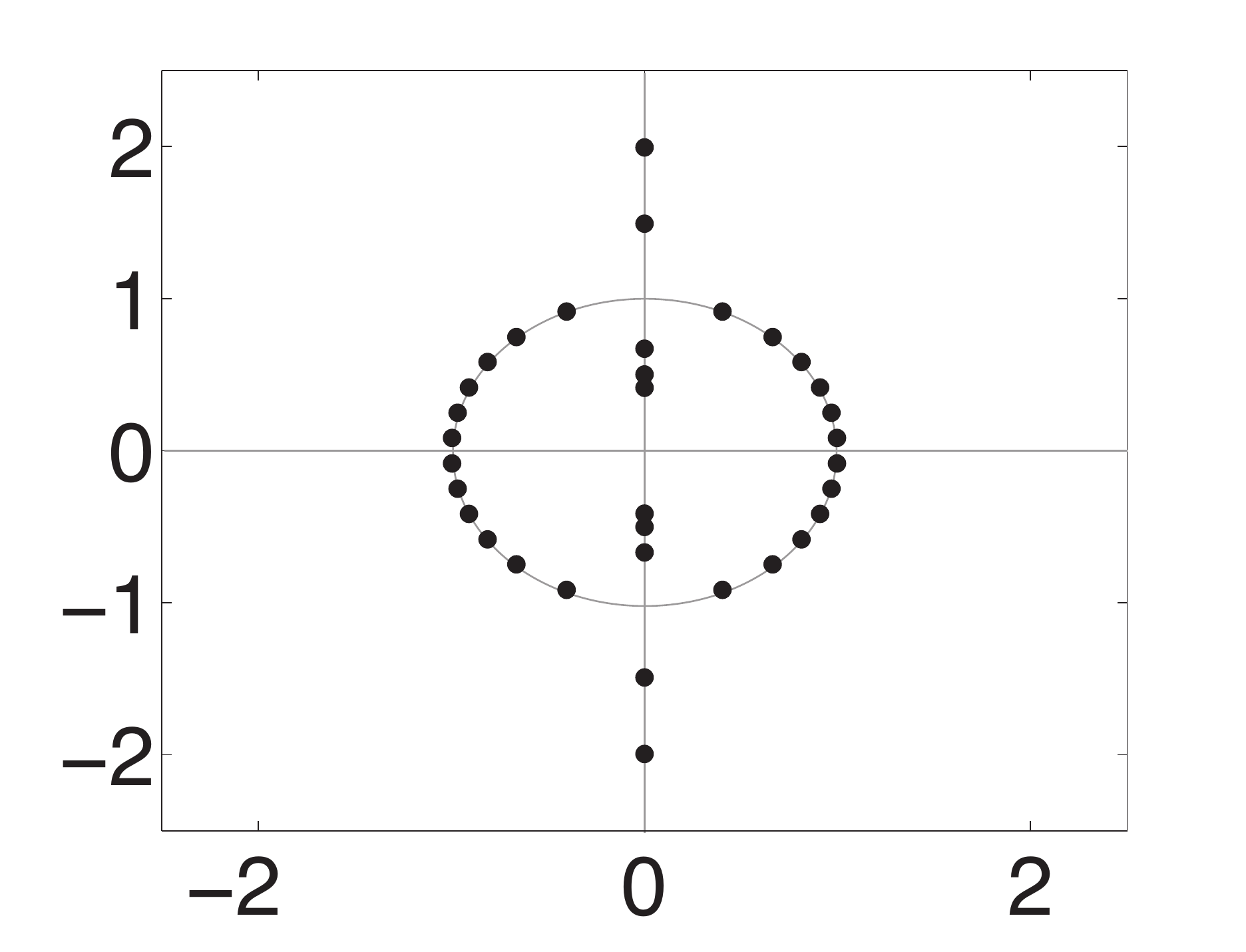}
                \end{minipage} }
        \subfigure[$\mu=1$, $N=16$]{
                \label{eigenvalues-N16-mu1}
                \begin{minipage}[b]{0.31\textwidth}
                  \centering
                        \includegraphics[width=2in,height=1.9in]{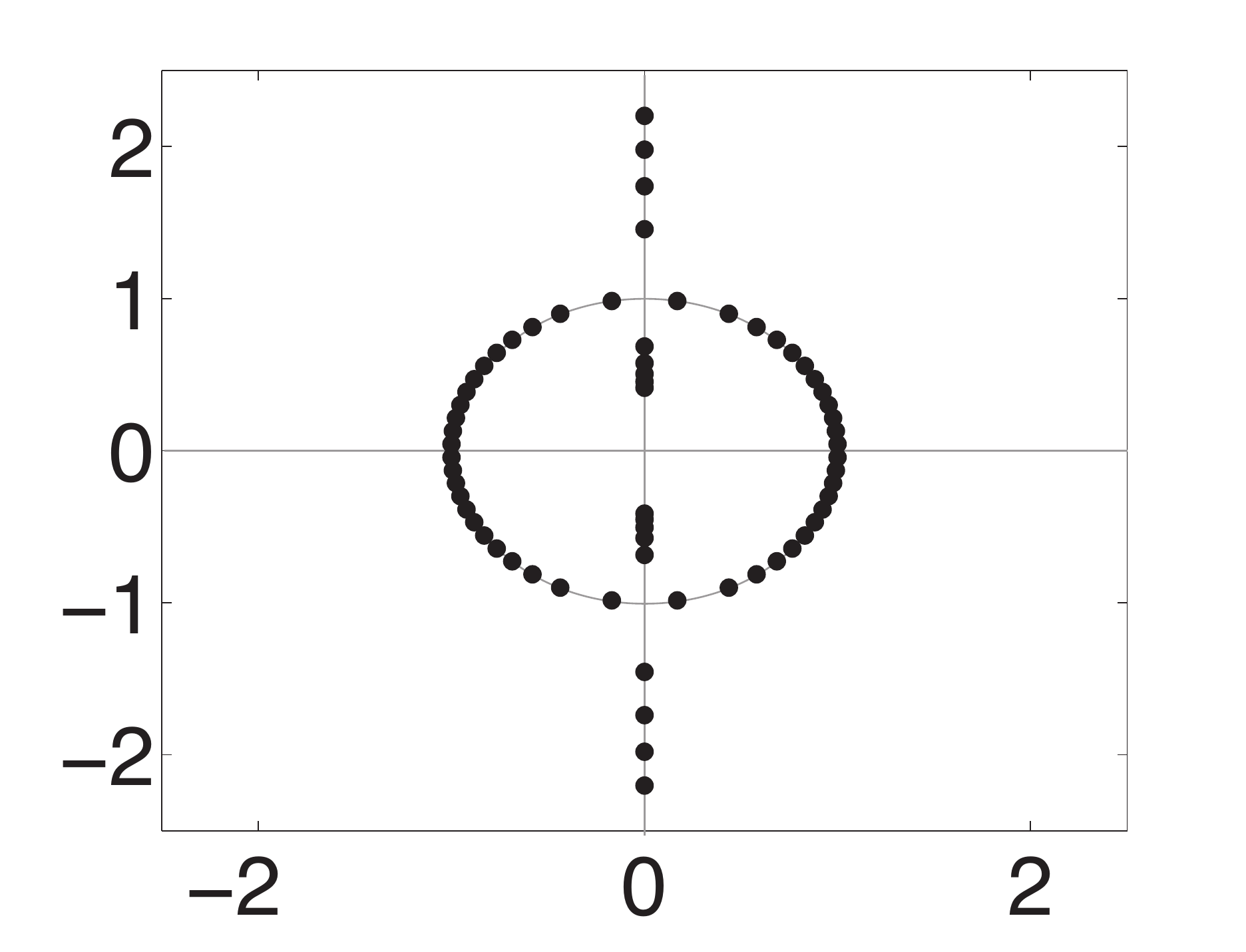}
                \end{minipage} }
              \caption{Eigenvalues for reflectionless cases
                $\e=\e_N(\mu)$ (see \eqref{eq:epsilonsequence}) with
                fixed $\mu=1$ and varying $N$.  The gray circle is
                $|z|=1$.}
        \label{eigenvalues-pic}
\end{figure}
The plots in figure \ref{eigenvalues-pic} illustrate the location of
eigenvalues for $\mu=1$ as $\e>0$ is varied.  If $\mu$ is varied as a
parameter, pairs of eigenvalues on the unit circle in the upper
half-plane corresponding to a single breather may collide at $z=i$ and
bifurcate off onto the imaginary axis, forming a kink-antikink pair
\begin{figure}[h]
        \subfigure[$\mu=0$, $N=4$] {
                \label{eigenvalues-N4-mu0}
                \begin{minipage}[b]{0.31\textwidth}
                  \centering
                        \includegraphics[width=2in,height=1.9in]{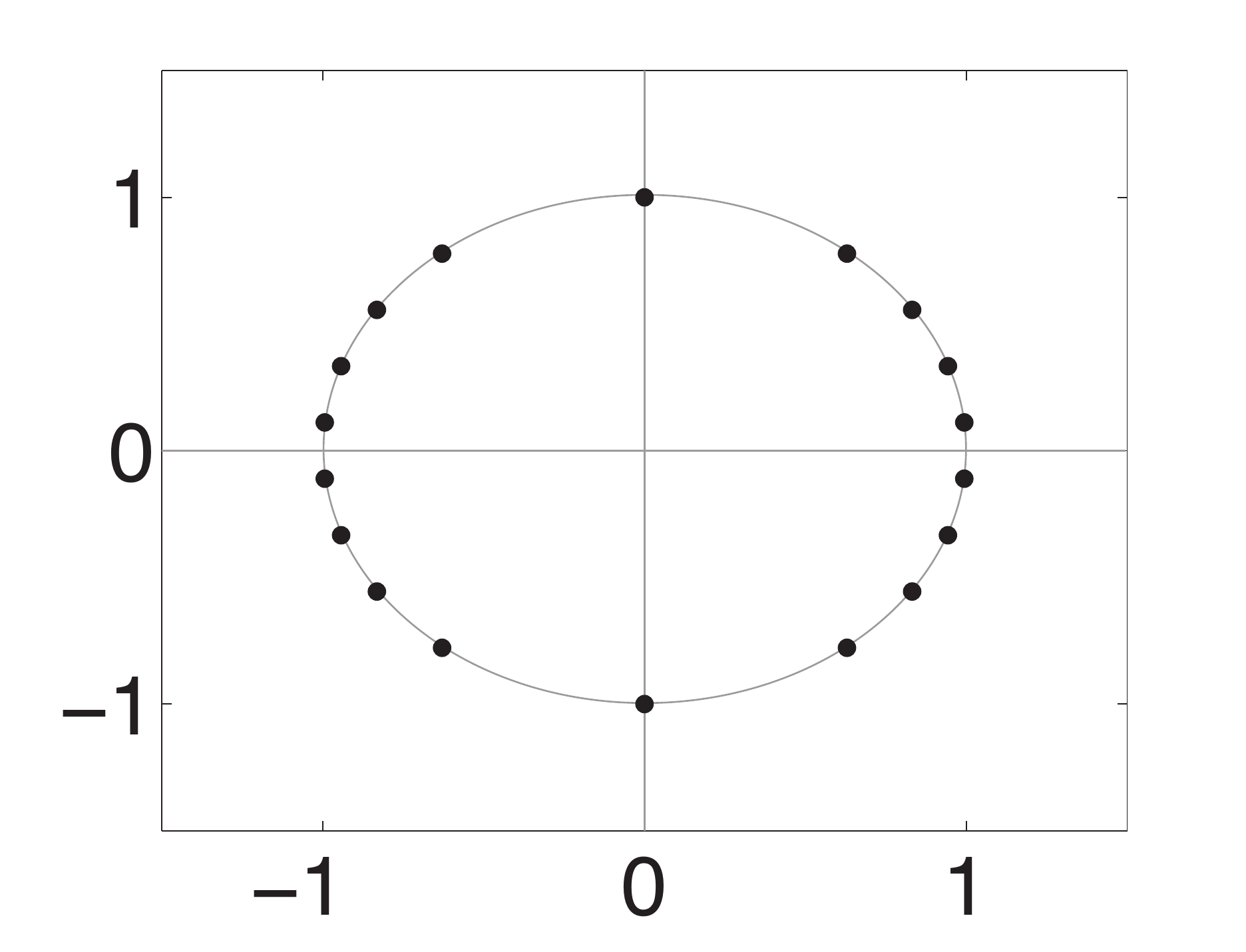}
                      \end{minipage} } 
        \subfigure[$\mu=0.8$, $N=4$]{
                \label{eigenvalues-N4-mu0p8}
                \begin{minipage}[b]{0.31\textwidth}
                  \centering
                        \includegraphics[width=2in,height=1.9in]{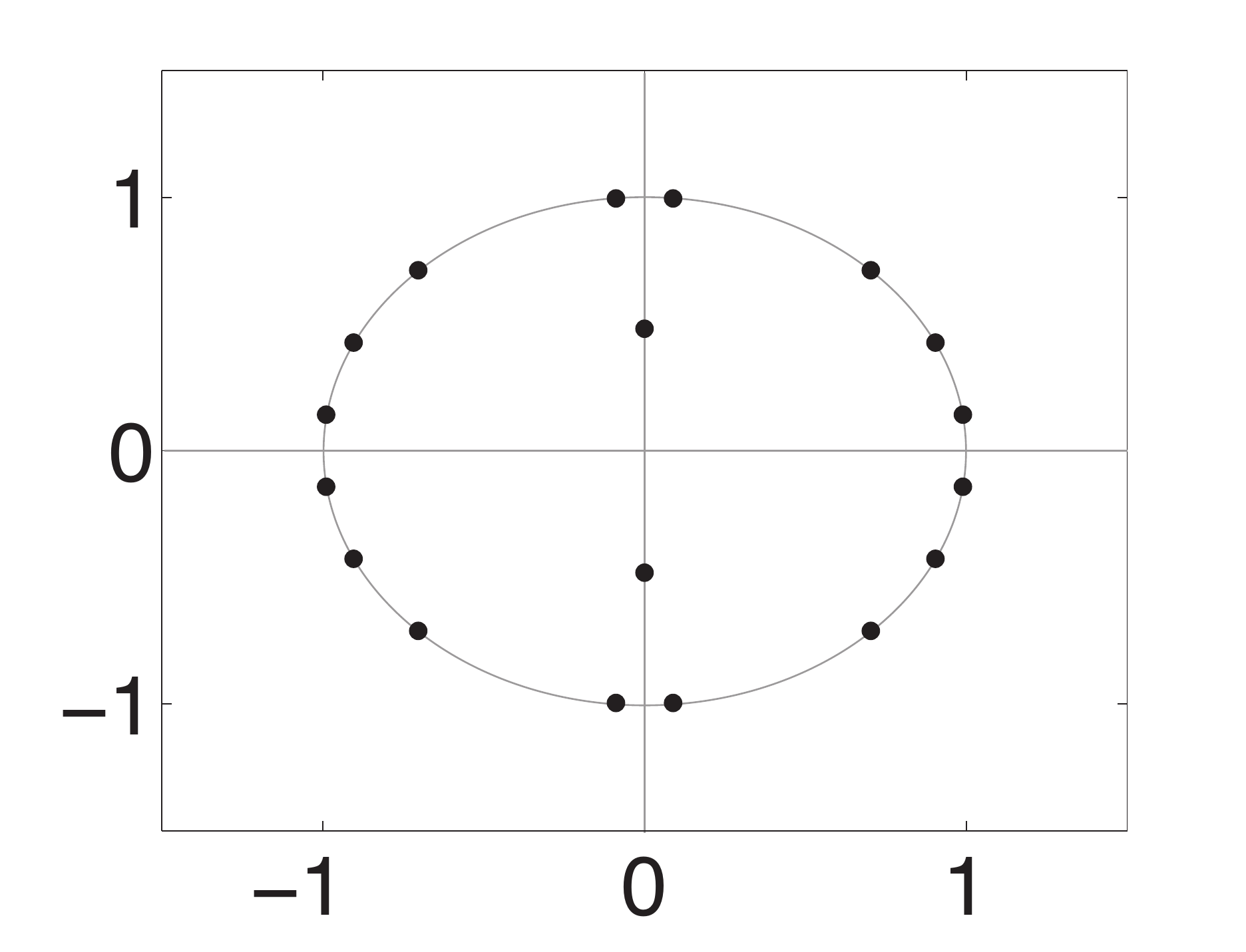}
                \end{minipage} }
        \subfigure[$\mu=0.82$, $N=4$]{
                \label{eigenvalues-N4-mu0p82}
                \begin{minipage}[b]{0.31\textwidth}
                  \centering
                        \includegraphics[width=2in,height=1.9in]{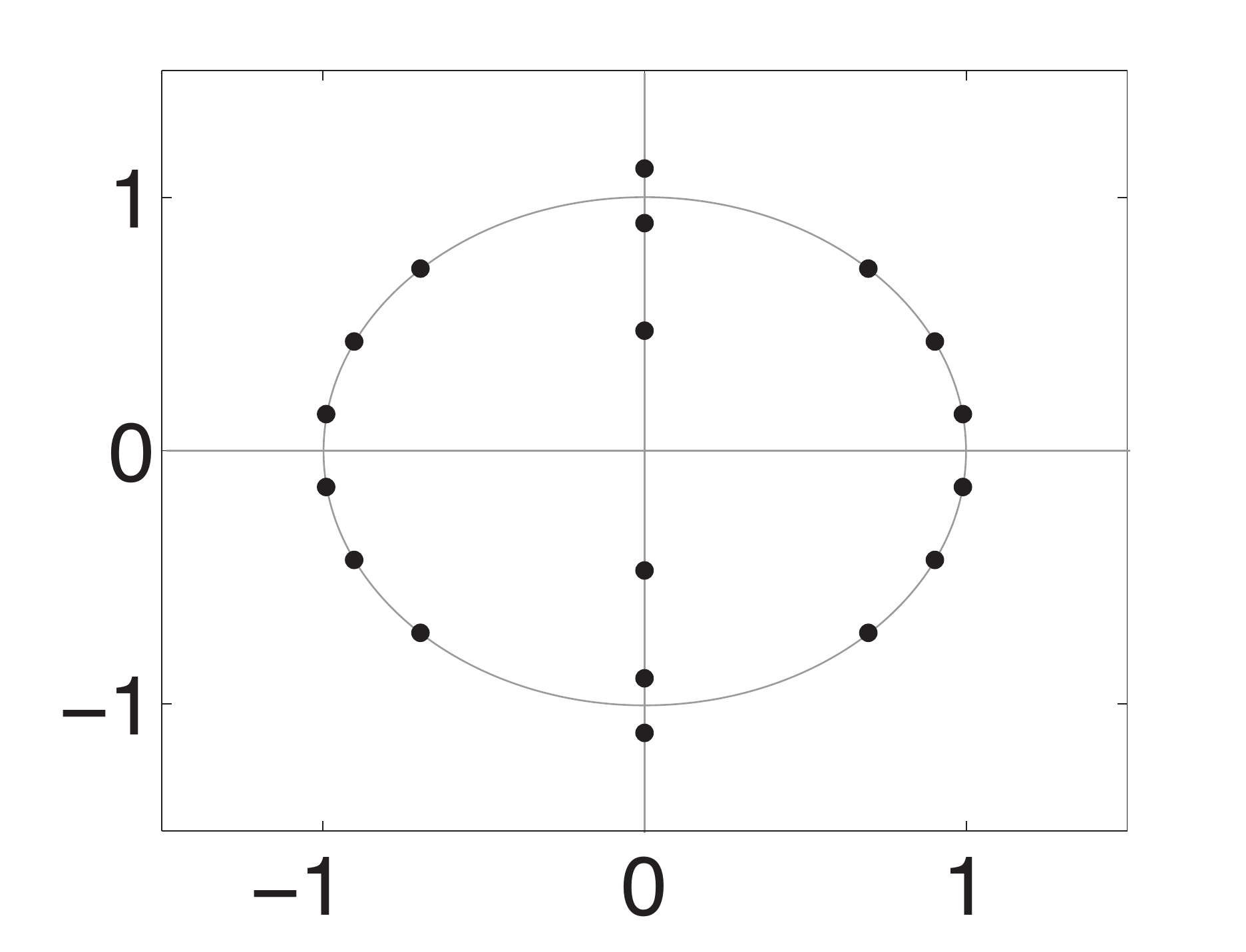}
                \end{minipage} }
              \caption{Bifurcation of a breather quartet into a
                kink-antikink pair in reflectionless cases with
                $\e=\e_N(\mu)$ (see \eqref{eq:epsilonsequence})
                holding $N$ fixed and varying $\mu$.  The gray circle
                is $|z|=1$.}
        \label{eigenvalue-bifurcation}
\end{figure}
(see figure \ref{eigenvalue-bifurcation}).  Note that this bifurcation
preserves the total topological charge of $u$.  Note also that for
$\e$ and $\mu$ such that $(\gamma-1)/2\e\in\{1,2,3,\dots\}$, there
exist double eigenvalues at $z=\pm i$.  The existence of eigenvalues
with algebraic multiplicity greater than one is worth noting.  For
instance, the self-adjoint Schr\"odinger eigenvalue problem associated
with the Korteweg-de Vries equation admits only simple eigenvalues.

\section{Inverse-scattering for reflectionless initial data}
\label{inverse-scattering}
We now reconstruct the matrix $\GG(z)=\GG(z;x,t,\e,\mu)$ (see equation
\eqref{m}) corresponding to the specific initial conditions
(\ref{init-cond}) from the exact scattering data given in
Theorem~\ref{scattering-data-theorem} in the reflectionless case when
$\rho(z):=S_{21}(z)/S_{22}(z)\equiv 0$.  We therefore fix
$N\in\mathbb{Z}^+$ and set $\e=\e_N(\mu)$ (see
\eqref{eq:epsilonsequence}).  We also assume the condition that
$(\gamma-1)/2\e\notin\{1,2,3,\dots\}$ so all the eigenvalues (poles of
$\GG(z)$) are simple.  Define 
\eq
\label{pairnumber}
M:=\left\lfloor\frac{1}{2\e}(\sqrt{\mu^2+1}-1)\right\rfloor 
\endeq
as the number of kink-antikink eigenvalue pairs.  Label 
the eigenvalues in the closed first quadrant as follows:
\begin{enumerate}
\item[1.] $z_0 = (\sqrt{\mu^2+1}-\mu)i$
\item[2.] $z_n = ie^b$ and $\widetilde{z}_n = ie^{-b}$ for $b=\text{arccosh}(\sqrt{\mu^2+1}-2n\e)$, $1\leq n \leq M$, $n\in\mathbb{Z}^+$ 
\item[3.] $z_n=e^{i\theta}$ for $\theta = \arcsin(\sqrt{\mu^2+1}-2n\e)$, $M+1 \leq n \leq N$, $n\in\mathbb{Z}^+$.
\end{enumerate}
Note that for the purely imaginary eigenvalues in case 2, the meaning of our 
notation is that $|z_n|>1$ while $|\widetilde{z}_n|<1$.  In this 
case, $-z_n$ and $-\widetilde{z}_n$ are also eigenvalues.  
In case 3 on the unit circle, if $z_n$ is an eigenvalue then $-z_n^*$, 
$-z_n$, and $z_n^*$ are also eigenvalues.  

\subsection{Numerical linear algebra algorithm for reflectionless potentials}
We use the conditions of the Riemann-Hilbert problem of inverse
scattering (see Appendix~\ref{appendix}) to determine the matrix
$\GG(z)$ for $(x,t)\in\mathbb{R}^2$.  In any reflectionless
inverse-scattering problem, $\GG(z)$ has no jump discontinuity across
the real $z$-axis, and is therefore a meromorphic function with poles
at the eigenvalues.  With the assumption that the poles are simple, we
may therefore expand $\GG(z)$ in partial fractions as
\begin{equation}
\label{M-partial-fractions}
\begin{split}
\GG(z) & = \mathbb{I} + \frac{1}{z-z_0}\KK_0^U + \frac{1}{z+z_0}\KK_0^L \\
 & \quad\quad{}+ \sum_{n=1}^M\left\{\frac{1}{z-z_n}\KK_n^U + 
\frac{1}{z-\widetilde{z}_n}\widetilde{\KK}_n^U + 
\frac{1}{z+z_n}\KK_n^L + 
\frac{1}{z+\widetilde{z}_n}\widetilde{\KK}_n^L\right\} \\
 & \quad\quad{}+ \sum_{n=M+1}^N\left\{\frac{1}{z-z_n}\BB_n^I + 
\frac{1}{z+z_n^*}\BB_n^{I\!I} + \frac{1}{z+z_n}\BB_n^{I\!I\!I} + 
\frac{1}{z-z_n^*}\BB_n^{IV}\right\}.
\end{split}
\end{equation}
The superscripts on the constant matrices $\KK_n^U$, 
$\widetilde{\KK}_n^U$, $\KK_n^L$, and $\widetilde{\KK}_n^L$ associated with 
the kink or antikink eigenvalues indicate if the 
associated eigenvalue is in the upper or 
lower half-plane, and the superscripts on the constant matrices $\BB_n^I$, 
$\BB_n^{I\!I}$, $\BB_n^{I\!I\!I}$, and $\BB_n^{IV}$ associated with the 
breather eigenvalues indicate the quadrant of 
the associated eigenvalue.  The residue conditions 
(\ref{residue-conditions}) show immediately that the second columns of 
$\KK_n^U$, $\widetilde{\KK}_n^U$, $\BB_n^I$, and $\BB_n^{I\!I}$ and the 
first columns of  $\KK_n^L$, $\widetilde{\KK}_n^L$, $\BB_n^{I\!I\!I}$, and 
$\BB_n^{IV}$ vanish for all $n$.  Write
\eq
\KK_n^U = \bbm p_n^U & 0 \\ q_n^U & 0 \ebm, \quad \widetilde{\KK}_n^U = \bbm \widetilde{p}_n^U & 0 \\ \widetilde{q}_n^U & 0 \ebm, \quad \KK_n^L = \bbm 0 & p_n^L \\ 0 & q_n^L \ebm, \quad \widetilde{\KK}_n^L = \bbm 0 & \widetilde{p}_n^L \\ 0 & \widetilde{q}_n^L \ebm
\endeq
for $1\leq n \leq M$ and $n=0$ when applicable, and 
\eq
\BB_n^I = \bbm p_n^I & 0 \\ q_n^I & 0 \ebm, \quad \BB_n^{I\!I} = \bbm p_n^{I\!I} & 0 \\ q_n^{I\!I} & 0 \ebm, \quad \BB_n^{I\!I\!I} = \bbm 0 & p_n^{I\!I\!I} \\ 0 & q_n^{I\!I\!I} \ebm, \quad \BB_n^{IV} = \bbm 0 & p_n^{IV} \\ 0 & q_n^{IV} \ebm.
\endeq
for $M+1\leq n \leq N$.  From the symmetries of equation 
(\ref{phi-symmetries}) it follows that
\begin{eqnarray}
\notag
\KK_n^U = -\sigma_2\KK_n^L\sigma_2, \quad 0\leq n \leq M \hspace{1in}\\
\widetilde{\KK}_n^U = -\sigma_2\widetilde{\KK}_n^L\sigma_2, \quad 1\leq n \leq M \hspace{1in}\\
\notag
\BB_n^I = -\BB_n^{I\!I*} = -\sigma_2\BB_n^{I\!I\!I}\sigma_2 = \sigma_2\BB_n^{IV*}\sigma_2, \quad M+1\leq n \leq N.
\end{eqnarray}
Note that the elements of $\KK_n^U$, $\widetilde{\KK}_n^U$, $\KK_n^L$,
and $\widetilde{\KK}_n^L$ are all imaginary.  These symmetries show
that the elements of the second row of $\GG(z)$ can be expressed in
terms of the elements of the first row, so to build $\GG(z)$ it
is sufficient to find the first row.  Moreover, according to
Proposition \ref{cos-u-from-m}, the potential $u$ may be recovered
from the first row of $\GG(z)$, and in terms of the partial-fraction
expansion \eqref{M-partial-fractions} this results in the formulae
\eq
\label{cosu-from-P}
\cos(u) = 1 - 2\left(\sum_{n=0}^M\frac{p_n^L}{z_n}+\sum_{n=M+1}^N\frac{p_n^{I\!I\!I}}{z_n}+\sum_{n=1}^M\frac{\widetilde{p}_n^L}{\widetilde{z}_n}+\sum_{n=M+1}^N\frac{p_n^{IV}}{-z_n^*}\right)^2
\endeq
\eq
\label{sinu-from-P}
\begin{split}
\sin(u) = -2\left(1+\sum_{n=0}^M\frac{p_n^U}{-z_n}+\sum_{n=M+1}^N\frac{p_n^I}{-z_n}+\sum_{n=1}^M\frac{\widetilde{p}_n^U}{-\widetilde{z}_n}+\sum_{n=M+1}^N\frac{p_n^{I\!I}}{z_n^*}\right)\hspace{.3in}\\
\cdot\left(\sum_{n=0}^M\frac{p_n^L}{z_n}+\sum_{n=M+1}^N\frac{p_n^{I\!I\!I}}{z_n}+\sum_{n=1}^M\frac{\widetilde{p}_n^L}{\widetilde{z}_n}+\sum_{n=M+1}^N\frac{p_n^{IV}}{-z_n^*}\right).
\end{split}
\endeq
Recall that each eigenvalue $z_n$ in the upper half-plane has an associated
modified proportionality constant $c_n$ which depends parametrically on $x$
and $t$ via an exponential factor.  We denote by $\{\widetilde{c}_n\}$ those
modified proportionality constants associated with the eigenvalues labeled
$\{\widetilde{z}_n\}$.
Define the vectors
\eq
\begin{split}
\mathbf{a} &:= 
[c_0,c_1,\dots,c_N,\widetilde{c}_1,\dots,
\widetilde{c}_M,-c_{M+1}^*,\dots,-c_N^*]^\mathsf{T}\,, \\
\mathbf{w} &:= [z_0,z_1,\dots,z_N,\widetilde{z}_1,\dots,
\widetilde{z}_M,-z_{M+1}^*,\dots,-z_N^*]^\mathsf{T}\,,\\
\mathbf{p}^{(1)} & := [p_0^U,\dots,p_M^U,p_{M+1}^I,\dots,
p_N^I,\widetilde{p}_1^U,\dots,\widetilde{p}_M^U,p_{M+1}^{I\!I},\dots,
p_N^{I\!I}]^\mathsf{T}\,,\\
\mathbf{p}^{(2)} &:= [p_0^L,\dots,p_M^L,p_{M+1}^{I\!I\!I},\dots,
p_N^{I\!I\!I},\widetilde{p}_1^L,\dots,\widetilde{p}_M^L,p_{M+1}^{IV},\dots,
p_N^{IV}]^\mathsf{T}\,. 
\end{split}
\endeq
Applying the residue conditions (\ref{residue-conditions}) to the partial 
fraction expansion \eqref{M-partial-fractions} yields a linear inhomogeneous 
system for $\mathbf{p}^{(1)}$ and $\mathbf{p}^{(2)}$:
\eq
\label{Phat-equation}
\bbm \mathbb{I}_{2N+1} & \mathbf{F} \\ -\mathbf{F} & \mathbb{I}_{2N+1} \ebm 
\bbm \mathbf{p}^{(1)}\\\mathbf{p}^{(2)}\ebm = 
\bbm \mathbf{0}_{2N+1} \\ \mathbf{a} \ebm\,, \quad\text{where}\quad 
F_{ij} := -\frac{a_i}{w_i+w_j}\,.
\endeq
Here $\mathbf{0}_{2N+1}$ is the vector of zeros of length $2N+1$ and 
$\mathbb{I}_{2N+1}$ is the $2N+1$ by $2N+1$ identity matrix.  The
$(x,t)$-dependence of the coefficient matrix and the right-hand side of
this linear system enters only through the modified proportionality constants
making up the vector $\mathbf{a}$.
Eliminating $\mathbf{p}^{(1)}$ using the first (block) row gives 
$\mathbf{p}^{(1)}=-\mathbf{F}\mathbf{p}^{(2)}$, and 
the resulting system for $\mathbf{p}^{(2)}$ is
\eq
\label{Phat2-equation}
(\mathbb{I}+\mathbf{F}^2)\mathbf{p}^{(2)} = \mathbf{a}.
\endeq

With the explicit use of the discrete scattering data, all of the
entries of $\mathbf{a}$ and $\mathbf{F}$ are known functions of $x$
and $t$.  Thus, for any choice of $x=x_0$ and $t=t_0$, the system 
(\ref{Phat-equation}) can be solved numerically, giving (via equations
(\ref{cosu-from-P}) and (\ref{sinu-from-P})) the value of $u(x_0,t_0)$
{\it independently} of the value of $u(x,t)$ at any other $x$ or $t$
values.

\subsection{Numerical results}
Here we apply this procedure to study the solution of the Cauchy
problem for the sine-Gordon equation \eqref{sine-Gordon} subject to
the initial data \eqref{init-cond} for various values of the
parameters $\mu$ and $\e$ that make the scattering data reflectionless
(so that $\e=\e_N(\mu)$ for some integer $N$).  We are especially
interested in the limit of large $N$, as this corresponds to the
semiclassical limit of $\e\downarrow 0$.  

For large $N$, the system \eqref{Phat2-equation} is poorly
conditioned, and it is therefore necessary to compute
$\mathbb{I}+\mathbf{F}^2$ and $\mathbf{a}$ with high
precision at a given pair $(x,t)\in\mathbb{R}^2$ to find
$\mathbf{p}^{(2)}$ and hence $u(x,t)$ to even a few decimal places of
accuracy.  For instance, for $N=16$, $\mu=1$, $x=0$, and $t=5$, the
condition number of $\mathbb{I}+\mathbf{F}^2$ is approximately
$3.5\times 10^{125}$, and it is necessary to use approximately 125-135
digit precision to accurately compute $u$.

\begin{figure}[h]
        \subfigure[$\mu=0$, $N=0$ ($\e=1$)] {
                \label{cosu_N0_mu0}
                \begin{minipage}[b]{0.45\textwidth}
                  \centering
                        \includegraphics[width=2.1in]{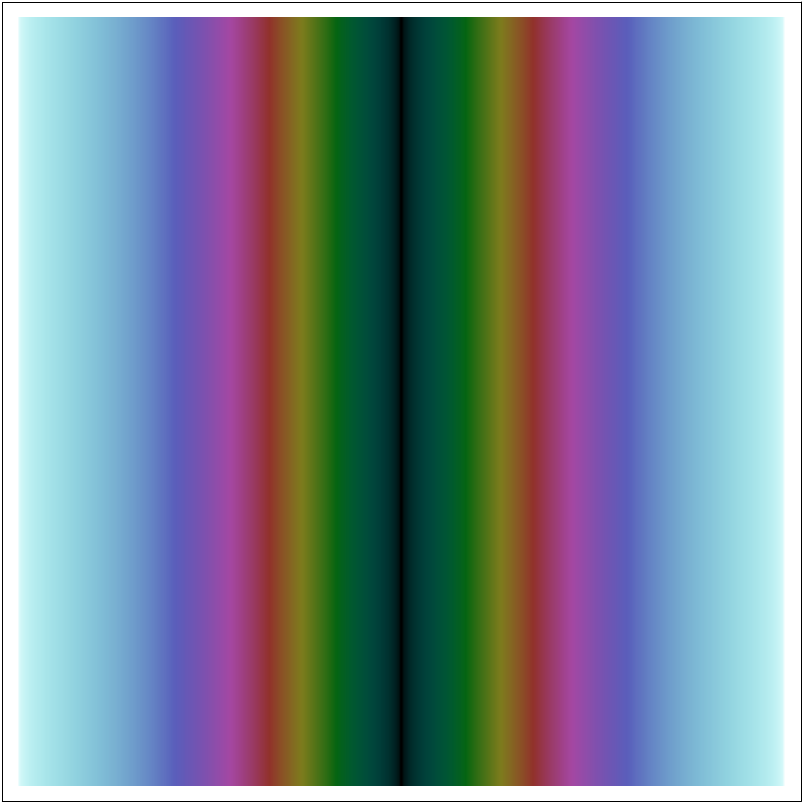}
                \end{minipage} } 
        \subfigure[$\mu=0$, $N=1$ $\left(\e=1/3\right)$]{
                \label{cosu_N1_mu0}
                \begin{minipage}[b]{0.45\textwidth} 
                  \centering
                        \includegraphics[width=2.1in]{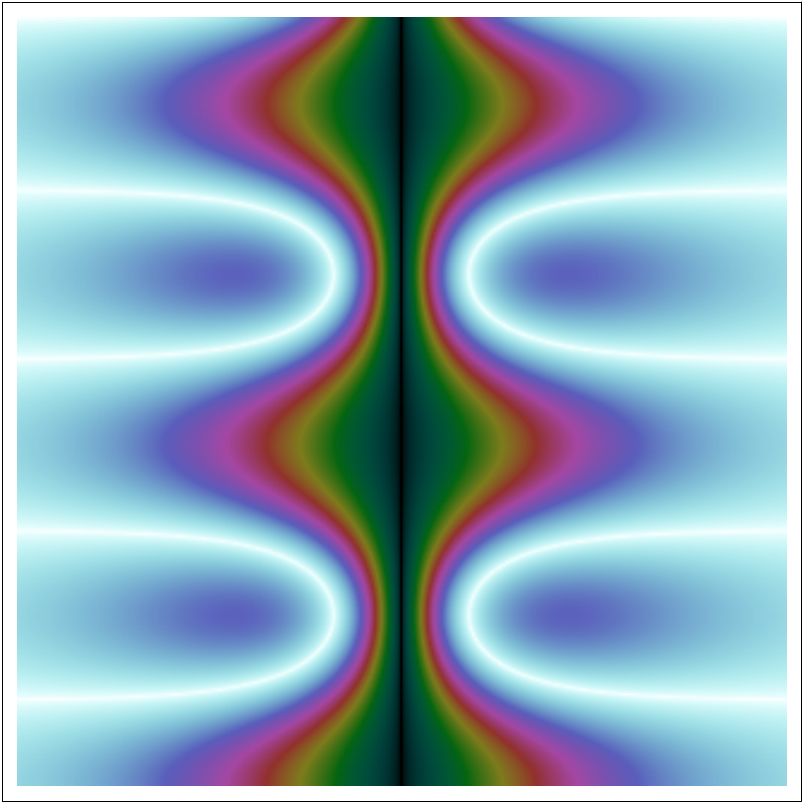}
                \end{minipage} }
        \subfigure[$\mu=0$, $N=2$ $\left(\e=1/5\right)$]{
                \label{cosu_N2_mu0}
                \begin{minipage}[b]{0.45\textwidth}
                  \centering
                  \includegraphics[width=2.1in]{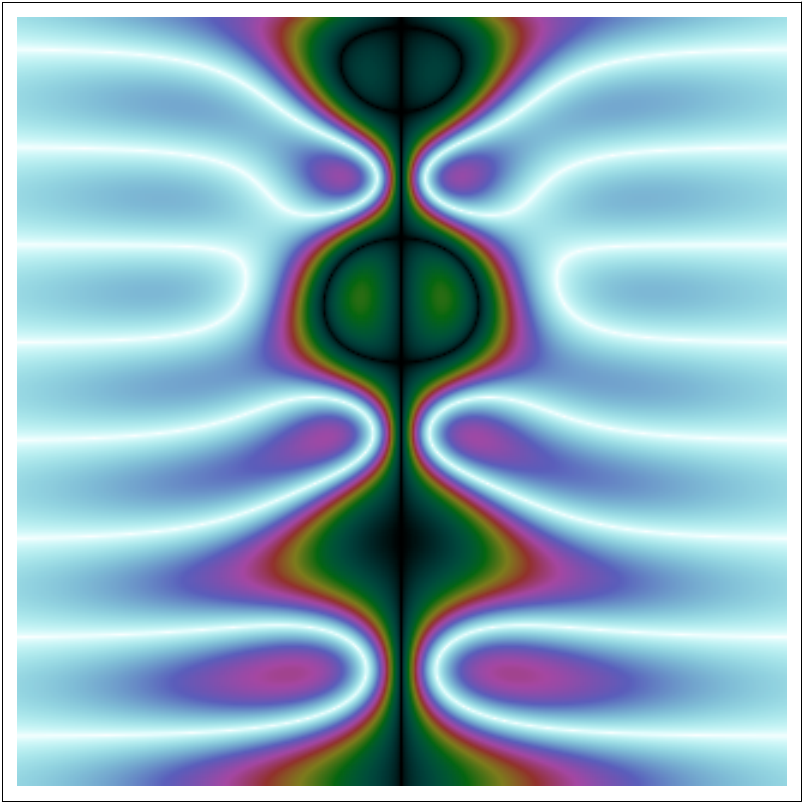}
                \end{minipage} }
        \subfigure[$\mu=0$, $N=4$ $\left(\e=1/9\right)$]{
                \label{cosu_N4_mu0}
                \begin{minipage}[b]{0.45\textwidth}
                  \centering
                  \includegraphics[width=2.1in]{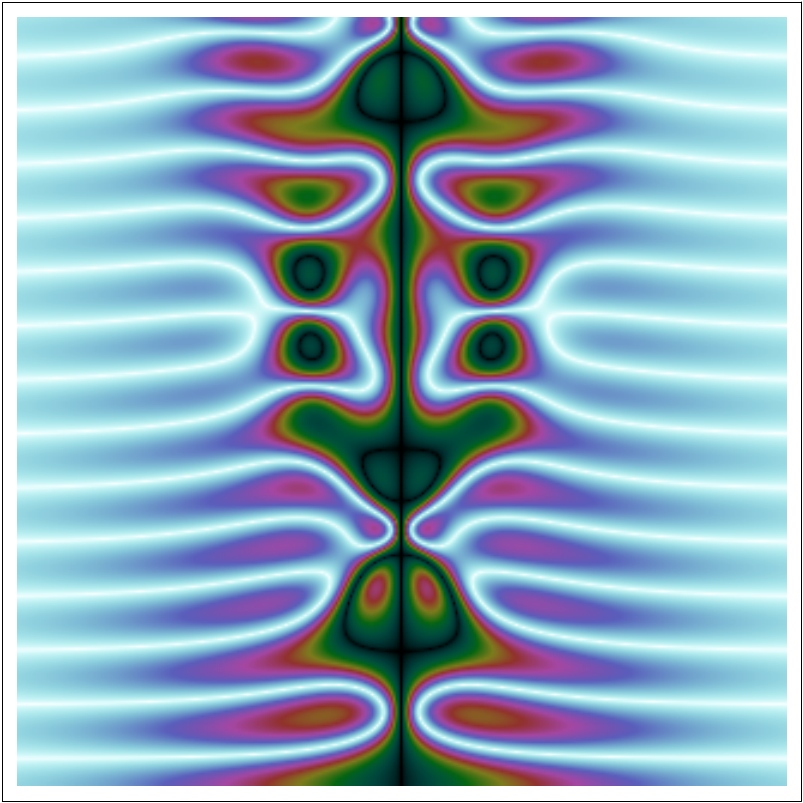}
                \end{minipage} }
\vspace{-.2in}
        \subfigure[$\mu=0$, $N=8$ $\left(\e=1/17\right)$]{
                \label{cosu_N8_mu0}
                \begin{minipage}[b]{0.45\textwidth}
                  \centering
                  \includegraphics[width=2.1in]{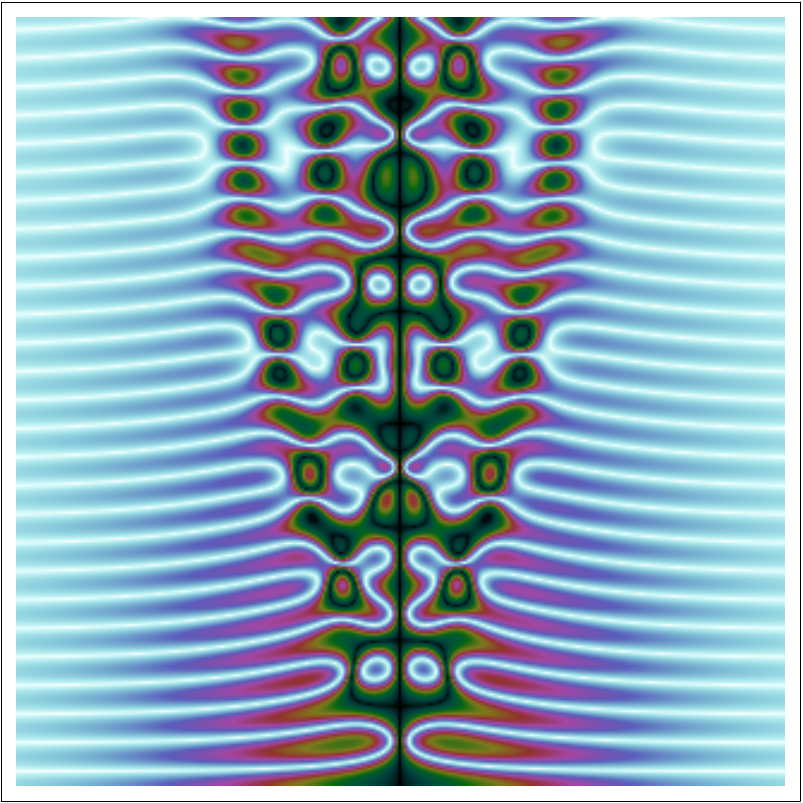}
                \end{minipage} }
        \subfigure[$\mu=0$, $N=16$ $\left(\e=1/33\right)$]{
                \label{cosu_N16_mu0}
                \begin{minipage}[b]{0.45\textwidth}
                  \centering
                  \includegraphics[width=2.1in]{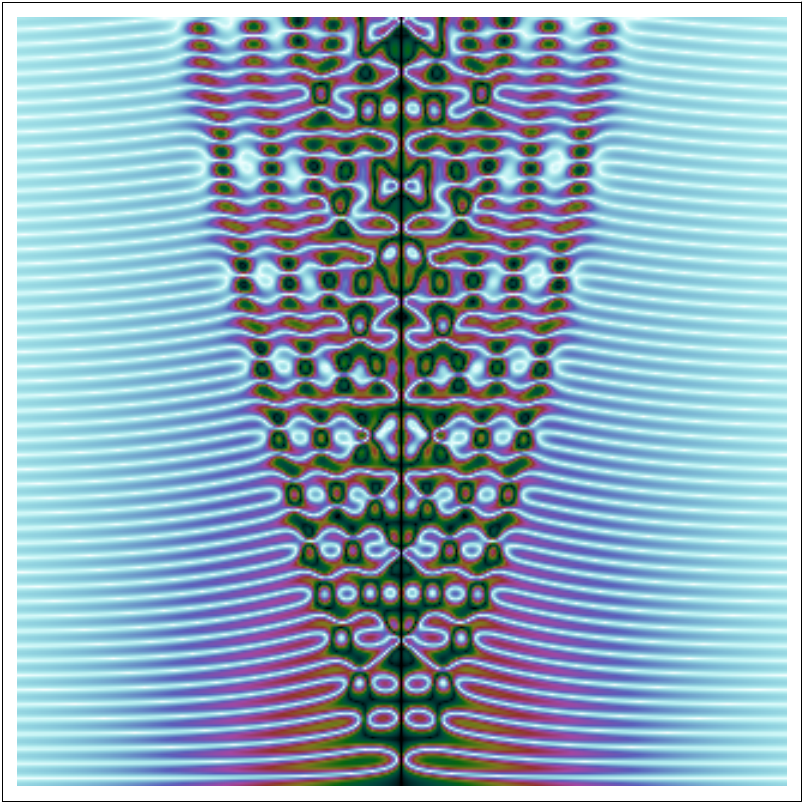}
                \end{minipage} }
  \caption{Plots of $\cos(u)$ for fixed $\mu=0$ and varying $N$ with $-2.5<x<2.5$ (horizontal axis) and $0<t<5$ (vertical axis).}
  \label{reflectionless-plots-mu0}
\end{figure}
We first study the special case of $\mu=0$.  Figure
\ref{reflectionless-plots-mu0} shows plots of the square region
$-2.5<x<2.5$ and $0<t<5$ with different colors indicating different
values of $\cos(u)$, with different plots corresponding to different
values of $N$ varying between $N=0$ and $N=16$ ($\e=\e_N(0)$ between
$1$ and $1/33$).  Lighter colors correspond to values of
$\cos(u)$ closer to $1$ and darker colors correspond to values of
$\cos(u)$ closer to $-1$.  The solutions of the sine-Gordon equation
\eqref{sine-Gordon} illustrated in these plots consist of a
``nonlinear superposition'' of $N$ breathers and one antikink.  As
each associated eigenvalue lies exactly on the unit circle in the
$z$-plane, the velocity of each of these soliton components, when
considered in absence of the others, is exactly zero.  In this sense,
the solution may be considered as a zero-velocity bound state of $N$
breathers and one antikink.  The most interesting phenomena are
associated with the semiclassical limit $\e\downarrow 0$ equivalent to
letting $N$ (the number of breathers) tend to infinity.  In this
limit, the plots suggest the asymptotic emergence of a fixed caustic
curve $t=t(x)$ in the space-time plane separating regions containing
different kinds of oscillatory behavior.  Indeed, for $|x|$
sufficiently large (that is, outside of the caustic), one observes
roll patterns characteristic of single-phase traveling waves. The latter
are simply the exact solutions of the sine-Gordon equation obtained by
substituting into \eqref{sine-Gordon} the traveling-wave ansatz $u(x,t)=
f((kx-\omega t)/\e)$, resulting in the ordinary differential equation
\begin{equation}
(\omega^2-k^2)\frac{d^2f}{d\xi^2} + \sin(f)=0\,,\quad\quad
\xi:=\e^{-1}(kx-\omega t)\,.
\label{eq:rollpatterns}
\end{equation}
Here $k$ is the wavenumber and $\omega$ is the frequency of the
traveling wave, and the waves appearing as the roll patterns in
figure~\ref{reflectionless-plots-mu0} correspond to phase velocities
$c:=\omega/k$ with $|c|>1$ (which makes \eqref{eq:rollpatterns} a
time-scaled version of the simple pendulum equation)\footnote{Of
  course, for these solutions of the hyperbolic sine-Gordon equation,
  the phase velocity exceeds the light speed of $|c|=1$.  In a sense,
  this fact does not contradict the hyperbolic nature of the equation,
  because the traveling wave is certainly not spatially localized, and
  moreover it has infinite energy.}.  The periodic solutions of
\eqref{eq:rollpatterns} are expressed in terms of elliptic functions,
and therefore we say that the roll patterns in
figure~\ref{reflectionless-plots-mu0} correspond to modulated waves of
genus $G=1$.  In the context of the phase portrait of the simple
pendulum, the roll-pattern oscillations outside of the central region
enclosed by the caustic curve correspond to librational motions of the
pendula, i.e.\@ orbits \emph{inside} the separatrix.  The sine-Gordon
equation \eqref{sine-Gordon} also has families of exact solutions
associated with hyperelliptic Riemann surfaces of arbitrarily large
genus $G$, and these solutions are represented in the form
$u(x,t)=f(\theta_1/\e,\dots,\theta_G/\e)$ where
$\theta_n=k_nx-\omega_nt$ and where $f$ is a multiperiodic function of
period $2\pi$ in each of its $G$ arguments.  In the case $G>1$, $u$ is
no longer a traveling wave, but rather is a \emph{multiphase wave}.
Reasoning by analogy with understood semiclassical limits of other
integrable equations, we may expect that the more complicated
oscillations evident in the plots of
figure~\ref{reflectionless-plots-mu0} for $t>t(x)$ (that is, inside of
the caustic curve) are modulated multiphase waves for some $G>1$.
Finally, we note that the caustic curve $t=t(x)$ appears to originate
from the point $x=t=0$.  As the velocity $u_t$ is zero at $t=0$ and
the pendulum angle is $u=-\pi$ at $x=t=0$, the point $x=0$ is the
unique point in the initial data corresponding to a point on the
separatrix of the phase portrait of the simple pendulum.

\begin{figure}[h]
        \subfigure[$\mu=1$, $N=0$ $(\e=\sqrt{2})$] {
                \label{cosu_N0_mu1}
                \begin{minipage}[b]{0.45\textwidth}
                  \centering
                        \includegraphics[width=2.1in]{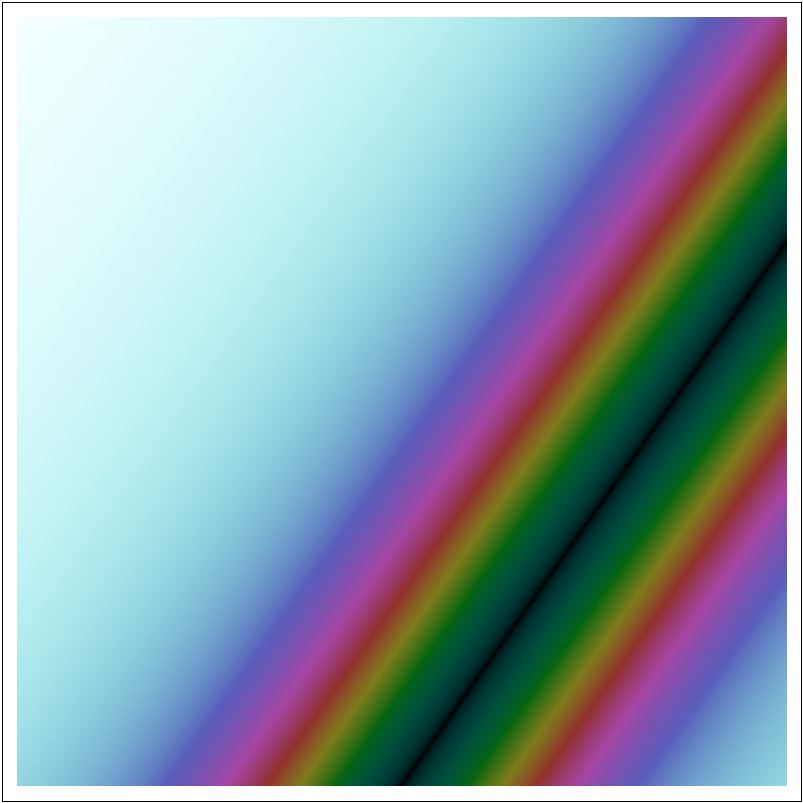}
                \end{minipage} } 
        \subfigure[$\mu=1$, $N=1$ $\left(\e=\sqrt{2}/3\right)$]{
                \label{cosu_N1_mu1}
                \begin{minipage}[b]{0.45\textwidth}
                  \centering
                        \includegraphics[width=2.1in]{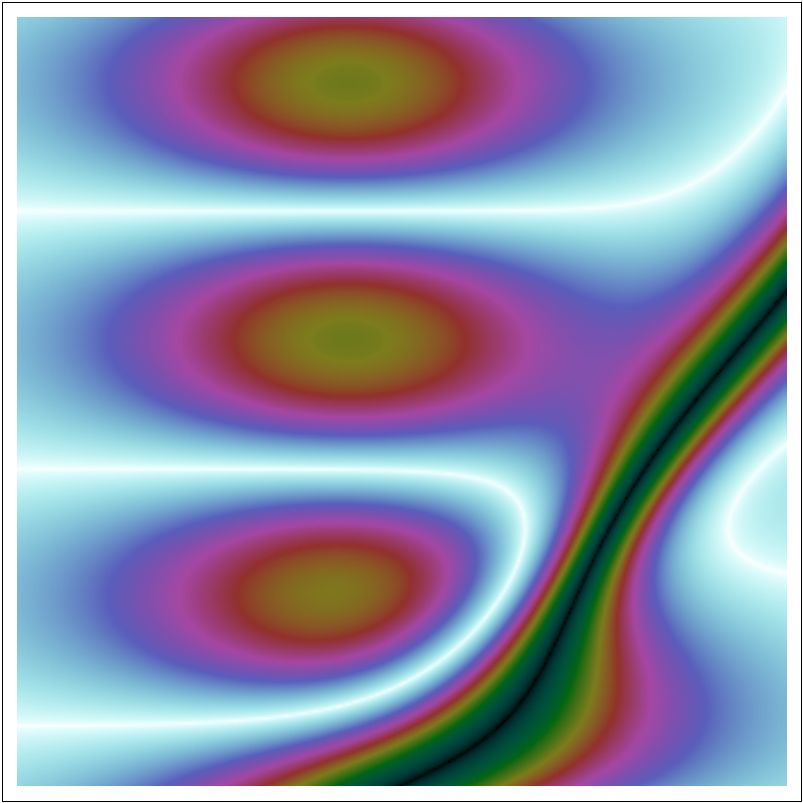}
                \end{minipage} }
        \subfigure[$\mu=1$, $N=2$ $\left(\e=\sqrt{2}/5\right)$]{
                \label{cosu_N2_mu1}
                \begin{minipage}[b]{0.45\textwidth}
                  \centering
                  \includegraphics[width=2.1in]{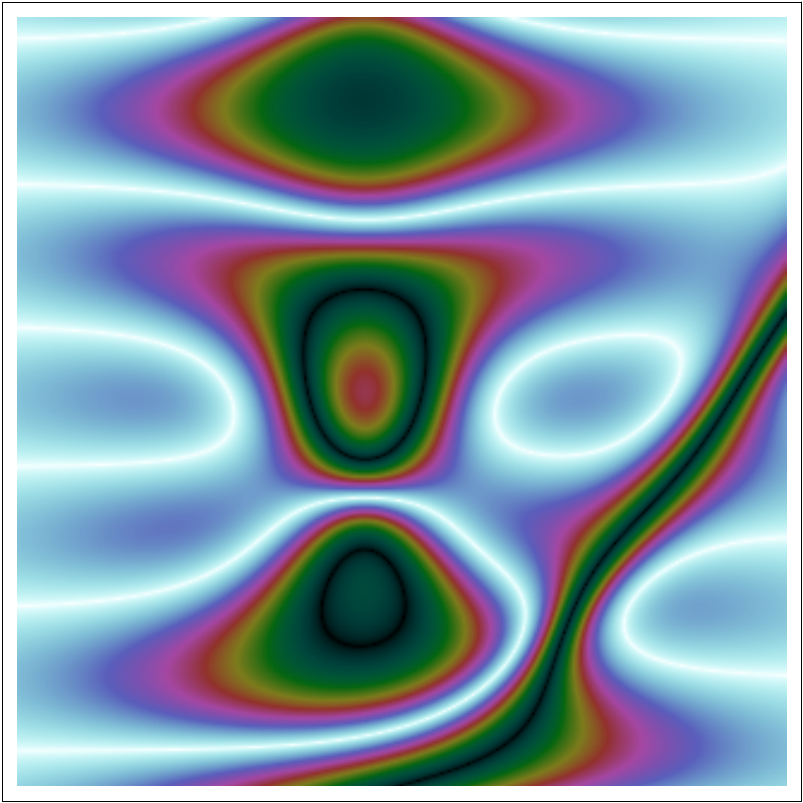}
                \end{minipage} }
        \subfigure[$\mu=1$, $N=4$ $\left(\e=\sqrt{2}/9\right)$]{
                \label{cosu_N4_mu1}
                \begin{minipage}[b]{0.45\textwidth}
                  \centering
                  \includegraphics[width=2.1in]{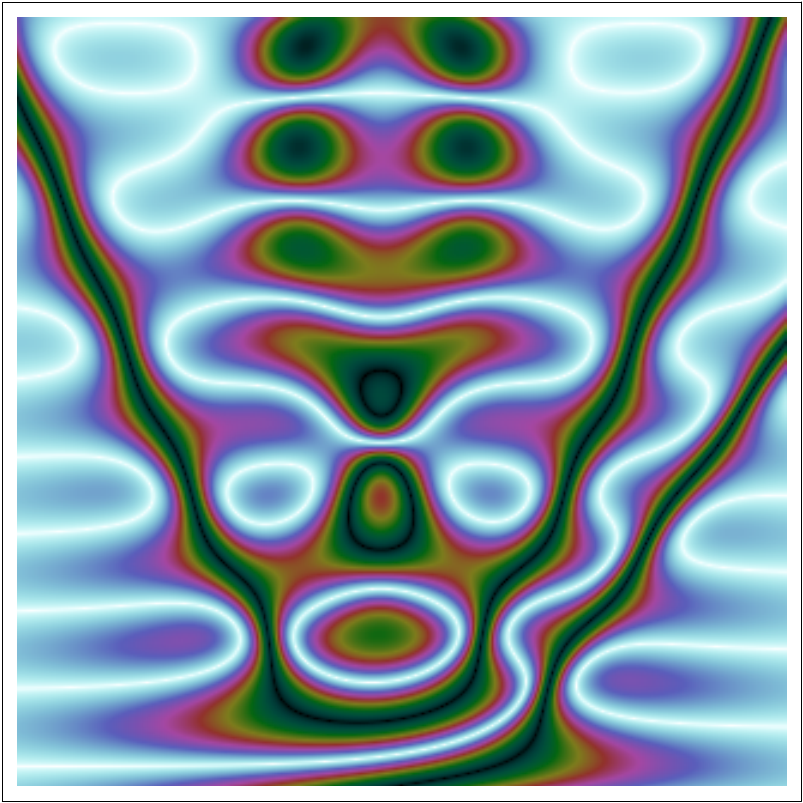}
                \end{minipage} }
\vspace{-.2in}
        \subfigure[$\mu=1$, $N=8$ $\left(\e=\sqrt{2}/17\right)$]{
                \label{cosu_N8_mu1}
                \begin{minipage}[b]{0.45\textwidth}
                  \centering
                  \includegraphics[width=2.1in]{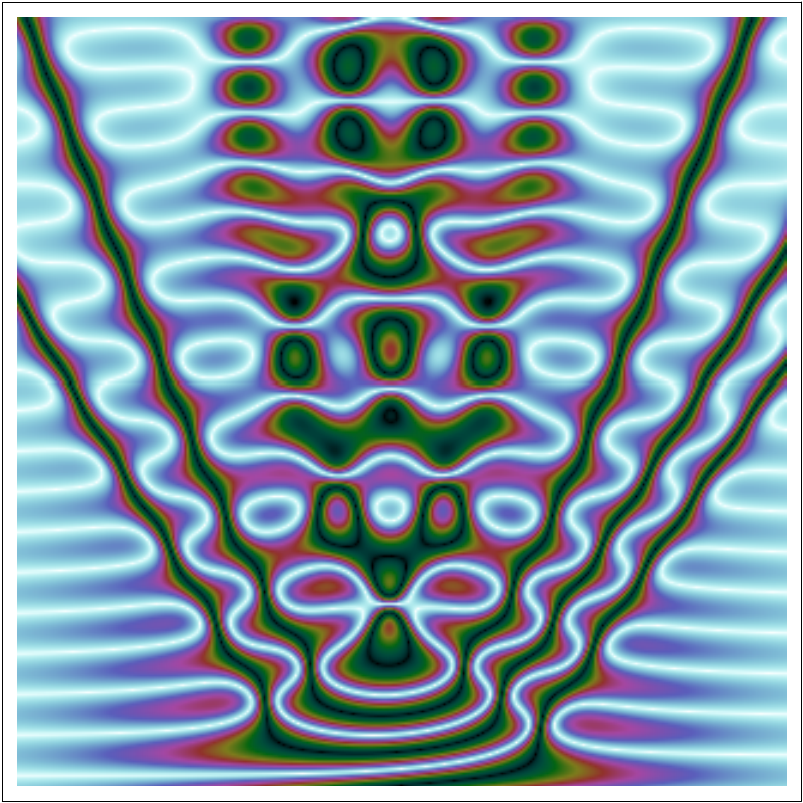}
                \end{minipage} }
        \subfigure[$\mu=1$, $N=16$ $\left(\e=\sqrt{2}/33\right)$]{
                \label{cosu_N16_mu1}
                \begin{minipage}[b]{0.45\textwidth}
                  \centering
                  \includegraphics[width=2.1in]{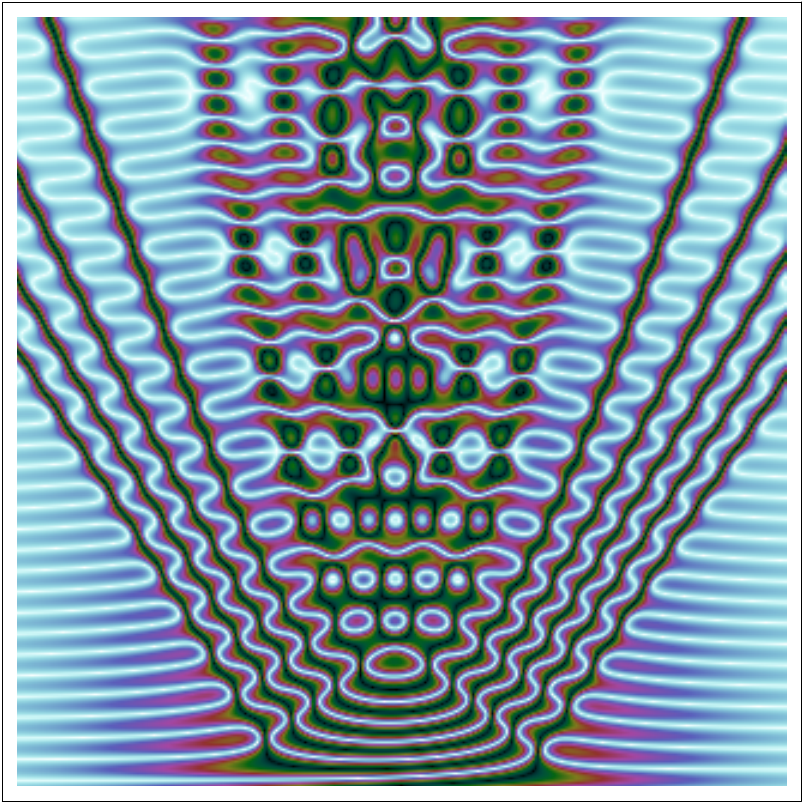}
                \end{minipage} }
\vspace{-.1in}
  \caption{Plots of $\cos(u)$ for fixed $\mu=1$ and varying $N$ for $-2.5<x<2.5$ (horizontal axis) and $0<t<5$ (vertical axis).}
  \label{reflectionless-plots-mu1}
\end{figure}
\begin{figure}[h]
\begin{center}
\includegraphics{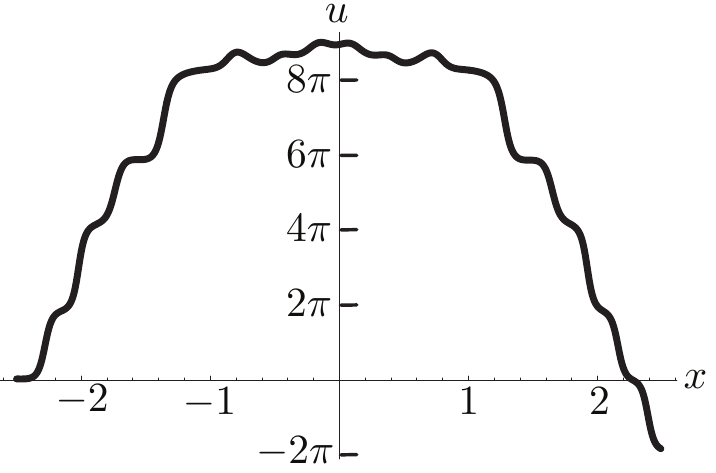}
\caption{Plot of $u$ for $t=2.5$, $-2.5<x<2.5$, $N=16$, $\mu=1$ 
$\left(\e=\sqrt{2}/33\right)$.}
\label{u_N16_mu1_t2p5}
\end{center}
\end{figure}

The evolution of the initial data \eqref{init-cond} for $\mu\neq 0$ is
depicted in the plots shown in figure \ref{reflectionless-plots-mu1}.
These plots are analogous to those in figure
\ref{reflectionless-plots-mu0}, except that we set $\mu=1$ and
considered $\e=\e_N(1)$.  The main effect of nonzero $\mu$ on the
discrete spectrum is to include, along with the quartets of
eigenvalues that correspond to breathers, an asymptotically (in the
limit $N\rightarrow\infty$) nonzero fraction of eigenvalues on the
imaginary axis that correspond to kinks and antikinks. The velocities
of the kinks and antikinks asymptotically fill out the entire range of
values $(-|\mu|/\sqrt{\mu^2+1},|\mu|/\sqrt{\mu^2+1})$.  There is
always one more antikink than there are kinks, and the ``excess''
antikink (corresponding to the eigenvalue $z_0=(\sqrt{2}-1)i$) carries
the topological charge.  This excess antikink always moves to the
right (this is a consequence of $\mu>0$, it turns out), and the
kink-antikink pairs corresponding to the other eigenvalues on the
imaginary axis are shed periodically in time and move to the left and
right.  As $\e\downarrow 0$, the outermost kinks or antikinks form a
caustic curve separating the modulated single-phase waves outside from
a region of the space-time containing the kink/antikink trains.  As
these trains propagate outwards over a field of modulated single-phase
waves, it seems reasonable to suppose that the pattern in this part of
the space-time would be described by a modulated multiphase wave of
genus $G=2$ that may be viewed as a nonlinear superposition of the
single-phase waves ($G=1$) and a kink or antikink train (also $G=1$,
although via orbits of \eqref{eq:rollpatterns} in the case
$\omega^2/k^2<1$ that lie outside of the separatrix).  That the
antikinks are moving to the right while the kinks are moving to the
left (for these plots corresponding to $\mu=1>0$) can be seen from a
plot of $u$ itself reconstructed from its sine and cosine subject to
the boundary condition $u(-\infty,t)=0$ as shown in figure
\ref{u_N16_mu1_t2p5}.
This plot corresponds to a horizontal slice of figure
\ref{cosu_N16_mu1} (or \ref{cosu_N16_mu1b} below), and it is
completely clear that the kinks occupy the left-hand portion of the
plot (in which from figure \ref{cosu_N16_mu1} we see that the waves
are propagating to the left) while the antikinks occupy the right-hand
portion (and by similar observations are propagating to the right).

The caustic curve simultaneously emerges at $t=0$ from two
(asymptotically) symmetric nonzero points $x$, and again these points
admit an interpretation in terms of the separatrix of the simple
pendulum equation (see below).  Between these $G=2$ regions there is a
triangular region containing pure single-phase oscillations that
persists for a time independent of $N$.  In the context of the phase
portrait of the simple pendulum, these oscillations correspond to
rotational motions of the pendula, i.e.\@ orbits \emph{outside} the
separatrix.  The collision of the two $G=2$ regions at the top of the
triangular $G=1$ (rotational) region results in a region containing more
complicated oscillations that resembles the region inside the caustic
curve for $\mu=0$ as seen in figure~\ref{reflectionless-plots-mu0}.
Note, however, that the oscillations occupying this central region may
be expected to be even more complicated than those present for $\mu=0$
because there are many kinks/antikinks with very small velocities, and
these will (if $N$ is sufficiently large) begin to interfere with the
bound state of breathers.  Finally, note that, while for $\mu\neq 0$
the exact solutions are not symmetric about $x=0$, the asymptotic
behavior evidently becomes symmetric as $\e\downarrow 0$.

\begin{figure}[h]
        \subfigure[$\mu=0$, $N=16$]{
                \label{cosu_N16_mu0b}
                \begin{minipage}[b]{0.45\textwidth}
                  \centering
                  \includegraphics[width=2.1in]{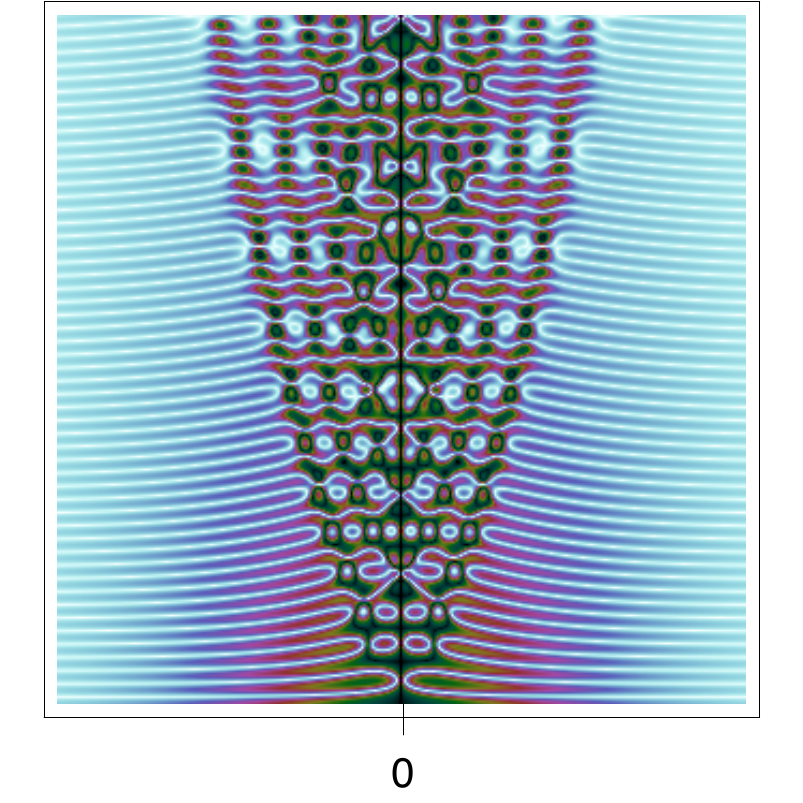}
                \end{minipage} }
        \subfigure[$\mu=1/2$, $N=16$]{
                \label{cosu_N16_mu0p5}
                \begin{minipage}[b]{0.45\textwidth}
                  \centering
                  \includegraphics[width=2.1in]{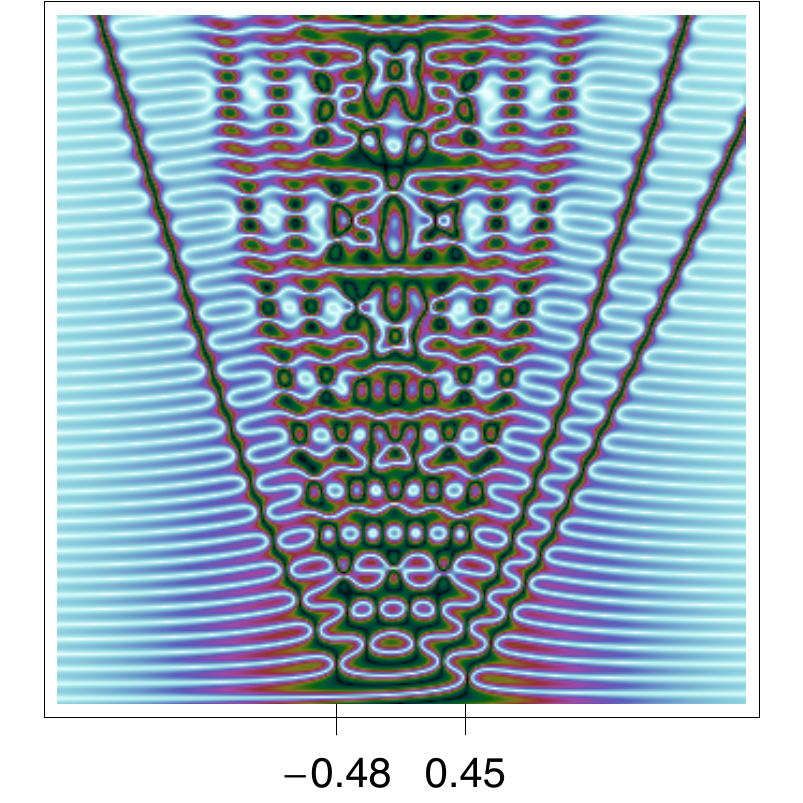}
                \end{minipage} }
        \subfigure[$\mu=1$, $N=16$]{
                \label{cosu_N16_mu1b}
                \begin{minipage}[b]{0.45\textwidth}
                  \centering
                  \includegraphics[width=2.1in]{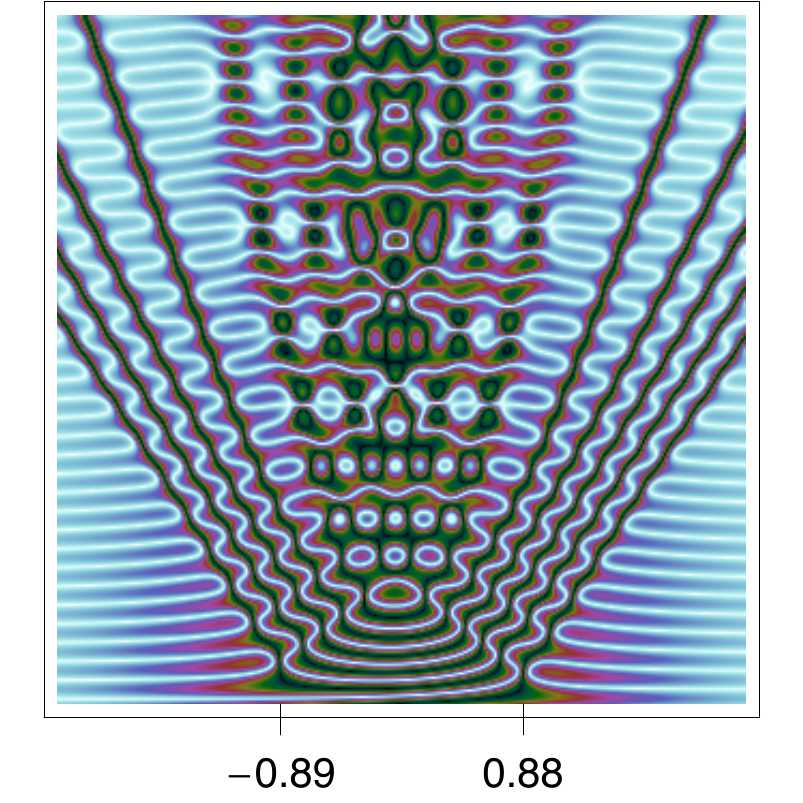}
                \end{minipage} }
        \subfigure[$\mu=2$, $N=16$]{
                \label{cosu_N16_mu2}
                \begin{minipage}[b]{0.45\textwidth}
                  \centering
                  \includegraphics[width=2.1in]{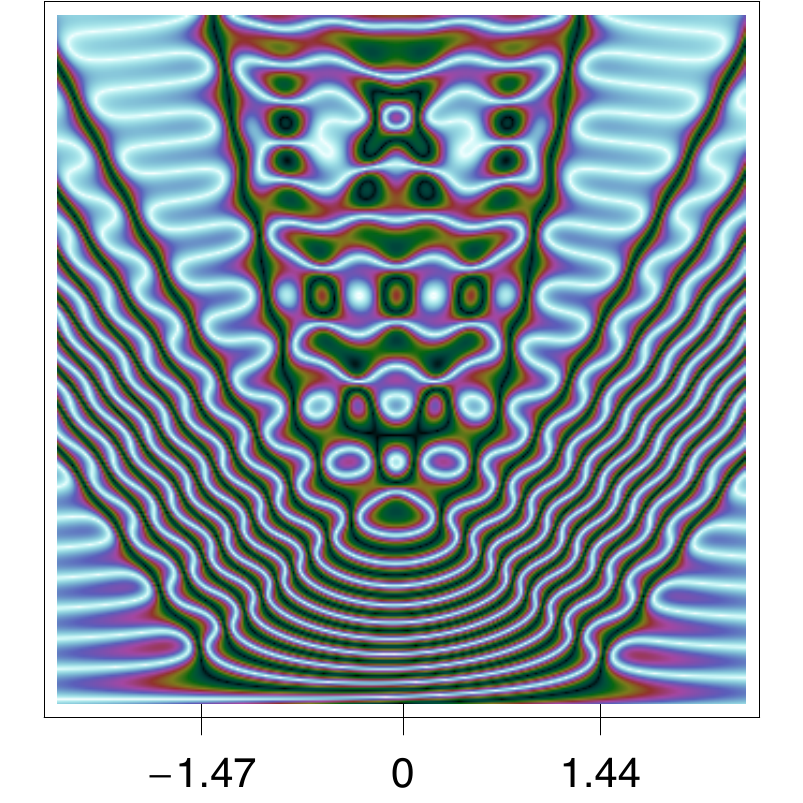}
                \end{minipage} }
  \caption{Plots of $\cos(u)$ for various $\mu$ with $-1.5<x<1.5$ (horizontal axis), $0<t<5$ (vertical axis), and $N=16$.}
  \label{reflectionless-plots-N16}
\end{figure}
The effect of varying $\mu$ can be seen from the plots shown in figure
\ref{reflectionless-plots-N16}.  Here, $N$ is fixed at the value
$N=16$ and $\mu$ is varied, with $\e=\e_{16}(\mu)$ holding to ensure
a reflectionless potential.  Note that the base of the triangular region of
the space-time containing single-phase rotational oscillations appears
to increase with $\mu$.  The ratio $M/(N-M)$ of eigenvalue quartets
corresponding to kink-antikink pairs to eigenvalue quartets
corresponding to breathers (see \eqref{pairnumber}) also increases as
$\mu$ increases for fixed $N$, an effect that is clearly visible in
the plots of figure \ref{reflectionless-plots-N16}.

\begin{figure}[h]
\begin{center}
\includegraphics{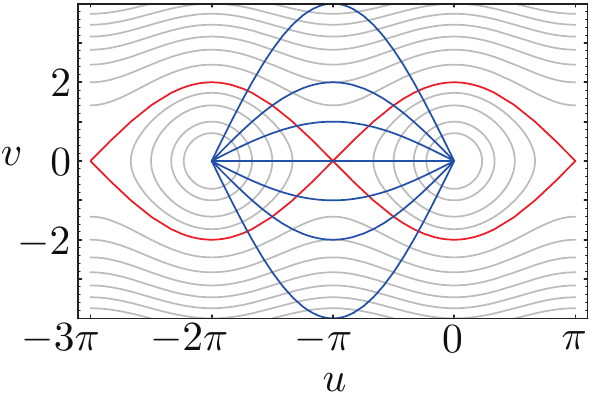}
\caption{The initial data \eqref{init-cond} plotted for $\mu=0$,
  $\mu=\pm1/2$, $\mu=\pm 1$, and $\mu=\pm 2$ in the phase portrait of
  the simple pendulum.}
\label{ppp}
\end{center}
\end{figure}

\begin{figure}[h]
\begin{center}
\includegraphics{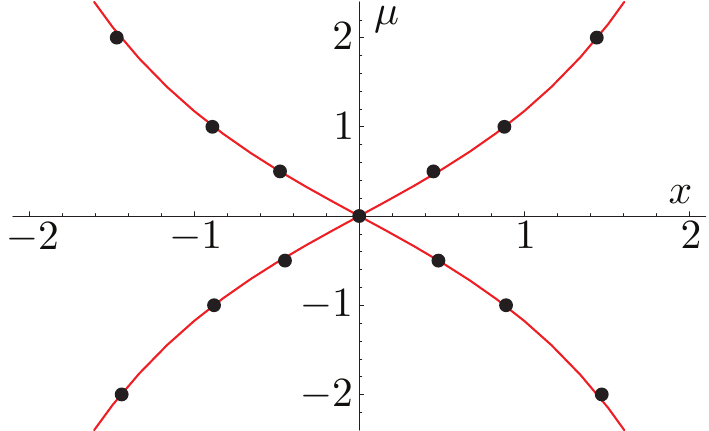}
\caption{The relation \eqref{eq:sinh} and the numerical data.  The
  data was collected only for $\mu\ge 0$, but by a natural symmetry
  (see below) we may include the point $(-x,-\mu)$ whenever we measure
  the point $(x,\mu)$. This plot suggests that an important role is
  played by the separatrix of the simple pendulum in the development
  of caustics for the semiclassical sine-Gordon equation.}
\label{arcsinh}
\end{center}
\end{figure}

The plots in figure \ref{reflectionless-plots-N16} also contain annotation
indicating our best guesses as to the values of $x$ from which the primary
caustic curve emerges at $t=0$.  These $x$-values may be predicted by
the following simple argument.  Let us rewrite the sine-Gordon equation
\eqref{sine-Gordon} as a perturbed simple pendulum equation in first-order
form:
\begin{equation}
\e\frac{du}{dt} = v\,,\quad\quad \e\frac{dv}{dt} = -\sin(u) + 
\e^2F(t;x)
\label{eq:pertpend}
\end{equation}
with forcing term $F(t;x):=u_{xx}(x,t)$.  We think of $u$ and $v$ as
the angle and angular velocity of a pendulum indexed by a parameter
$x$.  At the initial instant of time $t=0$, the function $u$ is smooth
and independent of $\e$, so the perturbation term $\e^2F(t;x)$ is very
small, and one expects $u(x,t)$ to evolve nearly independently for
different values of $x\in\mathbb{R}$.  This situation of independent
pendulum motions might be expected to persist until $u$ develops
rapidly-varying features of characteristic length proportional to
$\e$, for in such a situation we would have $u_{xx}\sim\e^{-2}$ and
hence the perturbation term is no longer negligible compared with
$\sin(u)$. Now, at any fixed time $t$, we may plot the phase points
$(u,v)$ in the phase portrait of the simple pendulum (that is, of
\eqref{eq:pertpend} with $F\equiv 0$), and this data will appear as a
curve parametrized by $x$.  Figure~\ref{ppp} shows the initial data
\eqref{init-cond} plotted parametrically in the phase portrait of the
simple pendulum for $\mu=0$, $\mu=\pm 1/2$, $\mu=\pm 1$, and
$\mu=\pm 2$ (blue curves).  
The separatrix for the simple pendulum equation is shown with red
curves.  It is clear that each blue curve intersects the separatrix at
exactly two points, and moreover, by unraveling the parametrization it
is easy to see that these two points correspond to two distinct values
of $x$.  Near these values of $x$, there are pendula undergoing
librational motions as well as pendula undergoing rotational motions.
This is the scenario under which the most rapid amplification of the
difference of angles $u$ for neighboring pendula is to be expected.
Therefore, we may make the prediction that the modulated single-phase
ansatz should break down immediately at $t=0$ at exactly the two
values of $x$ at which the initial data meets the separatrix.  These
values of $x$ are easily calculated.  Indeed, the separatrix is given
by the equation $v=\pm 2\cos(u/2)$, and the initial data satisfies
$\cos(u/2)=\tanh(x)$ and $v=2\mu\,\sech(x)$.  Therefore, the initial
data curve (blue) intersects the separatrix (red) at values $x$ for
which 
\begin{equation}
\mu=\pm \sinh(x)\,.  
\label{eq:sinh}
\end{equation}
To confirm this reasoning, we took our best guesses for the $x$-values
at which the phase transition occurs at $t=0$ as indicated on the
plots in figure~\ref{reflectionless-plots-N16} and created a data set
by combining these with the corresponding values of $\mu$.  The
ordered pairs $(x,\mu)$ making up this data set are plotted with black
dots in figure~\ref{arcsinh} along with the curves \eqref{eq:sinh}
plotted in red.
It is clear that this theory provides an accurate prediction of the
points $x$ from which the caustics emerge at time $t=0$.  While we
have only given a comparison with the theory for initial conditions of
the special form \eqref{init-cond}, it seems reasonable that the
principle should be the same for more general initial data.  That is,
one should locate the $x$-values at which the pair $(u=f(x),v=g(x))$
lies on the separatrix $v=\pm 2\cos(u/2)$ and expect complicated
oscillations to emerge from these points for $t>0$ in the
semiclassical limit.

We have only computed solutions corresponding to the initial data
\eqref{init-cond} for $\mu\geq0$.  That this is sufficient follows
from a simple symmetry between $\mu$ and $-\mu$.  Indeed, write
\eqref{sine-Gordon} and \eqref{init-cond} in first-order form as \eq
\begin{split}
\e u_t &= v \\
\e v_t& = \e^2 u_{xx}-\sin(u) 
\end{split}
\endeq
subject to the initial data
\begin{equation}
u(x,0) = f(x)\,, \quad\quad v(x,0) = 2\mu\,\sech(x)\,,
\end{equation}
where $\sin(f/2)=\sech(x)$, $\cos(f/2)=\tanh(x)$, and consider the 
substitutions 
\eq
V=-v, \quad U=2\pi-u, \quad X=-x, \quad T=t.
\endeq
Then the Cauchy problem for $U$ and $V$ consists of the first-order system
\eq
\begin{split}
\e U_T &= V \\
\e V_T &= \e^2 U_{XX}-\sin(U) 
\end{split}
\endeq
subject to the initial data
\begin{equation}
U(X,0) = F(X)\,, \quad\quad V(X,0) = -2\mu\,\sech(X)\,,
\end{equation}
where $\sin(F/2)=\sech(X)$, $\cos(F/2)=\tanh(X)$.  Therefore, $U$
satisfies the sine-Gordon equation with initial data of the form
\eqref{init-cond} but with $\mu$ replaced with $-\mu$.  In terms of
$\cos(u)$, replacing $\mu$ with $-\mu$ therefore simply amounts to
replacing $x$ with $-x$.

\section{Concluding Remarks}
\label{conclusion}
The main result of this paper is the exact calculation, via the theory
of hypergeometric functions, of the scattering data for the
noncharacteristic Cauchy problem for the semiclassical sine-Gordon
equation \eqref{sine-Gordon} subject to the initial data
\eqref{init-cond}.  That this calculation is valid for all
sufficiently small $\e$ means that the formulae for the scattering
data given in Theorem \ref{scattering-data-theorem} may be used to
formulate a corresponding inverse-scattering problem whose solution
will give detailed information about the semiclassical limit of the
sine-Gordon Cauchy problem.  Moreover, since for each value of the
parameter $\mu\in\mathbb{R}$ appearing in the initial data
\eqref{init-cond} there exists a sequence
$\{\e=\e_N(\mu)\}_{N=1}^\infty$ of values of $\e$ tending to zero for
which the scattering data are reflectionless, it is possible to
approach the semiclassical limit in such a way that the
inverse-scattering problem involves, for each $N$, only
finite-dimensional linear algebra.  As we have shown in Section
\ref{inverse-scattering}, this fact makes it quite feasible to use
numerical methods to solve the inverse-scattering problem for fairly
large values of $N$ and therefore study the semiclassical limit, at
least in a qualitative sense.  Our numerical reconstructions of the exact
solutions of the Cauchy problem indeed reveal marvelous structures
apparently emerging in the semiclassical limit.

Needless to say, a study of the semiclassical limit based solely on
numerics of the sort described in Section \ref{inverse-scattering} has
practical limitations.  To study the semiclassical limit really
requires allowing $N$ to become arbitrarily large, and the system
(\ref{Phat2-equation}) contains $2N+1$ equations and hence will
ultimately become numerically intractable for sufficiently large $N$.
This difficulty is compounded on the one hand by the fact that the
condition numbers of the matrices involved grow rapidly\footnote{One
  can see from the formula for the matrix elements of $\mathbf{F}$
  \eqref{Phat-equation} that $\mathbf{F}$ is proportional by the
  diagonal matrix $\mathrm{diag}(\mathbf{a})$ to a matrix of
  Cauchy/Hilbert type.  The latter is the classic example given in
  textbooks on numerical analysis of an ill-conditioned matrix.} with
$N$, and on the other by the necessity to use a grid spacing of order
$\e$ to resolve the microstructure of the solution.  In other words,
to study the semiclassical limit in this way, an asymptotically
badly-conditioned linear algebra problem in dimension proportional to
$N$ must be solved on a grid of approximately $\e^{-2}\sim N^2$ values
of $(x,t)$ in a fixed-size region.  

In our opinion, the main purpose of carrying out numerical experiments
like those in Section \ref{inverse-scattering} is to indicate
phenomena that would be of interest to study rigorously by other
(analytical) methods, and to motivate such a study.  For example,
figures \ref{reflectionless-plots-mu0} and
\ref{reflectionless-plots-mu1} clearly indicate the existence of a
limiting form of the $O(1)$ scale macrostructure independent of $\e$
in the semiclassical limit.  The (apparent) existence of caustic
curves separating different types of oscillations requires a careful
explanation, and such an explanation would be expected to also make
asymptotically accurate predictions for the locations of the caustics.
In integrable problems like the sine-Gordon equation, one expects the
microstructure of oscillations in between the caustic curves to be
described asymptotically by modulated exact multiphase solutions of
the equation associated with Riemann surfaces of genus $G$.  The
modulation itself is expected to be described by slowly-varying (that
is, independent of $\e$) fields satisfying an appropriate system of
quasilinear Whitham (modulation) equations.  The sine-Gordon problem
is quite different from other integrable problems for which the
semiclassical limit has been investigated in that it has Whitham
equations of both hyperbolic and elliptic type
\cite{Ercolani:1984,Ercolani:1986}.  To fully analyze these phenomena
from the starting point of the scattering data we give in
Theorem~\ref{scattering-data-theorem}, it is necessary to use very
precise methods of asymptotic analysis for Riemann-Hilbert problems to
find an asymptotic expansion for $u(x,t;\e,\mu)$ valid as
$\e\downarrow 0$.  Calculations of this sort, also in the discrete
spectral setting (that is, reflectionless inverse-scattering as is
available for this problem when $\e=\e_N(\mu)$), were carried out for
the semiclassical focusing NLS equation \eqref{NLS-N-soliton} for a
general class of initial data in \cite{Kamvissis:2003-book}.

True understanding of the semiclassical asymptotics of the Cauchy
problem for the sine-Gordon equation ultimately requires generalizing
the one-parameter family of initial data given by \eqref{init-cond}.
While the special initial data \eqref{init-cond} is quite natural,
satisfying the correct boundary conditions, and incorporating effects
such as nontrivial topological charge and tunable (via the parameter
$\mu$) initial velocity, one may certainly pose the Cauchy problem for
more general initial data $f(\cdot)$ and $g(\cdot)$ and ask for the
corresponding asymptotic behavior of $u(x,t;\e)$ as $\e\downarrow 0$.
One might hope that other initial conditions that are somehow close to
(\ref{init-cond}) might correspond to scattering data and dynamical
behavior of $u(x,t;\e)$ whose semiclassical asymptotics are similar to
those of the exactly solvable case.  This would indicate a kind of
stability of the semiclassical limit.  Furthermore, one may be
interested in the semiclassical asymptotics corresponding to initial
data that differ significantly from the special data
\eqref{init-cond}, for example by having a topological charge that is
different.  
Clearly, to begin to study more general initial data, it is necessary
to find quantitative approximations of the corresponding scattering
data.  A first step towards this goal is to seek conditions on general
initial data that force the eigenvalues to lie exactly on certain
contours in the complex $z$-plane.  For initial data satisfying such
conditions, WKB analysis can be used to find a leading-order estimate
in $\e$ for the scattering data, and with more work, the error of the
estimate can be analyzed.  For example, for the nonselfadjoint
Zakharov-Shabat eigenvalue problem relevant to the focusing NLS
equation \eqref{eq:unscaledNLS}, Klaus and Shaw \cite{Klaus:2002}
showed that if the initial condition $q(x,0)$ is real and monomodal,
then the discrete spectrum may only lie exactly on the imaginary axis.
In the semiclassical setting, this is an \emph{exact result} that
holds for all $\e>0$.  WKB calculations based on the Klaus-Shaw result
were used in \cite{Kamvissis:2003-book} to analyze certain so-called
semiclassical soliton ensembles.  As for sine-Gordon, Bronski and
Johnson \cite{Bronski:2007} recently found a result analogous to that
of Klaus and Shaw, showing in particular\footnote{Actually, they
  showed more: if $g\equiv 0$ and $\sin(f/2)$ is a Klaus-Shaw
  potential, then the discrete spectrum lies on the unit circle.  If
  the maximum value of this potential is unity, then the topological
  charge is $\pm 1$ and $f$ is monotone, but if the maximum value is
  smaller the topological charge is zero and $f$ is monomodal.}  that
if $g\equiv0$ and $f$ is monotone with topological charge $\pm1$, then
the eigenvalues must lie exactly on the unit circle in the $z$-plane.
A quantitative approach to the discrete spectrum for initial data of
the Bronski-Johnson type would be to map it onto a perturbation of the
specific initial data \eqref{init-cond} with $\mu=0$ (which is, of
course, a special case of a Bronski-Johnson potential) using a Langer
transformation, and to control the error introduced by the
perturbation for small $\e$.  This will be carried out in future work.

We conclude by drawing some comparisons between our results and those
of Tovbis and Venakides \cite{Tovbis:2000} for the nonselfadjoint
Zakharov-Shabat problem associated with the focusing NLS equation.
The class of Tovbis-Venakides potentials (see
\eqref{Tovbis-Venakides}) involves a parameter $\mu\in\mathbb{R}$
($\mu=0$ is the special case studied earlier by Satsuma and Yajima
\cite{Satsuma:1974}).  In fact, we chose to use the symbol $\mu$ for
the parameter in \eqref{init-cond} precisely because this parameter
plays a similar role.  One immediate observation is that only in the
case $\mu=0$ is the Tovbis-Venakides initial data of Klaus-Shaw type,
and similarly only in the case $\mu=0$ is the initial data
\eqref{init-cond} of Bronski-Johnson type.  Thus only for $\mu=0$ is
one guaranteed by general arguments\footnote{It turns out that for
  $\mu\neq 0$ the eigenvalues of the Tovbis-Venakides potentials (when
  they exist) lie exactly on the imaginary axis nonetheless.  However,
  for the initial data \eqref{init-cond} any nonzero value of $\mu$
  immediately introduces eigenvalues that are not confined to the unit
  circle.}  that the discrete spectrum is confined to a special curve
in the complex plane.  Another observation is that the parameter $\mu$
has a physical interpretation of velocity in both the Tovbis-Venakides
family of potentials (because in the hydrodynamic variables for
Schr\"odinger equations introduced long ago by Madelung, the velocity
of the quantum-corrected fluid motion is expressed in terms of
$\phi=A(x,t)e^{iS(x,t)/\e}$ by $S_x(x,t)$, and for the
Tovbis-Venakides potentials $S$ is proportional to $\mu$ at $t=0$) and
also in the family \eqref{init-cond} of initial data for sine-Gordon
(because the initial data is a solution of the advection equation with
velocity $\mu/\e$ as pointed out in the Introduction).  However, as
one important distinction, we note that the Tovbis-Venakides
potentials have the possibility of being reflectionless for certain
$\e$ only for $\mu=0$, while this possibility exists for the initial
data \eqref{init-cond} for every $\mu\in\mathbb{R}$.

\

\section*{Acknowledgments}  We are grateful to Jared Bronski for bringing 
to our attention the symmetric gauge for the eigenvalue problem and for 
sharing the results of his work with Mathew Johnson on eigenvalue confinement 
to the unit circle, and to James Colliander for suggesting an approach to 
study the well-posedness of the Cauchy problem for the sine-Gordon equation.  
We also thank the members of the 
integrable systems working group at the University of Michigan for their 
comments and feedback.  Both authors were partially supported by Focused 
Research Group grant DMS-0354373 from the National Science Foundation.

\appendix

\section{The Riemann-Hilbert Approach to Inverse Scattering for 
Sine-Gordon}
\label{appendix}
Our aim in this appendix is to present a completely self-contained
theory of inverse-scattering for the sine-Gordon equation in
laboratory coordinates.  In particular, we show how to represent the
solution of the Cauchy problem for the sine-Gordon equation with
$L^1$-Sobolev initial data (specifically, $1-\cos(u(x,0)),$
$\sin(u(x,0)),$ $u_x(x,0),$ $u_{xx}(x,0),$ $u_t(x,0),$ $u_{tx}(x,0)\in
L^1$) in terms of the solution of a certain matrix-valued
Riemann-Hilbert problem.  To ensure that various quantities used in
the inverse-scattering method are well defined with desirable
properties for all $t\ge 0$, we rely on a theory of the well-posedness
of the Cauchy problem that may be developed independently of any
inverse scattering methodology.  An outline of the relevant
well-posedness theory is given in Appendix \ref{app-well-posed}, in
which we show that the class of $L^1$-Sobolev potentials (in the sense
defined above) is preserved for all $t\ge 0$ under the evolution of
the sine-Gordon equation.  Many of the results to be described below
have appeared in the literature in one form or another.
For instance, the characterization of the Jost solutions assuming that
$1-\cos(u(x,t)),$ $\sin(u(x,t)),$ $u_x(x,t),$ $u_t(x,t)\in L^1$ at
time $t$ appeared in Kaup \cite{Kaup:1975}, and aspects of the
Riemann-Hilbert approach to inverse scattering were worked out for
initial data $f$ and $g$ in the Schwartz space by Zhou
\cite{Zhou:1995} and Cheng et al. \cite{Cheng:1999, Cheng:1997}.  The
well-posedness theory we present in Appendix~\ref{app-well-posed}
appears to be a new contribution to the subject.

The starting point for our analysis is the observation
\cite{Kaup:1975} that the sine-Gordon equation 
(\ref{sine-Gordon}) is the compatibility condition for the Lax pair 
\eq
\label{eigenvalue-eqn2}
4i\e\overline{\mathbf{w}}_x = \overline{\mathbf{L}}\overline{\mathbf{w}} := \bbm 4E+z^{-1}(1-\cos(u)) & -z^{-1}\sin(u)-i\e(u_x+u_t) \\ -z^{-1}\sin(u)+i\e(u_x+u_t) & -4E-z^{-1}(1-\cos(u)) \ebm \overline{\mathbf{w}}
\endeq
\eq
\label{time-evolution-eqn}
4i\e \overline{\mathbf{w}}_t = \overline{\mathbf{B}}\overline{\mathbf{w}}:= \bbm 4D-z^{-1}(1-\cos(u)) & z^{-1}\sin(u)-i\e (u_x+u_t) \\ z^{-1}\sin(u)+i\e (u_x+u_t) & -4D+z^{-1}(1-\cos(u)) \ebm \overline{\mathbf{w}}
\endeq
with $D(z)$ and $E(z)$ given in \eqref{D-and-E}.  In other words,
there exists a basis (determined, say, by specification of two
linearly independent vectors $\overline{\mathbf{w}}$ at $x=t=0$) of
simultaneous solutions of \eqref{eigenvalue-eqn2} and
\eqref{time-evolution-eqn} if and only if $u=u(x,t)$ is a solution of
the sine-Gordon equation \eqref{sine-Gordon}.

The Lax pair (\ref{time-evolution-eqn})--(\ref{eigenvalue-eqn2})
appears to have a singularity at $z=0$.  However, it is possible to
use a gauge transformation to move the singularity from $z=0$ to $z=\infty$
and in this way analysis for large $z$ can be continued to appropriate
sets with limit point $z=0$.  This gauge transformation will play an
important role in our analysis.

\subsection{Jost solutions of the scattering problem}
We now attempt to define the Jost solutions
$\JJi^\pm(x)=\JJi^\pm(x;z,t,\e)$ for $z\in\mathbb{R}$ as the
fundamental solution matrices of the eigenvalue equation
(\ref{eigenvalue-eqn2}) normalized as \eq
\label{Psi-infinity-normalizations}
\begin{split}
\JJi^+(x) = \bbm e^{-iEx/\e} & 0  \\ 0 & e^{iEx/\e} \ebm + o(1) \quad \text{ as } x\to +\infty, \\
\JJi^-(x) = \bbm e^{-iEx/\e} & 0 \\ 0 & e^{iEx/\e} \ebm + o(1) \quad \text{ as } x\to-\infty. 
\end{split}
\endeq
We denote the columns of $\JJi^\pm(x)$ as
\eq 
\JJi^\pm(x)=:[\jji_1^\pm(x),\jji_2^\pm(x)].
\endeq
The issue at hand is to determine whether these conditions uniquely determine
$\JJi^\pm(x)$ when $z$ is a real number, and then to further determine
what can be said for complex $z$.

To begin, we rewrite \eqref{eigenvalue-eqn2} in the form
\eq
\label{JJi-diffeq}
4i\e\JJi^\pm_x = (4E\sigma_3 + \QQi)\JJi^\pm
\endeq
with
\eq
\QQi(x;z,\e) := \bbm z^{-1}(1-\cos(f)) & -z^{-1}\sin(f)-i(\e f'+g) \\ -z^{-1}\sin(f)+i(\e f'+g) & -z^{-1}(1-\cos(f)) \ebm.
\endeq
Here $f$ is the value of $u$, and $g$ is that of $\e u_t$ at some fixed time 
$t$.  The purpose of this decomposition is to separate the part of the 
coefficient matrix that decays (in a certain sense) as $x\to\pm\infty$ ($\QQi$) 
from a constant term ($4E\sigma_3$).  Defining matrices
\eq
\label{m-plusminus}
\MMi^\pm = \left[\mmi_1^\pm,\mmi_2^\pm\right] := \JJi^\pm e^{iEx\sigma_3/\e},
\endeq
or equivalently in terms of the columns,
\eq
\label{m-ito-Phi-minus-and-Phi-plus}
\mmi_1^+ = \jji_1^+ e^{iEx/\e}, \quad \mmi_2^+ = \jji_2^+ e^{-iEx/\e}, \quad \mmi_1^- = \jji_1^- e^{iEx/\e}, \quad \mmi_2^- = \jji_2^- e^{-iEx/\e},
\endeq
one may easily translate the differential equation \eqref{JJi-diffeq} and
boundary conditions \eqref{Psi-infinity-normalizations} for $z\in\mathbb{R}$
into integral equations for the matrices $\MMi^\pm(x;z,\e)$:
\eqarr
\label{m-plus}
\MMi^+(x) & = & \bbm 1 & 0 \\ 0 & 1 \ebm - \frac{1}{4i\e}\int_x^{+\infty} e^{-iE(x-y)\sigma_3/\e}\QQi(y)\MMi^+(y)e^{iE(x-y)\sigma_3/\e}dy, \\
\label{m-minus}
\MMi^-(x) & = & \bbm 1 & 0 \\ 0 & 1 \ebm + \frac{1}{4i\e}\int_{-\infty}^x e^{-iE(x-y)\sigma_3/\e}\QQi(y)\MMi^-(y)e^{iE(x-y)\sigma_3/\e}dy.
\endeqarr
While these integral equations are formulated to correspond to
\eqref{eigenvalue-eqn2} and \eqref{Psi-infinity-normalizations} for
$z\in\mathbb{R}$, we may also consider them for complex $z$.
Proposition \ref{m-LHP-at-infinity} shows that the columns
$\mmi_1^+(x)$ and $\mmi_2^-(x)$ are well-defined by \eqref{m-plus} and
\eqref{m-minus} respectively as long as $\Im(z)\le 0$, and moreover
for each $x\in\mathbb{R}$ they are analytic for $\Im(z)<0$, and
continuous in the closed lower half $z$-plane for $z$ bounded away
from $z=0$.  Then Proposition \ref{m-LHP-at-zero} uses an alternate
gauge to extend continuity to small $z$.
\begin{prop}
\label{m-LHP-at-infinity}
Suppose $1-\cos(f)$, $\sin(f)$, $\e f'+g\in L^1$.  If $\Im(z)\le 0$,
then the first column of \eqref{m-plus} and the second column of
\eqref{m-minus} uniquely define solutions $\mmi_1^+(x;z)$ and
$\mmi_2^-(x;z)$.  These functions are, for each $x\in\mathbb{R}$,
analytic for $\Im(z)<0$ and continuous for
$z\in\{|z|\geq\delta\}\cap\{\Im(z)\leq0\}$ for each $\delta>0$.
\end{prop}
\begin{proof}
  The function $\mmi_1^+(x;z)$ is constructed from equation
  (\ref{m-plus}) via an iterative argument.  Define the $0^\text{th}$
  iterate for $\mmi_1^+$ as $\mmi_{1,0}^+(x) := [ 1 , 0]^\mathsf{T}$.  Then
  define the $n^\text{th}$ iterate inductively by 
\eq 
\mmi_{1,n}^+(x) :=
  \bbm 1 \\ 0 \ebm - \int_x^{+\infty}
  \KKi^+_1(y;x,z)\mmi_{1,n-1}^+(y)dy
\endeq
with
\eq
\label{KKi-1-plus}
\KKi^+_1(y) = \KKi^+_1(y;x,z) := \frac{1}{4i\e}\bbm z^{-1}(1-\cos(f)) & -z^{-1}\sin(f) -i(\e f'+g) \\ \left(-z^{-1}\sin(f)+i(\e f'+g)\right)e^{2i(x-y)E/\e} & -z^{-1}(1-\cos(f))e^{2i(x-y)E/\e} \ebm.
\endeq
Here $f$, $f'$, and $g$ are functions of $y$.  It follows that
\eq
\begin{split}
\mmi_{1,n}^+(x) = & \bbm 1 \\ 0 \ebm - \int_x^{+\infty} \KKi^+_1(x_1)\bbm 1 \\ 0 \ebm dx_1 + \cdots \\ 
&  + (-1)^n\int_x^{+\infty}\int_{x_1}^{+\infty}\cdots\int_{x_{n-1}}^{+\infty}\KKi^+_1(x_1)\cdots \KKi^+_1(x_n)\bbm 1 \\ 0 \ebm dx_n\cdots dx_1.
\end{split}
\endeq
If the sequence $\{\mmi_{1,n}^+(x)\}$ converges, then $\mmi_1^+(x)$ will be 
defined as its limit, which clearly has the form of an infinite series.  

Consider the $n^\text{th}$ term in this series.  Let
$\left\|\mathbf{v}\right\| := |v_1|+|v_2|$ be the $\ell^1$ vector norm
and $\left\|\mathbf{M}\right\| =
\max(|M_{11}|+|M_{21}|,|M_{12}|+|M_{22}|)$ be the induced matrix norm.
The key observation is that (because $\Im(E)\le 0$ for $\Im(z)\le 0$) the
assumption $\Im(z)\le 0$ implies that if $y>x$ then $\|\KKi^+_1(y)\|$ is
bounded by a linear combination of $1-\cos(f)$, $|\sin(f)|$, and $|\e
f'+g|$ with constant coefficients independent of $y$ and uniformly
bounded for $|z|\ge\delta>0$.  Therefore whenever $\Im(z)\le 0$ with
$z\neq 0$ we may define a function in $L^\infty(\mathbb{R})$ by \eq
\nu(x) := \int_x^{+\infty}\|\KKi^+_1(y)\|dy.
\endeq
Furthermore, $\|\nu(x)\|_{L^\infty}$ is uniformly bounded in $z$ for 
$z\in\{|z|\geq\delta\}\cap\{\Im(z)\leq0\}$ for every $\delta>0$.   Then
\eq
\begin{split}
\left\|\int_x^{+\infty}\int_{x_1}^{+\infty}\cdots\int_{x_{n-1}}^{+\infty}\right.&\left.\KKi^+_1(x_1)\cdots \KKi^+_1(x_n)\bbm 1 \\ 0 \ebm \,dx_n\cdots dx_1 \right\| \\ 
& \leq  \int^{+\infty}_x\int^{+\infty}_{x_1}\cdots\int^{+\infty}_{x_{n-1}}\|\KKi^+_1(x_1)\|\cdots \|\KKi^+_1(x_n)\|\,dx_n\cdots dx_1 \\
& =  \int_0^{\nu(x)}\int_0^{\nu(x_1)}\cdots\int_0^{\nu(x_{n-1})}d\nu(x_n)
\cdots d\nu(x_1) \\ 
& =  \frac{\nu(x)^n}{n!}\,.
\end{split}
\endeq
It follows that the partial sums $\mmi_{1,n}^+(x)$ are majorized by
those of an exponential series, and so the sequence of partial sums
converges and the limit furnishes the unique solution of the first
column of the integral equation \eqref{m-plus}.  By uniformity of the
convergence, analyticity for $\Im(z)<0$ and continuity for
$z\in\{|z|\geq\delta\}\cap\{\Im(z)\leq0\}$ for each $\delta>0$
extend from the partial sums to the limit $\mmi_1^+(x;z)$.  We
also have the estimate 
\eq \|\mmi_1^+(x)\| \leq e^{\nu(x)} \leq
\exp\left(\int_{-\infty}^{+\infty}\|\KKi^+_1(y)\|dy\right)<\infty,
\endeq
which is uniform for $\delta>0$.  The argument for $\mmi_2^-(x;z)$ is
similar.
\end{proof}
\noindent
The argument in Proposition \ref{m-LHP-at-infinity} fails for $z$ near
$z=0$ because of the coefficient $z^{-1}$ in the matrix entries of
$\KKi^-_1(y)$.  The use of an alternate gauge, which we call the
\emph{zero gauge}, circumvents this problem.  We define a new set of
functions in terms of the Jost solutions $\JJi^\pm(x)$ by \eq
\label{zero-gauge}
\JJz^\pm(x) := \bbm  \displaystyle \cos\left(\frac{f(x)}{2}\right) & 
\displaystyle \sin\left(\frac{f(x)}{2}\right) \\ \\
\displaystyle -\sin\left(\frac{f(x)}{2}\right) & 
\displaystyle \cos\left(\frac{f(x)}{2}\right) \ebm \JJi^\pm(x)\,,
\endeq
with columns $\JJz^\pm(x) =: [ \jjz^\pm_1(x),  \jjz^\pm_2(x)]$.  Note  that 
this gauge transformation can be interpreted as a rotation of the Jost 
solution column vectors by an angle $-f(x)/2$.  It follows by direct
calculation that the gauge-transformed
matrices $\JJz^\pm(x)$ satisfy the modified eigenvalue equation
\eq
4i\e\JJz^\pm_x = \left(4E\sigma_3+\QQz\right)\JJz^\pm
\label{eq:modee}
\endeq
where
\eq
\QQz := \bbm -z(1-\cos(f)) & -z\sin(f)+i(\e f'-g) \\ -z\sin(f)-i(\e f'-g) & z(1-\cos(f)) \ebm.
\endeq
Assuming the boundary conditions
\begin{equation}
\lim_{x\to-\infty}u(x,0)=0 \quad\quad\text{and}\quad\quad 
\lim_{x\to +\infty}u(x,0)=2\pi n
\end{equation}
hold in a suitable sense, the required behavior of 
$\JJz^\pm(x)$ as $x\to\pm\infty$ is derived from \eqref{zero-gauge}
and \eqref{Psi-infinity-normalizations}:
\eq
\begin{split}
&\JJz^+(x) = \bbm (-1)^{n+1}e^{-iEx/\e} & 0 \\ 0 & (-1)^{n+1}e^{iEx/\e} \ebm +o(1) \quad \text{as }x\to +\infty \text{ for }z\in\mathbb{R},\\
&\JJz^-(x) = \bbm e^{-iEx/\e} & 0 \\ 0 & e^{iEx/\e} \ebm +o(1) \quad \text{as } x\to-\infty \text{ for }z\in\mathbb{R}.
\end{split}
\label{eq:J0norm}
\endeq
Analogous to equation (\ref{m-plusminus}), define
\eq
\MMz^\pm(x) := \JJz^\pm(x) e^{iEx\sigma_3/\e}
\endeq
with columns $\MMz^\pm(x)=:[\mmz_1^\pm(x), \mmz_2^\pm(x)]$.  It follows 
by integrating \eqref{eq:modee} using the boundary conditions \eqref{eq:J0norm}
that $\MMz^\pm(x)$ satisfy the integral equations
\eq
\label{m-zero-plus}
\MMz^+(x) = \bbm (-1)^{n+1} & 0 \\ 0 & (-1)^{n+1} \ebm - \frac{1}{4i\e}\int_x^{+\infty} e^{-iE(x-y)\sigma_3/\e}\QQz(y)\MMz^+(y)e^{iE(x-y)\sigma_3/\e}dy,
\endeq
\eq
\label{m-zero-minus}
\MMz^-(x) = \bbm 1 & 0 \\ 0 & 1 \ebm  + \frac{1}{4i\e}\int_{-\infty}^x e^{-iE(x-y)\sigma_3/\e}\QQz(y)\MMz^-(y)e^{iE(x-y)\sigma_3/\e}dy.
\endeq
Now these modified integral equations for the gauge-transformed
solutions $\MMz^\pm(x)$ are used to show that the columns of
$\MMi^\pm(x)$ are continuous in a neighborhood of $z=0$ in appropriate
half-planes.
\begin{prop}
\label{m-LHP-at-zero}
Suppose $1-\cos(f)$, $\sin(f)$, $\e f'-g\in L^1$.  Then for each
$x\in\mathbb{R}$ and for each $\delta>0$, $\mmi_1^+(x;z)$ and
$\mmi_2^-(x;z)$ are continuous functions of $z$ in the region
$z\in\{|z|<\delta\}\cap\{\Im(z)\leq0\}$.
\end{prop}
\begin{proof}
The function $\mmz_1^+(x)$ is constructed iteratively from equation 
(\ref{m-zero-plus}), 
similar to the construction of $\mmi_1^+(x)$ in Proposition 
\ref{m-LHP-at-infinity}.  Set $\mmz_{1,0}^+(x) := [(-1)^{n+1}, 0]^\mathsf{T}$.
Define the $n^\text{th}$ iterate by
\eq
\mmz_{1,n}^+(x):= \bbm (-1)^{n+1} \\ 0 \ebm - 
\int_x^{+\infty}\KKz^+_1(y;x,z)\mmz_{1,n-1}^+(y)\,dy\,,
\endeq
where
\eq
\label{KKz+1}
\KKz^+_1(y;x,z) := \frac{1}{4i\e}\bbm -z(1-\cos(f)) & -z\sin(f)+i(\e f'-g) \\ \left(-z\sin(f)-i(\e f'-g)\right)e^{2i(x-y)E/\e} & z(1-\cos(f))e^{2i(x-y)E/\e} \ebm.
\endeq
Aside from the exponential factors $e^{2i(x-y)E/\e}$, everywhere that
a factor of $z^{-1}$ occurred in $\KKi^+_1(y)$ there is in
$\KKz^+_1(y)$ a factor of $z$.  This allows parallel analysis as in
the proof of Proposition \ref{m-LHP-at-infinity} to go through with
the condition $|z|>\delta$ replaced by the condition $|z|<\delta$.
Thus, the iterates converge and $\mmz_1^+(x;z)$ is analytic in the
lower half $z$-plane and continuous in the closed lower half $z$-plane
for bounded $z$.  A similar argument works for $\mmz^-_2(x;z)$ as
well.  Finally, the gauge transformation (\ref{zero-gauge}) is
independent of $z$ and so does not affect the continuity, and
therefore $\mmi_1^+(x;z)$ and $\mmi_2^-(x;z)$ as defined by
Proposition \ref{m-LHP-at-infinity} are in fact continuous in the
whole closed lower half-plane.
\end{proof}
Together, Propositions \ref{m-LHP-at-infinity} and \ref{m-LHP-at-zero}
show $\mmi_1^+(x;z)$ and $\mmi_2^-(x;z)$ are analytic in the lower
half $z$-plane and continuous in the closed lower half $z$-plane.  An
analogous result holds for $\mmi_1^-(x;z)$ and $\mmi_2^+(x;z)$ in the
upper half-plane.
\begin{prop}
\label{m-UHP-at-zero}
Suppose $1-\cos(f)$, $\sin(f)$, $f'$, $g\in L^1$.  If $\Im(z)\ge 0$,
then the first column of \eqref{m-minus} and the second column of
\eqref{m-plus} uniquely define solutions $\mmi_1^-(x;z)$ and
$\mmi_2^+(x;z)$.  These functions are, for each $x\in\mathbb{R}$, 
analytic for $\Im(z)>0$ and continuous for $\Im(z)\geq 0$.
\end{prop}
\begin{proof}
The proof follows that of Propositions \ref{m-LHP-at-infinity} and 
\ref{m-LHP-at-zero}, taking into account the sign of the real part 
of the exponential factors in equations 
(\ref{m-plus}) and (\ref{m-minus}).
\end{proof}
Propositions \ref{m-LHP-at-zero} and \ref{m-UHP-at-zero} 
can be summarized as follows.
\begin{thm}[Kaup, \cite{Kaup:1975}]
\label{half-plane-extensions}
Suppose $1-\cos(f), \sin(f), f', g\in L^1$.  Then $\mmi_1^-(x;z)$ and
$\mmi_2^+(x;z)$ are well-defined and for each $x\in\mathbb{R}$ are
continuous for $z\in\mathbb{R}$ and extend continuously to analytic
functions in the upper half $z$-plane.  Similarly, $\mmi_1^+(x;z)$ and
$\mmi_2^-(x;z)$ are well-defined and for each $x\in\mathbb{R}$ are
continuous for $z\in\mathbb{R}$ and extend continuously to analytic
functions in the lower half $z$-plane.
\end{thm}
Next we establish a lemma showing that under the assumption of a
little more smoothness of the potentials, the $x$ derivatives of the
columns of $\MMi^\pm(x)$ are, for each fixed $x\in\mathbb{R}$,
uniformly bounded in the appropriate closed half-planes for $z$.
\begin{lemma}
\label{MMix-boundedness}
Suppose $1-\cos(f), \sin(f), f', f'', g, g' \in L^1$.  Then
$\mmi_{1x}^+(x;z)$ and $\mmi_{2x}^-(x;z)$ are uniformly bounded in $x$
for each fixed $z$ with $\Im(z)\le 0$, having $L^\infty$ norms that
are uniformly bounded for all such $z$.  Similarly, $\mmi_{1x}^-(x;z)$
and $\mmi_{2x}^+(x;z)$ are uniformly bounded in $x$ for each fixed $z$
with $\Im(z)\ge 0$, having $L^\infty$ norms that are uniformly bounded
for all such $z$.
\end{lemma}
\begin{proof}
  We show the result for $\mmi_{1x}^+$.  The proofs of the results for
  $\mmi_{2x}^-$, $\mmi_{1x}^-$, and $\mmi_{2x}^+$ are similar.  Choose
  $z$ such that $\Im(z)\leq0$.  From \eqref{m-plus}, $\mmi_1^+$
  satisfies the integral equation \eq
\label{mmi1+-integral-eqn}
\mmi_1^+(x;z) = \bbm 1 \\ 0 \ebm - \int_x^{+\infty} \KKi_1^+(y;x,z)
\mmi_1^+(y;z)\,dy
\endeq
with $\KKi_1^+$ defined by \eqref{KKi-1-plus}.  We write the entries of $\QQi$ 
and $\mmi_1^+$ as
\eq
\QQi=:\bbm \Qi_{11} & \Qi_{12} \\ \Qi_{21} & \Qi_{22} \ebm, \quad \mmi_1^+ =: \bbm \mi_{11}^+ \\ \mi_{21}^+ \ebm.
\endeq
The first entry of \eqref{mmi1+-integral-eqn} is 
\eq
\label{mi11+-integral-eqn}
\mi_{11}^+(x;z) = 1-\frac{1}{4i\e}\int_x^{+\infty}\Qi_{11}^+(y;z)\mi_{11}^+(y;z)+\Qi_{12}^+(y;z)\mi_{21}^+(y;z)\, dy,
\endeq
and differentiation in $x$ gives
\eq
\mi_{11x}^+(x;z) = \frac{1}{4i\e}[\Qi_{11}^+(x;z)\mi_{11}^+(x;z)+\Qi_{12}^+(x;z)\mi_{21}^+(x;z)].
\endeq
Thus $\mi_{11x}^+(x;z)$ is uniformly bounded in $x$ with $L^\infty$
norm uniformly bounded for $z\in\{|z|>\delta\}\cap\{\Im(z)\leq0\}$
because $\mmi_1^+$ and $\QQi$ are (we see that $f',g\in L^\infty$ by noting
$f'',g'\in L^1$ by assumption and applying the fundamental theorem of
calculus).  The second entry of \eqref{mmi1+-integral-eqn} is \eq
\mi_{21}^+(x;z) =
-\frac{1}{4i\e}\int_x^{+\infty}e^{2iE(x-y)/\e}[\Qi_{21}(y;z)\mi_{11}^+(y;z)
+ \Qi_{22}(y;z)\mi_{21}^+(y;z)]\,dy.
\endeq
Taking an $x$-derivative gives
\eq
\begin{split}
\mi_{21x}^+(x;z) = & \; \frac{1}{4i\e}[\Qi_{21}(x;z)\mi_{11}^+(x;z) + \Qi_{22}(x;z)\mi_{21}^+(x;z)] \\
 & \;-\frac{1}{4i\e}\int_x^{+\infty}\frac{2iE}{\e}e^{2iE(x-y)/\e}[\Qi_{21}(y;z)\mi_{11}^+(y;z) + \Qi_{22}(y;z)\mi_{21}^+(y;z)]\,dy \\
 = & \;\frac{1}{4i\e}[\Qi_{21}(x;z)\mi_{11}^+(x;z) + \Qi_{22}(x;z)\mi_{21}^+(x;z)] \\
 & \;+\frac{1}{4i\e}\int_x^{+\infty}\frac{d}{dy}\left(e^{2iE(x-y)/\e}\right)[\Qi_{21}(y;z)\mi_{11}^+(y;z) + \Qi_{22}(y;z)\mi_{21}^+(y;z)]\,dy.
\end{split}
\endeq
Now (for $y>x$ and in the indicated region of the $z$-plane) we have
$\exp(2iE(x-y)/\e),\mi_{11}^+,\mi_{21}^+\in L^\infty$.  Also, $\QQi
\to 0$ as $x\to\pm\infty$.  To see this, note that the
limiting value of $f_x$ exists as $x\rightarrow\pm\infty$ because $f_{xx}
\in L^1$, and moreover since $f_x\in L^1$ both limits must be zero.
The same reasoning holds for $1-\cos(f)$, $\sin(f)$, and $g$.
Therefore, \eq
\label{limits-of-Q-elements}
\lim_{x\to\pm\infty}1-\cos(f)=0, \quad \lim_{x\to\pm\infty}\sin(f)=0, \quad \lim_{x\to\pm\infty}f_x=0, \quad \lim_{x\to\pm\infty}g=0.
\endeq
Thus $\exp(2iE(x-y)/\e)[\Qi_{21}(y)\mi_{11}^+(y) + \Qi_{22}(y)\mi_{21}^+(y)]\to 0$ as 
$y\to+\infty$, and so integrating by parts and distributing the $y$-derivative 
gives
\eq
\begin{split}
\mi_{21x}^+(x;z) = & -\frac{1}{4i\e}\int_x^{+\infty}e^{2iE(x-y)/\e}[\Qi_{21y}(y;z)\mi_{11}^+(y;z)+\Qi_{21}(y;z)\mi_{11y}^+(y;z)+\Qi_{22y}(y;z)\mi_{21}^+(y;z)]dy \\
   & -\frac{1}{4i\e}\int_x^{+\infty}e^{2iE(x-y)/\e}\Qi_{22}(y;z)\mi_{21y}^+(y;z)dy \\
   =: & \; I(x) - \int_x^{+\infty}J(y;x,z)\mi_{21y}^+(y;z)dy.
\end{split}
\endeq
Note that $I(x)$ is uniformly bounded in $x$ with $L^\infty$ norm uniformly 
bounded for $z\in\{|z|<\delta\}\cap\{\Im(z)\leq0\}$ and that $J\in L^1$ 
with norm uniformly bounded for 
$z\in\{|z|<\delta\}\cap\{\Im(z)\leq0\}$.  Therefore, by an iteration 
argument as in the proof of Proposition \ref{m-LHP-at-infinity}, 
\eq
\|m_{21x}^+\|_{L^\infty} \leq \|I\|_{L^\infty} 
\exp\left(\int_{-\infty}^{+\infty}|J(y)|\,dy\right)<\infty
\endeq
where the bound is uniform for
$z\in\{|z|<\delta\}\cap\{\Im(z)\leq0\}$.  The uniform bound for
$z\in\{|z|>\delta\}\cap\{\Im(z)\leq0\}$ is shown similarly using the
zero gauge defined by \eqref{zero-gauge}.
\end{proof}

The assumption of additional smoothness of the potentials as above also
provides limiting values of the columns of $\MMi^\pm(x;z)$ in various
situations.
\begin{prop}
\label{m-limits}
Suppose $1-\cos(f), \sin(f), f', f'', g, g' \in L^1$.  Then the columns 
of $\MMi^\pm(x;z)$ have the following limits in $x$ and $z$:
\eq
\label{m-limit-large-x}
\mathop{\lim_{x\to+\infty}}_{\Im(z)\leq0}\mmi_1^+(x;z) = 
\bbm 1 \\ 0 \ebm, \quad 
\mathop{\lim_{x\to-\infty}}_{\Im(z)\leq0}\mmi_2^-(x;z) = 
\bbm 0 \\ 1 \ebm, \quad 
\mathop{\lim_{x\to-\infty}}_{\Im(z)\geq0}\mmi_1^-(x;z) = 
\bbm 1 \\ 0 \ebm, \quad 
\mathop{\lim_{x\to+\infty}}_{\Im(z)\geq0}\mmi_2^+(x;z) = \bbm 0 \\ 1 \ebm,
\endeq
\eq
\label{m-limit-large-z}
\mathop{\lim_{z\to\infty}}_{\Im(z)\leq0}\mmi_1^+(x;z) = 
\bbm 1 \\ 0 \ebm, \quad 
\mathop{\lim_{z\to\infty}}_{\Im(z)\leq0}\mmi_2^-(x;z) = 
\bbm 0 \\ 1 \ebm, \quad 
\mathop{\lim_{z\to\infty}}_{\Im(z)\geq0}\mmi_1^-(x;z) = 
\bbm 1 \\ 0 \ebm, \quad 
\mathop{\lim_{z\to\infty}}_{\Im(z)\geq0}\mmi_2^+(x;z) = \bbm 0 \\ 1 \ebm,
\endeq
\eq
\begin{split}
\label{m-limit-zero}
\mathop{\lim_{z\to 0}}_{\Im(z)\leq0}\mmi_1^+(x;z) = (-1)^{n+1}
\bbm \displaystyle \cos\left(\frac{f(x)}{2}\right) \vspace{0.1 in}\\ 
\displaystyle \sin\left(\frac{f(x)}{2}\right) \ebm, \quad 
\mathop{\lim_{z\to 0}}_{\Im(z)\leq0}\mmi_2^-(x;z) = \bbm 
\displaystyle -\sin\left(\frac{f(x)}{2}\right) \vspace{0.1 in}\\ 
\displaystyle \cos\left(\frac{f(x)}{2}\right) \ebm, \\
\mathop{\lim_{z\to 0}}_{\Im(z)\geq0}\mmi_1^-(x;z) = 
\bbm \displaystyle \cos\left(\frac{f(x)}{2}\right) \vspace{0.1 in}\\ 
\displaystyle \sin\left(\frac{f(x)}{2}\right) \ebm, \quad 
\mathop{\lim_{z\to 0}}_{\Im(z)\geq0}\mmi_2^+(x;z) = (-1)^{n+1}
\bbm \displaystyle -\sin\left(\frac{f(x)}{2}\right) \vspace{0.1 in}\\
\displaystyle \cos\left(\frac{f(x)}{2}\right) \ebm.
\end{split}
\endeq
\end{prop}
\begin{proof}
  We will prove the statements concerning $\mmi_1^+(x;z)$; the proofs
  of the corresponding limits for $\mmi_1^-(x;z)$, $\mmi_2^+(x;z)$,
  and $\mmi_2^-(x;z)$ are similar.

  We first establish the limit in $x$.  Fix
  $z\in\{|z|>\delta\}\cap\{\Im(z)\leq0\}$ for some fixed $\delta>0$.
  Consider the integral equation \eqref{mmi1+-integral-eqn} for
  $\mmi_1^+(x;z)$.  The product $\KKi_1^+\mmi_1^+\in L^1$ as a
  function of $y$ since $|z|>\delta$, since $\cos(f)-1, \sin(f), f', g \in
  L^1$, and since $\mmi_1^+\in L^\infty$ for
  $z\in\{|z|>\delta\}\cap\{\Im(z)\leq0\}$.  Furthermore,
  $\KKi_1^+\mmi_1^+\chi_{[x,+\infty)}$ tends to zero pointwise in $y$
  as $x\to +\infty$.  Therefore the limit for $\mmi_1^+(x;z)$ as
  $x\to+\infty$ holds by dominated convergence for
  $z\in\{|z|>\delta\}\cap\{\Im(z)\leq0\}$.  The result for
  $z\in\{|z|<\delta\}\cap\{\Im(z)\leq0\}$ holds by the same reasoning
  applied to the integral equation 
\eq \mmz_1^+(x;z) := \bbm (-1)^{n+1} \\ 0
  \ebm - \int_x^{+\infty}\KKz^+_1(y;x,z)\mmz_1^+(y;z)\,dy,
\endeq
written in the zero gauge with $\KKz_1^+$ given by \eqref{KKz+1}, and
the use of the gauge transformation \eqref{zero-gauge} to go back to
$\mmi_1^+(x;z)$.

Next consider the limit of $\mmi_1^+(x;z)$ as $z\to\infty$ for
$\Im(z)\leq0$.  The second entry of \eqref{mmi1+-integral-eqn} may be
written as \eq \mi_{21}^+(x;z) =
\frac{1}{4i\e}\int_x^{+\infty}\frac{d}{dy}\left(\frac{\e}{2iE}e^{2iE(x-y)/\e}\right)[\Qi_{21}(y;z)\mi_{11}^+(y;z)
+ \Qi_{22}(y;z)\mi_{21}^+(y;z)]\,dy.
\endeq
Integration by parts gives
\eq
\begin{split}
\mi_{21}^+(x;z) = & \frac{1}{8E}[\Qi_{21}(x;z)\mi_{11}^+(x;z)+\Qi_{22}(x;z)\mi_{21}^+(x;z)] \\
& +\int_x^{+\infty}\frac{1}{8E}e^{2iE(x-y)/\e}\frac{d}{dy}[\Qi_{21}(y;z)\mi_{11}^+(y;z)+\Qi_{22}(y;z)\mi_{21}^+(y;z)]\,dy.
\end{split}
\endeq
Since $f',g\in L^\infty$ (because $f'',g'\in L^1$), we have $\Qi_{21},
\Qi_{22} \in L^\infty$ for $|z|>\delta$.  Since also $\mi_{11}^+$ and
$\mi_{21}^+$ are uniformly bounded for $\Im(z)\leq0$ by Theorem
\ref{half-plane-extensions}, the boundary term $[\Qi_{21}\mi_{11}^+ +
\Qi_{22}\mi_{21}^+]/8E$ vanishes as $z\to\infty$ for $\Im(z)\leq0$
(and hence as $E\rightarrow\infty$).  As for the integral term,
$\Qi_{21}, \Qi_{21y}, \Qi_{22}, \Qi_{22y} \in L^1$ for $|z|>\delta$
and $e^{2iE(x-y)/\e},\mi_{11}^+,\mi_{11y}^+,\mi_{21}^+,\mi_{21y}^+$
are uniformly bounded for $y>x$ and $\Im(z)\leq0$.  Therefore
$e^{2iE(x-y)/\e}d/dy[\Qi_{21}\mi_{11}^++\Qi_{22}\mi_{21}^+] \in L^1$
for $z\in\{|z|>\delta\}\cap\{\Im(z)\leq 0\}$.  Since
$E\rightarrow\infty$ as $z\rightarrow\infty$, the integrand tends to
zero pointwise in $y$ almost everywhere as $z\to\infty$ for
$\Im(z)\leq0$.  By dominated convergence, \eq
\mathop{\lim_{z\to\infty}}_{\Im(z)\leq0}\mi_{21}^+(x;z) = 0.
\label{eq:m21pluslimit}
\endeq
To analyze $\mi_{11}^+(x;z)$ in the same limit, consider the integral
equation \eqref{mi11+-integral-eqn}.  The integrand is in $L^1$ for
$z\in\{|z|>\delta\}\cap\{\Im(z)\leq0\}$ since
$\Qi_{11}^+,\Qi_{12}^+\in L^1$ for $|z|>\delta$ and
$\mi_{11}^+,\mi_{21}^+\in L^\infty$ for $\Im(z)\leq0$.  In addition,
the integrand tends to zero pointwise as $z\to\infty$ for
$\Im(z)\leq0$ since $\Qi_{11}^+(y)$ tends to zero as $z\to\infty$ 
and \eqref{eq:m21pluslimit} holds, while
and $\mi_{11}^+, \Qi_{12}^+\in L^1$
for $\Im(z)\leq0$.  Thus, by dominated convergence, \eq
\mathop{\lim_{z\to\infty}}_{\Im(z)\leq0}\mi_{11}^+(x;z) = 1.
\endeq

Finally, we consider the asymptotic behavior in the limit $z\to 0$.  The
statement that
\eq
\mathop{\lim_{z\to0}}_{\Im(z)\leq0}\mmz_1^+(x;z) = \bbm (-1)^{n+1} \\ 0 \ebm
\endeq
holds may be shown as above using the zero gauge.  Then the limit of 
$\mmz_1^+(x;z)$ as $z\to0$ for $\Im(z)\leq0$ follows by inverting the gauge
transformation with the help of \eqref{zero-gauge}.
\end{proof}
Note that, from the asymptotic behavior of the columns of $\MMi^\pm(x;z)$ in
the limits $x\to\pm\infty$ and the fact that (Abel's theorem) Wronskians of
solutions of \eqref{eigenvalue-eqn2} are independent of $x$, we have
$\det(\JJi^\pm(x;z))\equiv 1$ for $x\in\mathbb{R}$ and $z\in\mathbb{R}$.

\subsection{Scattering data}
The Jost solution matrices $\JJi^+(x;z)$ and $\JJi^-(x;z)$ are both
fundamental solution matrices of the same system
\eqref{eigenvalue-eqn2}, so consequently the columns of $\JJi^+(x;z)$
are necessarily linear combinations (with coefficients independent of
$x$) of those of $\JJi^-(x;z)$.  Therefore, there exists a matrix
$\mathbf{S}(z)=\mathbf{S}(z;t,\e)$ such that \eq
\label{phi-plus-ito-phi-minus}
\JJi^+(x;z) = \JJi^-(x;z)\mathbf{S}(z)\,,\quad\quad
\mathbf{S}(z;t,\e) = \bbm S_{11}(z;t,\e) & S_{12}(z;t,\e) \\ S_{21}(z;t,\e) & S_{22}(z;t,\e) \ebm, \quad z\in\mathbb{R},
.
\endeq
The matrix $\mathbf{S}(z)$ is called the \emph{scattering matrix}.
The $t$-dependence of its elements comes from considering $f$ and $g$
to depend parametrically on $t$ (for example, if $f=u$ and $\e g= u_t$
come from a solution of the sine-Gordon equation \eqref{sine-Gordon}).
We will calculate this time dependence shortly (and in fact it will
turn out that the diagonal elements are independent of $t$).  Using
the fact that $\det(\JJi^\pm(x;z))=1$, we easily obtain the Wronskian
formulae
\begin{equation}\begin{split}
S_{11}(z)&=\det[\jji_1^+(x;z),\jji_2^-(x;z)]\,,\quad\quad
S_{12}(z)=\det[\jji_2^+(x;z),\jji_2^-(x;z)]\,,\\
S_{21}(z)&=\det[\jji_1^-(x;z),\jji_1^+(x;z)]\,,\quad\quad
S_{22}(z)=\det[\jji_1^-(x;z),\jji_2^+(x;z)]\,.
\end{split}
\label{eq:Wronskians}
\end{equation}
These formulae, in conjunction with Theorem
\ref{half-plane-extensions} and Proposition~\ref{m-limits}, provide a
proof of the following.
\begin{lemma}[Kaup, \cite{Kaup:1975}]
\label{half-plane-lemma}
Suppose $1-\cos(f), \sin(f), f', f'', g, g'\in L^1$.  Then $S_{22}(z)$
is continuous for $z\in\mathbb{R}$ and has a continuous extension into
the upper half $z$-plane as an analytic function, 
while $S_{11}(z)$ is continuous for
$z\in\mathbb{R}$ and has a continuous extension into the lower
half $z$-plane as an analytic function.  Moreover, 
\begin{equation}
\mathop{\lim_{z\to\infty}}_{\Im(z)\le 0}S_{11}(z)=
\mathop{\lim_{z\to\infty}}_{\Im(z)\ge 0}S_{22}(z)=1\,,
\end{equation}
and similarly
\begin{equation}
  \mathop{\lim_{z\to 0}}_{\Im(z)\le 0}S_{11}(z)=
\mathop{\lim_{z\to 0}}_{\Im(z)\ge 0}S_{22}(z)=
  (-1)^{n+1}\,.
\end{equation}
\end{lemma}
Next we record several important symmetries of the scattering matrix.
\begin{prop}[Kaup, \cite{Kaup:1975}]
\label{S22-and-S21-ito-S11-and-S12}
For $z\in\mathbb{R}$, the elements of the scattering matrix are related by $S_{11}(z) = S_{22}(-z) = S_{22}(z)^*$ and $S_{12}(z) = -S_{21}(-z) = -S_{21}(z)^*$.
\end{prop}
\begin{proof}
  Here it is essential that $z\in\mathbb{R}$ so that both columns of
  $\JJi^\pm(x;z)$ are simultaneously defined.  In the eigenvalue
  equation \eqref{eigenvalue-eqn2}, the coefficient matrix has the
  symmetry $\overline{\mathbf{L}}(x;z) = \sigma_2
  \overline{\mathbf{L}}(x;-z) \sigma_2$.  Therefore, \eq \sigma_2
  \JJi^\pm_x(x;-z) = \overline{\mathbf{L}}(x;z) \sigma_2
  \JJi^\pm(x;-z),
\endeq
and so $\JJi^\pm(x;z) = \sigma_2\JJi^\pm(x;-z)\mathbf{C}^\pm$ for 
some constant matrices $\mathbf{C}^\pm$.  Write 
\eq
\MMi^\pm(x;z) = \JJi^\pm(x;z)e^{iEx\sigma_3/\e} = \sigma_2\JJi^\pm(x;-z)\mathbf{C}^\pm e^{iEx\sigma_3/\e} = \sigma_2\MMi^\pm(x;-z)e^{iEx\sigma_3/\e}\mathbf{C}^\pm e^{iEx\sigma_3/\e}.
\endeq
Taking the limit as $x\to\pm\infty$ and using
Proposition~\ref{m-limits} shows that $\mathbf{C}^\pm = \sigma_2$.

Next, substituting the identity
\eq
\label{PsiI-at-z-ito-PsiI-at-minus-z}
\JJi^\pm(x;z) = \sigma_2\JJi^\pm(x;-z)\sigma_2
\endeq
into equation (\ref{phi-plus-ito-phi-minus}) gives
\eq
\mathbf{S}(z) = \sigma_2 \mathbf{S}(-z)\sigma_2,
\endeq
which shows $S_{11}(z) = S_{22}(-z)$ and $S_{12}(z) = -S_{21}(-z)$.
The matrix $\overline{\mathbf{L}}$ also has the symmetry
$\overline{\mathbf{L}}(x;z) = -\overline{\mathbf{L}}(x,z^*)^\dagger$
that holds for all $z\in\mathbb{C}$, in particular $z\in\mathbb{R}$.
Restricting to $z\in\mathbb{R}$, this implies \eq
\JJi_x^\pm(x;z)^\dagger =
-\JJi^\pm(x;z)^\dagger\overline{\mathbf{L}}(x;z).
\label{eq:Jdaggerx}
\endeq
Furthermore,
\eq
\frac{d}{dx}\JJi^\pm(x;z)^{-1} = -\JJi^\pm(x;z)^{-1}
\frac{d}{dx}\JJi^\pm(x;z)\cdot\JJi^\pm(x;z)^{-1} = -\JJi^\pm(x;z)^{-1}\overline{\mathbf{L}}(x;z),
\endeq
and so by comparison with \eqref{eq:Jdaggerx} $\JJi^\pm(x;z)^\dagger =
\mathbf{D}^\pm\JJi^\pm(x;z)^{-1}$ for some constant matrices
$\mathbf{D}^\pm$.  Now 
\eq \MMi^\pm(x;z)^\dagger =
e^{-iEx\sigma_3/\e}\JJi^\pm(x;z)^\dagger =
e^{-iEx\sigma_3/\e}\mathbf{D}^\pm\JJi^\pm(x;z)^{-1} =
e^{-iEx\sigma_3/\e}\mathbf{D}^\pm
e^{iEx\sigma_3/\e}\MMi^\pm(x;z)^{-1}.
\endeq
Again taking the limit as $x\to\pm\infty$ and using
Proposition~\ref{m-limits} shows that $\mathbf{D}^\pm =
\mathbb{I}$.  Substituting the identity \eq
\label{PsiI-at-z-ito-PsiI-at-zbar}
\JJi^\pm(x;z)^\dagger = \JJi^\pm(x;z)^{-1}
\endeq
into equation (\ref{phi-plus-ito-phi-minus}) yields
\eq
\mathbf{S}(z)^\dagger = \mathbf{S}(z)^{-1},
\endeq
from which it follows that $S_{11}(z) = S_{22}(z)^*$ and 
$S_{12}(z) = -S_{21}(z)^*$.
\end{proof}
By definition, the eigenvalues for the scattering problem
\eqref{eigenvalue-eqn2} are the complex numbers $z$ for which there is
a solution of (\ref{eigenvalue-eqn2}) in $L^2(\mathbb{R})$.  The Jost
solution $\jji_1^-(x;z)$ is defined for $\Im(z)\ge 0$ and, according
to Proposition~\ref{m-limits} and the relation
\eqref{m-ito-Phi-minus-and-Phi-plus} between $\jji_1^-(x;z)$ and
$\mmi_1^-(x;z)$, $\jji_1^-(x;z)$ decays exponentially to zero as
$x\to-\infty$ if and only if $\Im(z)>0$.  Similarly, the Jost solution
$\jji_2^+(x;z)$ is defined for $\Im(z)\ge 0$ and decays exponentially
to zero as $x\to +\infty$ if and only if $\Im(z)>0$.  All other
solutions blow up exponentially in these limits.  Therefore, the
eigenvalues $z$ in the open upper half-plane are exactly those values
of $z$ for which $\jji_1^-(x;z)$ is proportional to $\jji_2^+(x;z)$.
Recalling the representation \eqref{eq:Wronskians} of $S_{22}(z)$ as a
Wronskian, the eigenvalues with $\Im(z)>0$ are precisely the roots of
$S_{22}(z)$.  By similar arguments, the eigenvalues in the open lower
half-plane are precisely the roots of $S_{11}(z)$.  There are no real
eigenvalues, because according to Proposition~\ref{m-limits} all
solutions oscillate for large $|x|$ when $z$ is real.  Suppose that
$z$ is an eigenvalue in the upper half-plane.  Then it follows that
there is a nonzero \emph{proportionality constant} $\eta$ such that
\begin{equation}
\jji_1^-(x;z)=\eta\,\jj_2^+(x;z)\,.
\label{sigma-n-def}
\end{equation}

Let $I$, $I\!I$, $I\!I\!I$, and $IV$ be the four quadrants of the
plane.  The following corollary can be obtained from Proposition
\ref{S22-and-S21-ito-S11-and-S12} since $S_{11}(z)$ extends from the
real axis to the lower half-plane as $S_{22}(z^*)^*$.
\begin{cor}
If $z$ is an eigenvalue on the imaginary axis, then $-z$ 
is also an eigenvalue.  Similarly, if $z\in I$ is an eigenvalue, then 
$-z^*\in I\!I$, $-z\in I\!I\!I$, and $z^*\in IV$ are also 
eigenvalues.
\end{cor}
\noindent
As a result, the eigenvalues come either in pairs (on the imaginary
axis) or in quartets off the axes.  Using the symmetries  
\eq
\label{phi-symmetries}
\jji_1^\pm(x;z) = \bbm 0 & 1 \\ -1 & 0 \ebm \jji_2^\pm(x;-z) = \bbm 0 & 1 \\ -1 & 0 \ebm \jji_2^\pm(x;z^*)^*,
\endeq
(which follow from \eqref{PsiI-at-z-ito-PsiI-at-minus-z} and
\eqref{PsiI-at-z-ito-PsiI-at-zbar} upon extension to complex $z$)
the relation \eqref{sigma-n-def} holding for an eigenvalue $z$ with $\Im(z)>0$
implies that 
\eq
\jji_1^-(x;-z^*) = \eta^*\jji_2^+(x;-z^*),\quad
\jji_2^-(x;-z) = -\eta\,\jji_1^+(x;-z), \quad 
\jji_2^-(x;z^*) = -\eta^*\jji_1^+(x;z^*).
\label{eq:etasyms}
\endeq
If $z$ is an eigenvalue on the positive imaginary axis, then these 
symmetries show $\eta\in\mathbb{R}$.  Note that if $S_{21}(z)$ has a 
meromorphic extension into the upper half-plane and is finite and nonzero 
at an eigenvalue $z$ in the upper half-plane, then $\eta = S_{21}(z)$.

\begin{defn}
  Suppose that $S_{22}(z)$ has only simple zeros in the open upper
  half-plane.  The scattering data for the Cauchy problem consist of
  (i) the reflection coefficient
\begin{equation}
\rho(z):=\frac{S_{21}(z)}{S_{22}(z)}\,,\quad\quad 
z\in\mathbb{R}, 
\label{eq:refcoeff}
\end{equation}
(ii) the eigenvalues, or the zeros $\{z_n\}$ of $S_{22}(z)$ in the open
upper half-plane, and (iii) the modified proportionality constants
$\{c_n^0\}$ where
\begin{equation}
c_n^0:=\frac{\eta_n}{S_{22}'(z_n)}\,.  
\end{equation}
\end{defn}
\noindent It turns out that this information is sufficient to
reconstruct the potentials $f$ and $g$, assuming that $S_{22}(z)$ has
no real zeros or complex multiple zeros.

Up to this point, we have considered \eqref{eigenvalue-eqn2} (or
equivalently \eqref{JJi-diffeq}) for $u=f(x)$ and $\e u_t=g(x)$ as
fixed functions of $x$.  However, if $u(x,t)$ evolves in time $t$
according to the sine-Gordon equation, then $f$ and $g$ will depend
parametrically on $t$, and so will the Jost solutions of
\eqref{JJi-diffeq}.  We must therefore expect the scattering matrix
$\mathbf{S}(z)=\mathbf{S}(z;t)$ and the proportionality constants
$\{\eta_n=\eta_n(t)\}$ to vary with $t$.  Since the sine-Gordon
equation is the compatibility condition for \eqref{eigenvalue-eqn2}
and \eqref{time-evolution-eqn}, we may use \eqref{time-evolution-eqn}
to calculate the time dependence.  In order that all quantities of
interest remain well-defined as $t$ varies, we introduce a technical
condition on solutions $u(x,t)$ of the sine-Gordon equation
\eqref{sine-Gordon}.
\begin{defn} Let $p\ge 1$.  A solution $u(x,t)$ of the sine-Gordon
  equation \eqref{sine-Gordon} is said to have $L^p$-Sobolev
  regularity if $1-\cos(u)$ $\sin(u)$, $u_x$, $u_{xx}$, $u_t$, and
  $u_{tx}$ all exist in the sense of distributions and lie in the
  space $L^p(\mathbb{R})$ as functions of $x$ for all $t\ge 0$.
\end{defn}
Appendix~\ref{app-well-posed} contains a proof of the fact that as a
dynamical system, the sine-Gordon equation preserves $L^p$-Sobolev
regularity, so in fact, it is only a condition on initial data.  The
case of interest in inverse-scattering theory is $p=1$.
\begin{prop} \label{t-dep} Suppose that $u=u(x,t)$ is a solution of
  the sine-Gordon equation \eqref{sine-Gordon} having $L^1$-Sobolev
  regularity.  Then the corresponding time evolution of the scattering
  data computed by solving \eqref{JJi-diffeq} with potentials
  $f(x)=u(x,t)$ and $g(x)=\e u_t(x,t)$ is given by \eq
\label{S11-S22-time-evolution}
S_{11}(z) \text{ and } S_{22}(z) \text { (and thus the eigenvalues) are independent of } t,
\endeq
\eq
\label{S12-S21-time-evolution}
S_{12}(z;t) = S_{12}(z;0)e^{-2iD(z)t/\e}, \quad S_{21}(z;t) = 
S_{21}(z;0)e^{2iD(z)t/\e},
\endeq
\eq
\label{time-evolution-of-sigma}
\eta_n(t) = \eta_n(0)e^{2iD(z_n)t/\e},
\endeq
where $\eta_n$ is the proportionality constant associated to the
eigenvalue $z_n$ in the open upper half-plane.
\end{prop}
\begin{proof}
  Since $u(x,t)$ is a solution of the sine-Gordon equation, we can
  find functions $c^\pm_1(t;z)$, $c^\pm_2(t;z)$ (independent of $x$)
  such that $\mathbf{w}^\pm_1 := c^\pm_1\jji_1^\pm$ and
  $\mathbf{w}^\pm_2 := c^\pm_2\jji_2^\pm$ are simultaneous solutions
  of the Lax pair \eqref{eigenvalue-eqn2} and
  \eqref{time-evolution-eqn}.  Inserting $\mathbf{w}_1^+$ into
  \eqref{time-evolution-eqn} and using the relation
  \eqref{m-ito-Phi-minus-and-Phi-plus} between $\jji_1^+$ and
  $\mmi_1^+$, we find
\eq
\label{mmi-t-diffeq}
4i\e\frac{d}{dt}\log(c_1^+)\,\mmi_1^+ + 4i\e\mmi_{1t}^+ = \BBi\mmi_1^+.
\endeq
The limits of $\mmi_1^+$ and $\BBi$ as $x\to+\infty$ exist by Proposition 
\ref{m-limits} and equation \eqref{limits-of-Q-elements}.  We now show 
$\mmi_{1t}^+$ also has a limit as $x\to+\infty$.  
Taking a time derivative of \eqref{mmi1+-integral-eqn} shows
\eq
\mmi_{1t}^+(x;z,t) = -\int_x^{+\infty}\KKi_{1t}^+(y;z,t)\mmi_1^+(y;z,t)\,dy 
- \int_x^{+\infty}\KKi_1^+(y;z,t)\mmi_{1t}^+(y;z,t)\,dy\,.
\label{eq:timedepinteqn}
\endeq
For $z\in\{\Im(z)\le 0\}\cap\{|z|>\delta\}$ for any fixed $\delta$, we
have $\KKi_1^+,\KKi_{1t}^+\in L^1$ and $\mmi_1^+\in L^\infty$ as
functions of $y$ by the assumptions that $1-\cos(u), \sin(u), u_x, u_t,
u_{xx}, u_{tx}\in L^1$ (note $u_{tt} = u_{xx}-\sin(u)/\e^2$).
Therefore by an iteration argument, $\mmi_{1t}^+\in L^\infty$ as a
function of $x$ uniformly for $z\in\{\Im(z)\le 0\}\cap\{|z|>\delta\}$.
By an analogous argument using the zero gauge, one sees that
$\mmi_{1t}^+\in L^\infty$ as a function of $x$ for $\Im(z)\le 0$.  It
then follows from a dominated convergence argument applied to
\eqref{eq:timedepinteqn} with this new information in hand that \eq
\mathop{\lim_{x\to+\infty}}_{\Im(z)\le 0}\mmi_{1t}^+(x;z,t) = 0.
\endeq
Using this result to take the limit of \eqref{mmi-t-diffeq} as
$x\to+\infty$ gives 
\eq 4i\e\frac{d}{dt}\log(c_1^+)\bbm 1 \\ 0 \ebm =
4D\bbm 1 \\ 0 \ebm,
\endeq
and so up to a multiplicative constant (independent of $x$ and $t$),
$\mathbf{w}^+_1 = e^{-iDt/\e}\jji_1^+$ for $\Im(z)\le 0$.  Similar
arguments for the other Jost functions show that in the
respective closed half-planes of existence 
\eq
\label{W-ito-Psi}
\mathbf{w}^\pm_1 = e^{-iDt/\e}\jji_1^\pm \quad \text{and} \quad \mathbf{w}^\pm_2 = e^{iDt/\e}\jji_2^\pm\,,
\endeq
again up to multiplicative constants.
Substituting the expressions (\ref{W-ito-Psi}) into the time-evolution
equation \eqref{time-evolution-eqn} gives in particular 
\eq
\label{ODE-for-Psi}
\e\JJi^\pm_t = \e\overline{\mathbf{B}}\JJi^\pm + iD\JJi^\pm\sigma_3\,,
\quad\quad z\in\mathbb{R}.
\endeq
Differentiating equation (\ref{phi-plus-ito-phi-minus}) gives
\eq
\label{time-differentiated-S}
\JJi^+_t = \JJi^-_t\mathbf{S} + \JJi^- \mathbf{S}_t.
\endeq
Substituting (\ref{ODE-for-Psi}) into (\ref{time-differentiated-S})
and using \eqref{phi-plus-ito-phi-minus} gives \eq
\e\frac{d}{dt}\mathbf{S}(z)=iD(z)[\mathbf{S}(z),\sigma_3]=
iD(z)(\mathbf{S}(z)\sigma_3-\sigma_3\mathbf{S}(z))\,,
\endeq
which yields \eqref{S11-S22-time-evolution} and \eqref{S12-S21-time-evolution}.

For the time evolution of a proportionality constant $\eta$ associated to
an eigenvalue $z$ in the upper half-plane via \eqref{sigma-n-def},  
we differentiate the relation \eqref{sigma-n-def}
with respect to $t$ and obtain
\eq
\jji_{1t}^-(x;z,t) = \jji_2^+(x;z,t)\frac{d\eta}{dt} + \eta\,
\jji_{2t}^+(x;z,t)\,.
\endeq
Obtaining the time evolution of $\jji_1^-(x;z,t)$ by substituting
(\ref{W-ito-Psi}) into \eqref{time-evolution-eqn} gives \eq
\e\overline{\mathbf{B}}\jji_1^-(x;z,t) + iD(z)\jji_1^-(x;z,t) =
\e\jji_2^+(x;z,t)\frac{d\eta}{dt} + \e\eta \overline{\mathbf{B}}\jji_2^+(x;z,t)
- i\eta D(z)\jji_2^+(x;z,t)\,.
\endeq
Using \eqref{sigma-n-def} to eliminate $\jji_1^-(x;z,t)$ gives, 
since $\jji_2^+(x;z,t)\neq[0,0]^\mathsf{T}$,
\eq
\e\frac{d\eta}{dt} = 2iD(z)\eta,
\endeq
which gives equation (\ref{time-evolution-of-sigma}).
\end{proof}

\subsection{The matrix $\GG$ and its properties} We now introduce a
piecewise-meromorphic function $\GG(z)=\GG(z;x,t,\e)$ whose
singularities and jump discontinuities encode the scattering data.
Assuming that $u$ is a solution of the sine-Gordon equation with
$L^1$-Sobolev regularity, define a corresponding matrix $\GG(z)$ by \eq
\label{m}
\GG(z) := \begin{cases} \bbm 
\displaystyle \frac{1}{S_{22}(z)} \mmi_1^-(x;z,t), \mmi_2^+(x;z,t) \ebm = 
\bbm \displaystyle \frac{1}{S_{22}(z)}\jji_1^-(x;z,t) e^{iEx/\e}, 
\jji_2^+(x;z,t) e^{-iEx/\e} \ebm, & \Im(z)>0 \vspace{.05in} \\ 
\bbm \displaystyle \mmi_1^+(x;z,t), \frac{1}{S_{11}(z)}\mmi_2^-(x;z,t) 
\ebm = \bbm \displaystyle \jji_1^+(x;z,t) e^{iEx/\e}, 
\frac{1}{S_{11}(z)}\jji_2^-(x;z,t) e^{-iEx/\e} \ebm, & \Im(z)<0\,. \end{cases}
\endeq
From the symmetries of the scattering matrix 
(Proposition \ref{S22-and-S21-ito-S11-and-S12}) and of the Jost functions 
\eqref{phi-symmetries}, $\GG(z)$ has the symmetries
\eq
\label{MMi-symmetries}
\GG(z) = \sigma_2\GG(-z)\sigma_2, \quad \GG(z)^{-1} = -\GG(z^*)^\dagger.
\endeq
The matrix $\GG(z)$ will have poles at the eigenvalues $\{z_n\}$ due to the 
presence of $S_{jj}(z)$ in the denominator.  
\begin{prop}
\label{residue-prop}
Suppose that $u$ is a solution of the sine-Gordon equation with
$L^1$-Sobolev regularity, and let $\GG(z)$ be the corresponding matrix
defined by \eqref{m}.  Let $z_n\in I$ be a simple eigenvalue (that is,
a simple root of $S_{22}(z)$) and let $\eta_n$ be the corresponding
proportionality constant defined by equation (\ref{sigma-n-def}).
Then \eq
\label{residue-conditions}
\begin{split}
\mathop{\rm{Res}}_{z_n}\GG(z) = \lim_{z\to z_n}\GG(z) \bbm 0 & 0 \\ c_n & 0 \ebm, \quad \mathop{\rm{Res}}_{-z_n^*}\GG(z) = \lim_{z\to -z_n^*}\GG(z) \bbm 0 & 0 \\ -c_n^* & 0 \ebm, \\
\mathop{\rm{Res}}_{-z_n}\GG(z) = \lim_{z\to -z_n}\GG(z) \bbm 0 & c_n \\ 0 & 0 \ebm, \quad \mathop{\rm{Res}}_{z_n^*}\GG(z) = \lim_{z\to z_n^*}\GG(z) \bbm 0 & -c_n^* \\ 0 & 0 \ebm.
\end{split}
\endeq
where 
\eq
\label{cn}
c_n = c_n^0 e^{2iE(z_n)x/\e+2iD(z_n)t/\e}, \quad c_n^0:=\frac{\eta_n(0)}{S_{22}'(z_n)}
\endeq
These formulae also hold for eigenvalues $z_n$ on the positive imaginary 
axis, in which case we have $c_n\in i\mathbb{R}$.
\end{prop}
\begin{proof}
Consider $z_n$ in the first quadrant and let $\mathbf{0}_2:=[0,0]^\mathsf{T}$.
Using \eqref{sigma-n-def},
\eq
\begin{split}
\mathop{\rm{Res}}_{z_n}\GG(z) &= 
\bbm \displaystyle \frac{1}{S_{22}'(z_n)}\jji_1^-(x;z_n,t)e^{iE(z_n)x/\e}, 
\mathbf{0}_2\ebm \\
&= \bbm \displaystyle \frac{\eta_n(t) e^{2iE(z_n)x/\e}}{S_{22}'(z_n)}
\jji_2^+(x;z_n,t)e^{-iE(z_n)x/\e},\mathbf{0}_2\ebm \\
&= \lim_{z\to z_n}\GG(z) \bbm 0 & 0 \\ c_n & 0 \ebm.
\end{split}
\endeq
The other three proportionality constants are handled similarly with the
help of \eqref{eq:etasyms}.  
\end{proof}
The matrix $\GG(z)$ also generally has a jump discontinuity across the
real axis.  This discontinuity is expressed in terms of the reflection
coefficient defined by \eqref{eq:refcoeff}.  Note that, by definition,
$\rho(z)\equiv 0$ for
reflectionless potentials. Note also that if $\rho(z;0)\equiv 0$, then
$\rho(z;t)\equiv 0$ for all $t\ge 0$ according to Proposition~\ref{t-dep}. 
\begin{prop}[Cheng et al. \cite{Cheng:1999}]
\label{real-axis-jump-prop}
Suppose that $u$ is a solution of the sine-Gordon equation with
$L^1$-Sobolev regularity, and let $\GG(z)$ be the corresponding matrix
defined for $z\in\mathbb{C}\setminus\mathbb{R}$ by \eqref{m}.  Suppose
also that $S_{11}(z)$ and $S_{22}(z)$ do not vanish for any real $z$.
Then $\GG(z)$ takes continuous boundary values $\GG_+(z)$ and
$\GG_-(z)$ for $z\in\mathbb{R}$ from the upper and lower half-planes
respectively.  Moreover, these boundary values are related by
$\GG_+(z) = \GG_-(z) \mathbf{V}(z)$, where the jump matrix is defined
for $z\in\mathbb{R}$ as the continuous function \eq
\label{real-axis-jump}
\begin{split}
\mathbf{V}(z)=\mathbf{V}(z;x,t,\e) := &
\bbm 1+|\rho(z;0)|^2 & -\rho(z;0)^*e^{-2i(Ex+Dt)/\e} \\ -\rho(z;0) e^{2i(Ex+Dt)/\e} & 1 \ebm \\
{}= &\bbm 1 & -\rho(z;0)^*e^{-2i(Ex+Dt)/\e} \\ 0 & 1 \ebm 
\bbm 1 & 0 \\ -\rho(z;0) e^{2i(Ex+Dt)/\e} & 1 \ebm.
\end{split}
\endeq
The jump matrix satisfies
\begin{equation}
\lim_{z\to 0}\mathbf{V}(z) = \lim_{z\to\pm \infty}\mathbf{V}(z)=\mathbb{I}\,.
\label{eq:jumplims}
\end{equation}
\end{prop}
\begin{proof}
The continuity of the boundary values follows from the definition
\eqref{m} and established properties of the columns of $\MMi^\pm$, given
that the denominators $S_{jj}(z)$ do not vanish for $z\in\mathbb{R}$.

From equation \eqref{phi-plus-ito-phi-minus} and
$\det(\mathbf{S}(z))=1$, \eq \bbm \jji_1^-, \jji_2^- \ebm = \bbm
\jji_1^+, \jji_2^+ \ebm \bbm S_{22} & -S_{12} \\ -S_{21} & S_{11}
\ebm.
\endeq
Therefore, using \eqref{m-ito-Phi-minus-and-Phi-plus}, we obtain
\eq
\begin{split}
\frac{1}{S_{11}}\mmi_2^- = \frac{1}{S_{11}}\jji_2^- e^{-iEx/\e} = \frac{1}{S_{11}}(-S_{12}\jji_1^+ + S_{11}\jji_2^+)e^{-iEx/\e} & = -\frac{S_{12}}{S_{11}}e^{-2iEx/\e}\mmi_1^+ + \mmi_2^+ \\
 & = \frac{S_{21}^*}{S_{22}^*}e^{-2iEx/\e}\mmi_1^+ + \mmi_2^+
\end{split}
\endeq
and
\eq
\frac{1}{S_{22}}\mmi_1^- = \frac{1}{S_{22}}\jji_1^- e^{iEx/\e} = \frac{1}{S_{22}}(S_{22}\jji_1^+-S_{21}\jji_2^+)e^{iEx/\e} = \mmi_1^+ - \frac{S_{21}}{S_{22}}e^{2iEx/\e}\mmi_2^+.
\endeq
Combining these two equations in matrix form using the identities
\begin{equation}
\GG_+(z) = [S_{22}(z)^{-1}\mmi_1^-(x;z,t),\mmi_2^+(x;z,t)]\,,\quad\quad
\GG_-(z) = [\mmi_1^+(x;z,t),S_{11}(z)^{-1}\mmi_2^-(x;z,t)]
\end{equation}
gives \eq \GG_+(z) \bbm 1 & 0 \\ \rho(z;t)e^{2iEx/\e} & 1
\ebm = \GG_-(z)\bbm 1 & -\rho(z;t)^*e^{-2iEx/\e} \\ 0 & 1
\ebm,
\endeq
which, after taking into account Proposition~\ref{t-dep}, shows
(\ref{real-axis-jump}).  

Finally, using Proposition~\ref{m-limits}, the relations
\eqref{m-ito-Phi-minus-and-Phi-plus} between the columns of $\MMi^\pm$
and those of $\JJi^\pm$, and the Wronskian relations
\eqref{eq:Wronskians} shows that \eq
\label{S-limits-z}
\lim_{z\to0}\mathbf{S}(z) = (-1)^{n-1}\mathbb{I}, \quad 
\lim_{z\to\infty}\mathbf{S}(z) = \mathbb{I},
\endeq
which implies the limits \eqref{eq:jumplims} of the jump matrix.
\end{proof}
Therefore in any reflectionless case the condition $\rho\equiv0$
implies that $\mathbf{V}(z)\equiv \mathbf{0}$, and by continuity of
the boundary values holding in the absence of real zeros of
$S_{jj}(z)$, a Cauchy integral argument shows that $\GG(z)$ is
meromorphic in the entire $z$-plane.  The next proposition follows
immediately from \eqref{m-limit-large-z} in Proposition \ref{m-limits}
and \eqref{S-limits-z}.
\begin{prop}
\label{MMi-limit-z-prop}
Suppose that $u$ is a solution of the sine-Gordon equation with $L^1$-Sobolev
regularity.  Then the matrix $\GG(z)$ defined by \eqref{m} satisfies
\eq
\label{MMi-limit-z}
\lim_{z\to\infty}\GG(z)=\mathbb{I}\,.
\endeq
(The limit is independent of direction in the complex plane.)
\end{prop}
The next proposition describes how to 
recover $\cos(u)$ and $\sin(u)$ (and thus $u$) from $\GG(z)$.
\begin{prop}
\label{cos-u-from-m}
Suppose that $u$ is a solution of the sine-Gordon equation with
$L^1$-Sobolev regularity.  Let
$\GG^{(0)}:=\displaystyle\lim_{z\to0}\GG(z)$.  Then $\cos(u)$ and
$\sin(u)$ are expressed in terms of the elements $G_{ij}^{(0)}$ of
$\GG^{(0)}$ by \eq
\label{cos-from-m}
\cos(u) = G^{(0)}_{11}G^{(0)}_{22}+G^{(0)}_{12}G^{(0)}_{21} = 1+2G^{(0)}_{12}G^{(0)}_{21}
\endeq
and
\eq
\label{sin-from-m}
\sin(u) = 2G^{(0)}_{21}G^{(0)}_{22} = -2G^{(0)}_{11}G^{(0)}_{12}.
\endeq
\end{prop}
\begin{proof}
  From Proposition~\ref{real-axis-jump-prop} we have
  $\mathbf{V}(0)=\mathbb{I}$, and it follows that $\GG^{(0)}$ is
  well defined.  Also, from Lemma \ref{MMix-boundedness}, $\GG_x(z)$ remains
  bounded in the limit $z\to0$.  Since the columns of $\GG(z)$ satisfy
the differential equation \eqref{JJi-diffeq} where they are defined, 
\eq
\label{MMi-diffeq}
4i\e\GG_x = 4E\sigma_3\GG-4E\GG\sigma_3+\QQi\GG.
\endeq
Let 
\begin{equation}
\QQi^{(-1)}:=\lim_{z\to0}z\QQi = \bbm 1-\cos(u) & -\sin(u)\\
-\sin(u) & -(1-\cos(u))\ebm\,.
\end{equation}
Then multiplying 
\eqref{MMi-diffeq} by $z$ and taking the limit as $z\to0$ gives
\eq
0 = -\sigma_3\GG^{(0)}+\GG^{(0)}\sigma_3 + \QQi^{(-1)}\GG^{(0)}.
\endeq
Therefore
\eq
\QQi^{(-1)} = \sigma_3-\GG^{(0)}\sigma_3\GG^{(0)-1},
\endeq
which gives \eqref{cos-from-m} and \eqref{sin-from-m}.  The
consistency and reality follow from the holomorphic and
antiholomorphic symmetries \eqref{MMi-symmetries} at $z=0$.  The
Pythagorean identity follows from the identity $\det\GG(z)=1$, which
in turn follows from \eqref{m-ito-Phi-minus-and-Phi-plus} and
\eqref{eq:Wronskians}.
\end{proof}

\subsection{Riemann-Hilbert problem}
From this point onward, we assume that $u=u(x,t)$ is a solution of 
the sine-Gordon equation with the following properties:
\begin{itemize}
\item[(a)] The solution $u$ has $L^1$-Sobolev regularity.
\item[(b)] All eigenvalues corresponding to $u$ are simple.
\item[(c)] The entries $S_{11}(z)$ and $S_{22}(z)$ of the scattering matrix 
$\mathbf{S}(z)$ have no real zeros.
\end{itemize}
These are natural conditions\footnote{According to the arguments to be
  given in Appendix \ref{app-well-posed}, the solution to the Cauchy
  problem corresponding to the special initial data \eqref{init-cond}
  satisfies condition (a) for all $\mu\in\mathbb{R}$.  Note, however,
  that for fixed $\mu\neq0$, condition (b) does not hold for this
  data if $\e$ satisfies $(\sqrt{\mu^2+1}-1)/2\e\in\{1,2,3,\dots\}$.
  Furthermore, for any fixed $\mu$, condition (c) fails for this
  data if $\e$ satisfies $(\sqrt{\mu^2+1}-1)/2\e\in\{1,2,3,\dots\}$ as
  $\e\downarrow0$.  In other words, conditions (b) and (c) fail
  infinitely often in the limit $\e\downarrow0$.  Nevertheless,
  conditions (b) and (c) both hold generically with respect to
  $\e$.  Also, if conditions (b) and (c) hold at a fixed time (say
  $t=0$), they will hold for all time.}.

While $\GG(z)$ has been defined (see \eqref{m}) in terms of solutions
of differential equations with nonconstant coefficients involving
$u(x,t)$, we have shown directly that $\GG(z)$ has certain properties
that depend only on the scattering data for $u$, which may be
calculated at any convenient instant of time, say, $t=0$.  The
\emph{Riemann-Hilbert problem} asks whether it is in fact possible to
determine $\GG(z)$ (and hence $u$, via Proposition \ref{cos-u-from-m})
purely from the scattering data.  To pose the Riemann-Hilbert problem,
we seek a matrix $\HH(z)=\HH(z;x,t,\e)$ with the following properties:
\begin{enumerate}
\item[(I)] $\HH(z)\to\mathbb{I}$ as $z\to\infty$ uniformly in \emph{all} 
directions.
\item[(II)] $\HH(z)$ is meromorphic in the upper and lower half-planes with
simple poles only.  
The residues of $\HH(z)$ are required to satisfy
\eqref{residue-conditions} and \eqref{cn}.
\item[(III)] On the real axis $z\in\mathbb{R}$, $\HH(z)$ obtains continuous 
boundary values and satisfies the multiplicative jump relation
\eq
\HH_+(z) = \HH_-(z)\mathbf{V}(z)
\endeq
with $\mathbf{V}(z)$ given by \eqref{real-axis-jump}.
\end{enumerate}
Note that the problem description only involves the scattering data, and also
that $x$ and $t$ enter in a simple explicit manner.
\begin{thm}
  Assume conditions (a), (b), and (c).  Then the Riemann-Hilbert
  problem has a unique solution for all $x\in\mathbb{R}$ and $t\geq0$.
\end{thm}
\begin{proof}
  As a consequence of the $L^1$-Sobolev regularity of the solution to
  the Cauchy problem, the matrix $\GG(z)$ defined by the formula
  \eqref{m} can in principle be constructed for any $x\in\mathbb{R}$
  and $t\geq0$.  From Propositions \ref{residue-prop},
  \ref{real-axis-jump-prop}, and \ref{MMi-limit-z-prop}, $\GG(z)$ is,
  therefore, a particular solution of the Riemann-Hilbert problem.  

  It remains to determine whether the solution is unique.  First,
  suppose that $\HH(z) $ is any solution to the Riemann-Hilbert
  problem.  We will show that $\det(\HH(z))\equiv1$.  Indeed, suppose
  $z_0$ is a (simple) pole of $\HH(z)$ with associated residue
  parameter $c_0$.  From the residue condition (II), $\HH(z)$ has a
  Laurent series of the form
\eq \HH(z) =
  \left[\frac{c_0\hh_2^{(0)}}{z-z_0}+\sum_{n=0}^\infty
    \hh_1^{(n)}(z-z_0)^n, \,\,\,\sum_{n=0}^\infty \hh_2^{(n)}(z-z_0)^n
  \right]\,,
\endeq
where $\hh_j^{(n)}$ are constant vectors.  Expanding the determinant
by columns shows $\det(\HH(z)) = O(1)$ near $z=z_0$, that is,
$\det(\HH(z))$ has no poles, and so for all $z\not\in\mathbb{R}$
$\det(\HH(z))$ is a scalar analytic function of $z$.  Moreover, from
the jump condition (II), $\det(\HH_+(z)) = \det(\HH_-(z))$, and also
$\det(\HH(z))$ obtains continuous boundary values along the real axis.
Since $\det(\HH(z))$ is analytic in the open upper and lower
half-planes, Cauchy's theorem shows $\det(\HH(z))$ is analytic on the
real axis as well.  Therefore $\det(\HH(z))$ is an entire function,
and from the normalization condition (III) we have $\det(\HH(z))\to 1$
as $z\to\infty$.  Therefore $\det(\HH(z))\equiv 1$ by Liouville's theorem.

Now assume $\HHt(z)$ is another solution to the Riemann-Hilbert problem
and consider the matrix $\EE(z):=\HH(z)\HHt(z)^{-1}$.  Using 
$\det(\HHt(z))\equiv
1$, direct multiplication shows that all singularities of $\EE(z)$ 
are removable.
For $z\in\mathbb{R}$, 
\eq \EE_+(z) = \HH_+(z)\HHt_+(z)^{-1} =
\HH_-(z)\mathbf{V}(z)\mathbf{V}(z)^{-1}\HHt_-(z)^{-1} = 
\HH_-(z)\HHt_-(z)^{-1} = \EE_-(z),
\endeq
so $\EE(z)$ has no jump discontinuity across the real axis.
Furthermore, $\EE(z)$ also achieves continuous boundary values on the
real axis, and therefore, by a Cauchy argument, is an entire function.
In addition, $\EE(z)\to\mathbb{I}$ as $z\to\infty$ as this is true of
both $\HH(z)$ and $\HHt(z)$.  Therefore, up to removable
singularities, $\EE(z)$ is entire and bounded in the complex plane,
and again by Liouville's theorem $\EE(z)=\mathbb{I}$, or put another
way, $\HH(z) \equiv \HHt(z)$.
\end{proof}
The significance of this result is that the solution of the Cauchy
problem for the sine-Gordon equation corresponding to a broad class of
initial data can be completely characterized through the solution of
the Riemann-Hilbert problem given the scattering data calculated at
$t=0$ (which therefore encode the Cauchy data $f$ and $g$).  In
particular, this point of view is well-suited to asymptotic analysis
in various limits of interest, for example, $\e\downarrow 0$.

Note that, in formulation presented above, the existence of a
classical solution $\HH(z)\equiv \GG(z)$ to the Riemann-Hilbert
problem follows from the well-posedness theory for the sine-Gordon
Cauchy problem explained in Appendix~\ref{app-well-posed}.  In the
absence of such an independently developed theory of well-posedness,
it is still possible to prove the existence of a solution to the
Riemann-Hilbert problem by direct means in various situations.  See,
for example, \cite{Zhou:1998}.

To close this appendix, we comment on the semiclassical asymptotic
analysis of the Riemann-Hilbert problem corresponding to the special
initial data \eqref{init-cond}, which is work in progress.  The
so-called \emph{steepest descent method} for matrix-valued
Riemann-Hilbert problems developed by Deift, Zhou, and their coworkers
is a powerful method of asymptotic analysis in which a basic technique
is to introduce a sequence of carefully-chosen piecewise meromorphic
changes of the dependent variable $\HH(z)$ (frequently these are
called \emph{deformations} of a Riemann-Hilbert problem).  The
ultimate aim of these deformations is to move the jump discontinuities
from one contour to another in order to exchange oscillations for
exponential decay.  A simple example of such a deformation arises from
the factorization \eqref{real-axis-jump} of the jump matrix
$\mathbf{V}(z)$ (see, for example, \cite{Tovbis:2004, Tovbis:2006}).
The idea is to replace $\HH(z)$ with another unknown $\HH^{(1)}(z)$
defined as follows.  Let $D_\pm$ be bounded subsets of $\mathbb{C}_\pm$
both adjacent to the same interval $I$ of the real axis.  Then we set
\begin{equation}
\begin{split}
\HH^{(1)}(z)&:=\HH(z)\bbm 1 & 0\\\rho(z;0)e^{2i(Ex+Dt)/\e} & 1\ebm
\,,\quad\quad z\in D_+\,,\\
\HH^{(1)}(z)&:=\HH(z)\bbm 1 & -\rho(z^*;0)^*e^{-2i(Ex+Dt)/\e}\\
0 & 1\ebm\,,\quad\quad z\in D_-\,,
\end{split}
\end{equation}
and, for all other $z$ in the upper and lower half-planes, we simply
set $\HH^{(1)}(z):=\HH(z)$.  Note, in particular, that as a result of
the factorization \eqref{real-axis-jump}, the matrix $\HH^{(1)}(z)$
extends continuously to the real interval $I$, that is, there is no
longer any jump discontinuity across $I$.  Typically, the residues of
the poles of $\HH(z)=\GG(z)$ are exactly cancelled in the regions
$D_\pm$ by this deformation.  That is, a deformation based on
\eqref{real-axis-jump} has the added benefit of removing the
poles\footnote{Of course, in the (nongeneric) reflectionless cases for
  which $\e=\e_N(\mu)$ and hence $\rho\equiv 0$ making
  $\mathbf{V}(z)\equiv\mathbb{I}$, both factors of $\mathbf{V}(z)$ as
  written in \eqref{real-axis-jump} are trivial, so the poles may not
  be removed in this way.  However, the poles may indeed be removed by
  finding an appropriate analytic interpolant of the proportionality
  constants $\{\eta_n\}$ at the corresponding eigenvalues $\{z_n\}$
  and using this interpolant along with the formula
  \eqref{eq:S22-prime-general} to construct a meromorphic function
  whose residues at the eigenvalues are the modified proportionality
  constants $\{c_n\}$.  In \cite{Kamvissis:2003-book}, this approach
  was used to remove the poles from the reflectionless
  inverse-scattering problem associated with Klaus-Shaw initial data
  for the focusing NLS equation.}  from the Riemann-Hilbert problem.
However, if $S_{21}(z)$ has (phantom) poles (as in the case
corresponding to the initial data \eqref{init-cond}; see the
discussion just before the statement of
Proposition~\ref{proportionality-t-zero}) and if any of these poles lie
in the domains $D_\pm$, then the above deformation will also introduce
new poles into the matrix $\HH^{(1)}(z)$ where there were none in
$\HH(z)$.  Avoiding the phantom poles places additional conditions on
the regions $D_\pm$ in which the change of variables can be made, and
these conditions further complicate the steepest-descent analysis.

\section{The $L^p$-Sobolev theory of the Cauchy problem for the
  sine-Gordon equation}
\label{app-well-posed}

Consider the Cauchy problem corresponding to topological charge $n$:
\begin{equation}
\begin{cases}
\text{PDE:} \quad & \displaystyle 
\e^2u_{tt}-\e^2u_{xx} + \sin(u)=0\,,\quad
x\in\mathbb{R}\,,\quad t>0\\
\text{Initial Conditions:}  \quad &\displaystyle  u(x,0)=f(x)\,,\quad \e
u_t(x,0)=
g(x)\\
\text{Boundary Conditions:}  \quad &\displaystyle u(-\infty,t)\equiv 0\,,\quad
u(+\infty,t)\equiv 2\pi n\,,\quad n\in\mathbb{Z}\,.
\end{cases}
\label{eq:Cauchy}
\end{equation}
Here $f(\cdot)$ and $g(\cdot)$ are given functions; $f$ satisfies the
given boundary conditions and $g$ vanishes as $|x|\rightarrow\infty$
in a sense to be prescribed later.  Regardless of the value of the topological
charge, we may easily transform this problem into one for which the
corresponding boundary conditions are homogeneous by making the substitutions
\begin{equation}
u(x,t)=U(x,t)+b(x)\,,\quad\quad
f(x)=F(x)+b(x)\,,
\end{equation}
where $b(\cdot)$ is function that satisfies the (typically
nonhomogeneous) boundary conditions.  For convenience, we will suppose
that $b'(\cdot)$ is in the Schwartz space, and moreover, that
$b(-x)-2\pi n = -b(-x)$.  A concrete example of a function $b$ of this type
is given by 
\begin{equation}
b(x):=\pi n(1+\tanh(x))\,,
\end{equation}
but our arguments will not rely on this particular definition.
Thus, $U$ and $F$ satisfy
homogeneous boundary conditions, and so we arrive at the equivalent
Cauchy problem:
\begin{equation}
\begin{cases}
\text{PDE:} \quad & \displaystyle 
U_{tt}-U_{xx} =Q(x,t)\,,\quad
x\in\mathbb{R}\,,\quad t>0\\
\text{Initial Conditions:}  \quad &\displaystyle  U(x,0)=F(x)\,,
\quad U_t(x,0)=G(x)
\\
\text{Boundary Conditions:}  \quad &\displaystyle U(-\infty,t)\equiv 0\,,\quad
U(+\infty,t)\equiv 0\,,
\end{cases}
\label{eq:CauchyII}
\end{equation}
where
\begin{equation}
Q(x,t):=b''(x)-\frac{1}{\epsilon^2}\sin(U(x,t)+b(x))
\end{equation}
and, for symmetry of notation, we have defined
\begin{equation}
G(x):=\frac{1}{\epsilon}g(x)\,.
\end{equation}

We may express $U$ in terms of the initial data $F$, $G$, and the
source term $Q$ with the help of Green's function:
\begin{equation}
U(x,t)=\frac{1}{2}F(x+t)+\frac{1}{2}F(x-t) +\frac{1}{2}
\int_{x-t}^{x+t}G(x_0)\,dx_0 + \frac{1}{2}\int_0^t\int_{x-(t-t_0)}^{x+(t-t_0)}Q(x_0,t_0)\,dx_0\,dt_0\,.
\label{eq:CauchyIIweak}
\end{equation}
Since $Q$ depends on $U$, this formula (Duhamel's formula) does not represent
the solution in closed form, but as derivatives of $U$ are not involved 
it amounts to a useful reformulation of the Cauchy problem 
\eqref{eq:CauchyII} in weak form.  We introduce a nonlinear operator
$\mathcal{T}$ by 
\begin{equation}
\mathcal{T}[U](x,t):=A(x,t)+\mathcal{S}[U](x,t)\,,
\end{equation}
where
\begin{equation}
A(x,t):=\frac{1}{2}F(x+t)+\frac{1}{2}F(x-t) +\frac{1}{2}
\int_{x-t}^{x+t}G(x_0)\,dx_0 +\frac{1}{2}\int_0^t\int_{x-(t-t_0)}^{x+(t-t_0)}
b''(x_0)\,dx_0\,dt_0
\label{eq:Fdefine}
\end{equation}
and
\begin{equation}
\mathcal{S}[U](x,t):=-\frac{1}{2\epsilon^2}\int_0^t\int_{x-(t-t_0)}^{x+(t-t_0)}
\sin(U(x_0,t_0)+b(x_0))\,dx_0\,dt_0\,.
\end{equation}
Thus, the weak formulation \eqref{eq:CauchyIIweak} of the Cauchy problem 
\eqref{eq:CauchyII}
takes the form of a fixed-point equation in function space:  
$U=\mathcal{T}[U]$. 

We want to study this problem in the $L^p$ spaces with respect to $x$.
For each $T>0$ and $p\ge 1$, define a norm by
\begin{equation}
\|U\|_{X_T^p}:=\sup_{0\le t\le T}\|U(\cdot,t)\|_{L^p}=\sup_{0\le t\le T}\left[
\int_{-\infty}^{+\infty}|U(x,t)|^p\,dx\right]^\frac{1}{p}\,,
\end{equation}
and a corresponding Banach space of functions $U$:
\begin{equation}
X_T^p:=\left\{\text{$U(x,t)$, $x\in\mathbb{R}$, $t\in [0,T]$ such that
$\|U\|_{X_T^p}<\infty$}\right\}\,.
\end{equation}

\begin{lemma}
\label{lem:1}
Suppose that $F\in L^p(\mathbb{R})$ and $G\in L^p(\mathbb{R})$.  Then
the function $A$ defined by \eqref{eq:Fdefine} lies in the space
$X_T^p$ for each $T>0$, and
\begin{equation}
\|A\|_{X_T^p}\le \|F\|_{L^p} + T\|G\|_{L^p} + \frac{T^2}{2}
\|b''\|_{L^p}\,.
\end{equation}
\end{lemma}
\begin{proof}
By the Minkowski inequality and translation invariance of $L^p$, 
\begin{equation}
\begin{split}
\|A(\cdot,t)\|_{L^p}&\le \|F\|_{L^p} + \frac{1}{2}
\left[\int_{-\infty}^{+\infty}\left|\int_{x-t}^{x+t}G(x_0)\,dx_0\right|^p\,dx\right]^{\tfrac{1}{p}} \\
&\quad\quad\quad{}+ \frac{1}{2}\left[\int_{-\infty}^{+\infty}\left|
\int_0^t\int_{x-(t-t_0)}^{x+(t-t_0)}b''(x_0)\,dx_0\,dt_0
\right|^p\,dx\right]^{\tfrac{1}{p}}\,.
\end{split}
\end{equation}
The integral triangle inequality gives
\begin{equation}
\begin{split}
\|A(\cdot,t)\|_{L^p}&\le \|F\|_{L^p} + \frac{1}{2}
\left[\int_{-\infty}^{+\infty}\left(\int_{x-t}^{x+t}|G(x_0)|\,dx_0\right)^p\,dx\right]^{\tfrac{1}{p}} \\
&\quad\quad\quad{}+ \frac{1}{2}\left[\int_{-\infty}^{+\infty}\left(
\int_0^t\int_{x-(t-t_0)}^{x+(t-t_0)}|b''(x_0)|\,dx_0\,dt_0
\right)^p\,dx\right]^{\tfrac{1}{p}}\,.
\end{split}
\end{equation}
Now, by H\"older's inequality we have
\begin{equation}
\begin{split}
\left(\int_{x-t}^{x+t}|G(x_0)|\,dx_0\right)^p&\le
\left(\left[\int_{x-t}^{x+t}1^q\,dx_0\right]^{\tfrac{1}{q}}\left[\int_{x-t}^{x+t}
|G(x_0)|^p\,dx_0\right]^{\tfrac{1}{p}}\right)^p\\
&{}=(2t)^{\tfrac{p}{q}}\int_{x-t}^{x+t}
|G(x_0)|^p\,dx_0
\end{split}
\end{equation}
and
\begin{equation}
\begin{split}
\left(\int_0^t\int_{x-(t-t_0)}^{x+(t-t_0)}|b''(x_0)|\,dx_0\,dt_0\right)^p
&\le
\left(\left[\int_0^t\int_{x-(t-t_0)}^{x+(t-t_0)}1^q\,dx_0\,dt_0\right]^{\tfrac{1}{q}}
\left[\int_0^t\int_{x-(t-t_0)}^{x+(t-t_0)}
|b''(x_0)|^p\,dx_0\,dt_0\right]^{\tfrac{1}{p}}\right)^p\\
&{}=t^{\tfrac{2p}{q}}\int_0^t
\int_{x-(t-t_0)}^{x+(t-t_0)}
|b''(x_0)|^p\,dx_0\,dt_0
\end{split}
\end{equation}
where $q$ satisfies $1/p+1/q=1$.  Therefore,
\begin{equation}
\begin{split}
\|A(\cdot,t)\|_{L^p}&
\le \|F\|_{L^p}+\frac{1}{2}(2t)^{\tfrac{1}{q}}
\left[\int_{-\infty}^{+\infty}\int_{x-t}^{x+t}|G(x_0)|^p\,dx_0\,dx
\right]^{\tfrac{1}{p}}\\
&{}+\frac{1}{2}t^{\tfrac{2}{q}}\left[\int_{-\infty}^{+\infty}\int_0^t\int_{x-(t-t_0)}^{x+(t-t_0)}|b''(x_0)|^p\,dx_0\,dt_0\,dx\right]^{\tfrac{1}{p}}\,.
\end{split}
\end{equation}
Using Fubini's theorem to exchange the order of integration, we have
\begin{equation}
\begin{split}
\int_{-\infty}^{+\infty}\int_{x-t}^{x+t}|G(x_0)|^p\,dx_0\,dx &= 
\int_{-\infty}^{+\infty}\int_{x_0-t}^{x_0+t}|G(x_0)|^p\,dx\,dx_0\\
&{}=
\int_{-\infty}^{+\infty}|G(x_0)|^p\int_{x_0-t}^{x_0+t}\,dx\,dx_0\\
&{}=
2t\|G\|_{L^p}^p\,,
\end{split}
\end{equation}
and
\begin{equation}
\begin{split}
\int_{-\infty}^{+\infty}\int_0^t\int_{x-(t-t_0)}^{x+(t-t_0)}|b''(x_0)|^p
\,dx_0\,dt_0\,dx &= \int_0^t\int_{-\infty}^{+\infty}\int_{x-(t-t_0)}^{x+(t-t_0)}
|b''(x_0)|^p\,dx_0\,dx\,dt_0 \\
&{}= 
\int_0^t\int_{-\infty}^{+\infty}\int_{x_0-(t-t_0)}^{x_0+(t-t_0)}
|b''(x_0)|^p\,dx\,dx_0\,dt_0 \\
&{}=t^2\|b''\|_{L^p}^p\,.
\end{split}
\end{equation}
Therefore, using $1/p+1/q=1$, we have
\begin{equation}
\|A(\cdot,t)\|_{L^p}\le \|F\|_{L^p} + t\|G\|_{L^p}
 + \frac{1}{2}t^2\|b''\|_{L^p}\,,
\end{equation}
and taking a supremum over $t\in [0,T]$ completes the proof.
\end{proof}

In terms of the function $b(\cdot)$, let another function $c(\cdot)$ be defined
as follows:
\[
c(x):=|b(x)|\chi_-(x) + |b(-x)|\chi_+(x)\,,
\]
where $\chi_+(x)$ and $\chi_-(x)$ are the characteristic (indicator)
functions of the sets $x>0$ and $x<0$, respectively.  This function is
bounded and rapidly decaying as $|x|\rightarrow\infty$.
\begin{lemma}
\label{lem:2}
Let $T>0$, and suppose that $U\in X_T^p$.  Then $\mathcal{S}[U]\in X_T^p$, and 
\begin{equation}
\|\mathcal{S}[U]\|_{X_T^p}\le\frac{T^2}{2\epsilon^2}\|U\|_{X_T^p} + 
\frac{T^2}{2\epsilon^2}\|c\|_{L^p}\,.
\end{equation}
\end{lemma}
\begin{proof}
By the triangle inequality for integrals we have
\begin{equation}
\begin{split}
\|\mathcal{S}[U](\cdot,t)\|_{L^p} &= \frac{1}{2\epsilon^2}
\left[\int_{-\infty}^{+\infty}\left|\int_0^t\int_{x-(t-t_0)}^{x+(t-t_0)}
\sin(U(x_0,t_0)+b(x_0))\,dx_0\,dt_0\right|^p\,dx\right]^{\tfrac{1}{p}}\\
&{}\le 
\frac{1}{2\epsilon^2}
\left[\int_{-\infty}^{+\infty}\left(\int_0^t\int_{x-(t-t_0)}^{x+(t-t_0)}
|\sin(U(x_0,t_0)+b(x_0))|\,dx_0\,dt_0\right)^p\,dx\right]^{\tfrac{1}{p}}\,.
\end{split}
\end{equation}
For any real $x$ and $t$ we have, using periodicity of the sine
function, the inequality $|\sin(x)|\le |x|$, and the property
$b(x)-2\pi n=-b(-x)$ that
\begin{equation}
\begin{split}
\left|\sin(U(x,t)+b(x))\right|&=\left|
\sin(U(x,t)+b(x))\chi_-(x) +
\sin(U(x,t)-b(-x))\chi_+(x)\right|\\
&\le \left|\sin(U(x,t)+b(x))\right|\chi_-(x) + 
\left|\sin(U(x,t)-b(-x))\right|
\chi_+(x)\\
&\le \left|U(x,t)+b(x)\right|\chi_-(x) + 
\left|U(x,t)-b(-x)\right|\chi_+(x)\\
&\le
|U(x,t)|+|b(x)|\chi_-(x)
+|b(-x)|\chi_+(x)\\
&= |U(x,t)| + c(x)\,.
\end{split}
\label{eq:sinebreakup}
\end{equation}
Therefore, by the Minkowski inequality, 
\begin{equation}
\begin{split}
\|\mathcal{S}[U](\cdot,t)\|_{L^p}&\le\frac{1}{2\epsilon^2}
\left[\int_{-\infty}^{+\infty}\left(\int_0^t\int_{x-(t-t_0)}^{x+(t-t_0)}
|U(x_0,t_0)|\,dx_0\,dt_0\right)^p\,dx\right]^\frac{1}{p} \\
&\quad\quad {}+ \frac{1}{2\epsilon^2}
\left[\int_{-\infty}^{+\infty}\left(\int_0^t\int_{x-(t-t_0)}^{x+(t-t_0)}
c(x_0)\,dx_0\,dt_0\right)^p\,dx\right]^\frac{1}{p} \,.
\end{split}
\end{equation}
Applying H\"older's inequality, we then find
\begin{equation}
\begin{split}
\|\mathcal{S}[U](\cdot,t)\|_{L^p}&\le \frac{t^{\tfrac{2}{q}}}{2\epsilon^2}
\left[\int_{-\infty}^{+\infty}\int_0^t\int_{x-(t-t_0)}^{x+(t-t_0)}|U(x_0,t_0)|^p\,
dx_0\,dt_0\,dx\right]^{\tfrac{1}{p}}\\
&\quad\quad{}+\frac{t^{\tfrac{2}{q}}}{2\epsilon^2}
\left[\int_{-\infty}^{+\infty}\int_0^t\int_{x-(t-t_0)}^{x+(t-t_0)}c(x_0)^p\,
dx_0\,dt_0\,dx\right]^{\tfrac{1}{p}}\,,
\end{split}
\end{equation}
where $1/p+1/q=1$.  The order of integration may be exchanged by
Fubini's theorem:
\begin{equation}
\begin{split}
\int_{-\infty}^{+\infty}\int_0^t\int_{x-(t-t_0)}^{x+(t-t_0)}|U(x_0,t_0)|^p\,dx_0\,
dt_0\,dx &= \int_0^t\int_{-\infty}^{+\infty}\int_{x-(t-t_0)}^{x+(t-t_0)}
|U(x_0,t_0)|^p\,dx_0\,dx\,dt_0\\
&=\int_0^t\int_{-\infty}^{+\infty}\int_{x_0-(t-t_0)}^{x_0+(t-t_0)}
|U(x_0,t_0)|^p\,dx\,dx_0\,dt_0\\
&=\int_0^t\|U(\cdot,t_0)\|_{L^p}^p\cdot 2(t-t_0)\,dt_0\\
&\le t^2\sup_{0\le t_0\le t}\|U(\cdot,t_0)\|_{L^p}^p\,.
\end{split}
\end{equation}
Similarly,
\begin{equation}
\int_{-\infty}^{+\infty}\int_0^t\int_{x-(t-t_0)}^{x+(t-t_0)}c(x_0)^p\,dx_0\,
dt_0\,dx = t^2\|c\|_{L^p}^p\,.
\end{equation}
The proof is therefore complete upon taking a supremum over $0\le t\le T$.
\end{proof}

\begin{lemma}
  Let $T>0$.  Then whenever $U$ and $V$ are in $X_T^p$,
\begin{equation}
\|\mathcal{T}[U]-\mathcal{T}[V]\|_{X_T^p}\le\frac{T^2}{2\epsilon^2}
\|U-V\|_{X_T^p}\,.
\end{equation}
\label{lem:3}
\end{lemma}
\begin{proof}
Clearly, we have $\mathcal{T}[U]-\mathcal{T}[V]=\mathcal{S}[U]-\mathcal{S}[V]$.
(We could have assumed further that $F$ and $G$ lie in $L^p$,
so that $\mathcal{T}[U]$ and $\mathcal{T}[V]$ are individually well-defined
as elements of $X_T^p$, but as $A(x,t)$ cancels out of the difference, it
is not necessary to make such an assumption here.)  Now by the triangle
inequality for integrals,
\begin{equation}
\begin{split}
\|\mathcal{S}[U](\cdot,t)-\mathcal{S}[V](\cdot,t)\|_{L^p}&=
\frac{1}{2\epsilon^2}\left[\int_{-\infty}^{+\infty}\left|\int_0^t
\int_{x-(t-t_0)}^{x+(t-t_0)}\left\{\sin(U(x_0,t_0)+b(x_0))
\right.\right.\right.\\
&\quad\quad{}-\left.\left.\left.
\sin(V(x_0,t_0)+b(x_0))\right\}\vphantom{\int}
\,dx_0\,dt_0\right|^p\,dx
\right]^{\tfrac{1}{p}}\\
&\le 
\frac{1}{2\epsilon^2}\left[\int_{-\infty}^{+\infty}\left(\int_0^t
\int_{x-(t-t_0)}^{x+(t-t_0)}\left|\sin(U(x_0,t_0)+b(x_0))
\right.\right.\right.\\
&\quad\quad{}-\left.\left.\left.
\sin(V(x_0,t_0)+b(x_0))\right|\vphantom{\int}
\,dx_0\,dt_0\right)^p\,dx
\right]^{\tfrac{1}{p}}\,.
\end{split}
\end{equation}
Now, since by the mean value theorem,
$|\sin(x)-\sin(y)|=|\cos(\xi)|\cdot|x-y|\le|x-y|$, we have simply
\begin{equation}
\|\mathcal{S}[U](\cdot,t)-\mathcal{S}[V](\cdot,t)\|_{L^p}\le
\frac{1}{2\epsilon^2}\left[\int_{-\infty}^{+\infty}
\left(\int_0^t\int_{x-(t-t_0)}^{x+(t-t_0)}|U(x_0,t_0)-V(x_0,t_0)|\,dx_0\,dt_0
\right)^p\,dx\right]^{\tfrac{1}{p}}\,.
\end{equation}
The proof is then finished upon using H\"older's inequality and Fubini's
theorem in exactly the same way as in the proofs of Lemma~\ref{lem:1}
and Lemma~\ref{lem:2}.
\end{proof}

\begin{thm}[Local Existence and Uniqueness in $L^p$]
  Let $p\ge 1$ and suppose that $F,G\in
  L^p$. Define $T:=\min\{\epsilon,1\}$. Then there exists a unique
  weak solution of the sine-Gordon Cauchy problem \eqref{eq:CauchyII}
  (that is a unique solution of \eqref{eq:CauchyIIweak}) in the space
  $X_T^p$.
\label{thm:local}
\end{thm}
\begin{proof}
Combining the results of Lemma~\ref{lem:1} and Lemma~\ref{lem:2}, we have
that for any $U\in X_T^p$,
\begin{equation}
\|\mathcal{T}[U]\|_{X_T^p}\le \|F\|_{L^p} + \|G\|_{L^p}T
+\frac{1}{2}\|b''\|_{L^p}T^2+ \frac{1}{2\epsilon^2}\|c\|_{L^p}T^2 + \frac{T^2}{2\epsilon^2}\|U\|_{X_T^p}\,.
\end{equation}
Since $T\le 1$, it is also true that
\begin{equation}
\|\mathcal{T}[U]\|_{X_T^p}\le \|F\|_{L^p} + \|G\|_{L^p}
+\frac{1}{2}\|b''\|_{L^p}+ \frac{1}{2\epsilon^2}\|c\|_{L^p}
+ \frac{T^2}{2\epsilon^2}\|U\|_{X_T^p}\,.
\label{eq:embedding1}
\end{equation}
Let 
\begin{equation}
R=R_p[F,G]:=2\left\{
\|F\|_{L^p} + \|G\|_{L^p}
+\frac{1}{2}\|b''\|_{L^p}+ 
\frac{1}{2\epsilon^2}\|c\|_{L^p}\right\}\,,
\end{equation}
and let $B_R$ denote the bounded subset of $X_T^p$ given by
\begin{equation}
B_R:=\left\{\text{$U\in X_T^p$ such that $\|U\|_{X_T^p}\le R$}\right\}\,.
\end{equation}
Then, \eqref{eq:embedding1} takes the form
\begin{equation}
\|\mathcal{T}[U]\|_{X_T^p}\le \frac{R}{2}+\frac{T^2}{\epsilon^2}\cdot
\frac{\|U\|_{X_T^p}}{2}\le \frac{R}{2}+\frac{\|U\|_{X_T^p}}{2}\,,
\end{equation}
where we have used $T\le\epsilon$.  Therefore, $U\in B_R$ implies that
$\mathcal{T}[U]\in B_R$ as well; that is, $\mathcal{T}$ maps the bounded
set $B_R$ to itself.  Furthermore, combining the inequality $T\le\epsilon$
with Lemma~\ref{lem:3} we obtain 
\begin{equation}
\|\mathcal{T}[U]-\mathcal{T}[V]\|_{X_T^p}\le\frac{1}{2}\|U-V\|_{X_T^p}\,,\quad\quad
U,V\in B_R\,.
\end{equation}
(Actually, this holds for all $U,V\in X_T^p$.)  Therefore,
$\mathcal{T}$ defines a contraction mapping on $B_R$, and so there exists
a unique fixed point $U\in B_R$ of the mapping $\mathcal{T}$, that is,
a unique solution in $B_R$ of the equation $U=\mathcal{T}[U]$ equivalent
to the weak form \eqref{eq:CauchyIIweak} of the Cauchy problem.  It is easy
to see that the number $R$ could also have been replaced by any larger number,
and therefore the solution guaranteed by the contraction mapping principle
is actually unique in the whole space $X_T^p$.
\end{proof}

\begin{thm}[Global Existence and Uniqueness in $L^p$]
  Let $p\ge 1$ and suppose that $F$, $G$, and
  $F'$ all lie in $L^p$.  Then for each finite $T>0$ there
  exists a unique weak solution of the sine-Gordon Cauchy problem
  \eqref{eq:CauchyII} in the space $X_T^p$.
\label{thm:global}
\end{thm}
\begin{proof}
  We wish to iterate the argument in the proof of
  Theorem~\ref{thm:local} by restarting the Cauchy problem at time $T$
  with new initial data for which $F(x)$ is replaced by
  $F_1(x):=U(x,T)$ and $G(x)$ is replaced by the
  distributional derivative $G_1(x):=U_t(x,T)$.  Since
\begin{equation}
\|F_1\|_{L^p}=\|U(\cdot,T)\|_{L^p}\le\|U\|_{X_T^p}<\infty\,,
\end{equation}
the new initial data satisfies $F_1\in L^p$.  However, the norm of
$X_T^p$ does not directly provide us with any control of
$t$-derivatives needed to place $G_1$ in the space $L^p$ together with
$F_1$.

To analyze $G_1$, differentiate \eqref{eq:CauchyIIweak} with
respect to $t$:
\begin{equation}
\begin{split}
U_t(x,t)&=\frac{1}{2}F'(x+t)-\frac{1}{2}F'(x-t) +
\frac{1}{2}G(x+t)+\frac{1}{2}G(x-t)\\
&\quad\quad{}+\frac{1}{2}
\int_0^t\left[b''(x+(t-t_0))
+b''(x-(t-t_0))\right]\,dt_0\\
&\quad\quad{}-\frac{1}{2\epsilon^2}\int_0^t
\left[\sin(U(x+(t-t_0),t_0)+b(x+(t-t_0)))\right.\\
&\quad\quad\quad\quad\quad\quad
\left.{}+\sin(U(x-(t-t_0),t_0)+b(x-(t-t_0)))
\right]\,dt_0\,.
\end{split}
\label{eq:Ut}
\end{equation}
It follows that
\begin{equation}
\begin{split}
\|U_t(\cdot,t)\|_{L^p}&\le \|F'\|_{L^p} + \|G\|_{L^p}\\
&\quad\quad{}+\frac{1}{2}\left\|\int_0^tb''(\cdot+(t-t_0))\,dt_0\right\|_{L^p}\\
&\quad\quad{}+\frac{1}{2}\left\|\int_0^tb''(\cdot-(t-t_0))\,dt_0\right\|_{L^p}\\
&\quad\quad{}+\frac{1}{2\epsilon^2}\left\|\int_0^t
\sin(U(\cdot+(t-t_0),t_0)+b(\cdot+(t-t_0)))\,dt_0\right\|_{L^p}\\
&\quad\quad{}+\frac{1}{2\epsilon^2}\left\|\int_0^t
\sin(U(\cdot-(t-t_0),t_0)+b(\cdot-(t-t_0)))\,dt_0\right\|_{L^p}\,.
\end{split}
\end{equation}
Now, by H\"older and Fubini arguments,
\begin{equation}
\begin{split}
\left\|\int_0^tb''(\cdot\pm (t-t_0))
\,dt_0\right\|_{L^p}&=
\left[\int_{-\infty}^{+\infty}\left|\int_0^tb''(x\pm(t-t_0))
\,dt_0\right|^p\,dx\right]^{\tfrac{1}{p}}\\
&\le \left[\int_{-\infty}^{+\infty}\left(\int_0^t|b''(x\pm(t-t_0))|\,
dt_0\right)^p\,dx\right]^{\tfrac{1}{p}}\\
&\le t^{\tfrac{1}{q}}\left[\int_{-\infty}^{+\infty}\int_0^t|b''(x\pm(t-t_0))|^p\,
dt_0\,dx\right]^{\tfrac{1}{p}}\\
&= t^{\tfrac{1}{q}}\left[\int_0^t\int_{-\infty}^{+\infty}
|b''(x\pm(t-t_0))|^p\,dx\,dt_0\right]^{\tfrac{1}{p}}\\
&=t^{\tfrac{1}{p}+\tfrac{1}{q}}\|b''\|_{L^p}\\
&=t\|b''\|_{L^p}\,.
\end{split}
\end{equation}
Using as well \eqref{eq:sinebreakup} and the Minkowski inequality,
\begin{multline}
\left\|\int_0^t\sin(U(\cdot\pm(t-t_0),t_0)+b(\cdot\pm(t-t_0)))\,dt_0\right\|_{L^p}\\
\begin{aligned}
&=
\left[\int_{-\infty}^{+\infty}\left|\int_0^t\sin(U(x\pm(t-t_0),t_0)+
b(x\pm(t-t_0)))\,dt_0\right|^p\,dx\right]^{\tfrac{1}{p}}\\
&\le
\left[\int_{-\infty}^{+\infty}\left(\int_0^t|\sin(U(x\pm(t-t_0),t_0)+
b(x\pm(t-t_0)))|\,dt_0\right)^p\,dx\right]^{\tfrac{1}{p}}\\
&\le
\left[\int_{-\infty}^{+\infty}\left(\int_0^t|U(x\pm(t-t_0),t_0)|\,dt_0+
\int_0^tc(x\pm(t-t_0))\,dt_0\right)^p\,dx\right]^{\tfrac{1}{p}}\\
&\le
\left[\int_{-\infty}^{+\infty}\left(\int_0^t|U(x\pm(t-t_0),t_0)|\,dt_0\right)^p
\,dx\right]^{\tfrac{1}{p}}+
\left[\int_{-\infty}^{+\infty}\left(\int_0^tc(x\pm (t-t_0))\,dt_0\right)^p
\,dx\right]^{\tfrac{1}{p}}\\
&\le
t^{\tfrac{1}{q}}
\left[\int_{-\infty}^{+\infty}\int_0^t|U(x\pm(t-t_0),t_0)|^p\,dt_0
\,dx\right]^{\tfrac{1}{p}}+t^{\tfrac{1}{q}}
\left[\int_{-\infty}^{+\infty}\int_0^tc(x\pm (t-t_0))^p\,dt_0
\,dx\right]^{\tfrac{1}{p}}\\
&=
t^{\tfrac{1}{q}}
\left[\int_0^t\int_{-\infty}^{+\infty}|U(x\pm(t-t_0),t_0)|^p\,dx\,dt_0
\right]^{\tfrac{1}{p}}+t^{\tfrac{1}{q}}
\left[\int_0^t\int_{-\infty}^{+\infty}c(x\pm (t-t_0))^p\,dx
\,dt_0\right]^{\tfrac{1}{p}}\\
&\le t\sup_{0\le t_0\le t}\|U(\cdot,t_0)\|_{L^p} + t\|c\|_{L^p}\,.
\end{aligned}
\end{multline}
Therefore, 
\begin{equation}
\|U_t(\cdot,t)\|_{L^p}\le\|F'\|_{L^p}+\|G\|_{L^p}
+ t\|b''\|_{L^p}+ \frac{t}{\epsilon^2}\|c\|_{L^p}+ \frac{t}{\epsilon^2}\sup_{0\le t_0\le t}\|U(\cdot,t_0)\|_{L^p}\,.
\label{eq:UtLpbound}
\end{equation}
Thus, if the initial data (already assumed to satisfy $F\in L^p$
and $G\in L^p$ to guarantee the existence of $U\in L^p$ for $t\in [0,T]$
according to Theorem~\ref{thm:local}) also satisfy $F'\in L^p$,
then $U_t$ is uniformly in $L^p$ for all $t\in [0,T]$, and in particular the 
new initial condition satisfies $G_1\in L^p$.  This is sufficient
to iterate the argument in the proof of Theorem~\ref{thm:local} an arbitrary
number of times, with a fixed time step $T$, and the proof is complete.
\end{proof}

\begin{thm}[Global $L^p$-Sobolev Regularity]
\label{thm:regularity}
Suppose the same conditions as in Theorem~\ref{thm:global}, namely
that $F$, $F'$, and $G$ are all in $L^p(\mathbb{R})$.  Then the unique
global weak solution of the Cauchy problem \eqref{eq:CauchyII}
satisfies $U\in L^\infty_\mathrm{loc}(L^p(\mathbb{R}))$, $U_x\in
L^\infty_\mathrm{loc}(L^p(\mathbb{R}))$, and $U_t\in
L^\infty_\mathrm{loc}(L^p(\mathbb{R}))$.  Moreover, if the initial
data have one more derivative in $L^p$ (that is, $F''$ and $G'$ are in
$L^p$ as well as $F$, $F'$, and $G$), then this further regularity is
preserved as well: one also has $U_{xx}\in
L^\infty_\mathrm{loc}(L^p(\mathbb{R}))$ and $U_{tx}\in
L^\infty_\mathrm{loc}(L^p(\mathbb{R}))$.
\end{thm}
\begin{proof}
  The fact that $U\in L^\infty_\mathrm{loc}(L^p(\mathbb{R}))$ follows
  from the statement of Theorem~\ref{thm:global}, and the fact that
  $U_t\in L^\infty_\mathrm{loc}(L^p(\mathbb{R}))$ follows from the
  estimate \eqref{eq:UtLpbound} in the corresponding proof.  The fact
  that $U_x\in L^\infty_\mathrm{loc}(L^p(\mathbb{R}))$ under the same
  hypotheses follows from the representation (obtained by
  differentiating \eqref{eq:CauchyIIweak} with respect to $x$)
\begin{equation}
\begin{split}
U_x(x,t)&=\frac{1}{2}F'(x+t)+\frac{1}{2}F'(x-t) +
\frac{1}{2}G(x+t)-\frac{1}{2}G(x-t)\\
&\quad\quad{}+\frac{1}{2}\int_0^t\left[b''(x+(t-t_0))
-b''(x-(t-t_0))\right]\,dt_0\\
&\quad\quad{}-\frac{1}{2\epsilon^2}\int_0^t
\left[\sin(U(x+(t-t_0),t_0)+b(x+(t-t_0)))\right.\\
&\quad\quad\quad\quad\quad\quad
\left.{}-\sin(U(x-(t-t_0),t_0)+b(x-(t-t_0)))
\right]\,dt_0\,,
\label{eq:Ux}
\end{split}
\end{equation}
which is analyzed precisely in the same way as $U_t$ was in the proof
of Theorem~\ref{thm:global}, leading to an estimate of exactly the
same form as \eqref{eq:UtLpbound}.

Now we suppose further that $F''\in L^p(\mathbb{R})$ and $G'\in
L^p(\mathbb{R})$, and consider the formula (obtained by differentiating
\eqref{eq:Ut} and \eqref{eq:Ux} with respect to $x$:
\begin{equation}
\begin{split}
\frac{\partial}{\partial x}U_{\stackrel{\scriptstyle t}{x}}(x,t)&=
\frac{1}{2}F''(x+t)\mp\frac{1}{2}F''(x-t) +\frac{1}{2}G'(x+t)\pm
\frac{1}{2}G'(x-t)\\
&\quad\quad{}+\frac{1}{2}\int_0^t\left[b'''(\xi_+)\pm b'''(\xi_-)
\right]\,dt_0\\
&\quad\quad
{}-\frac{1}{2\epsilon^2}\int_0^t\left[\cos(U(\xi_+,t_0)+b(\xi_+))\cdot(U_x(\xi_+,t_0)+b'(\xi_+))\right.\\
&\quad\quad\quad\quad\quad\quad
\left.{}\pm\cos(U(\xi_-,t_0)+b(\xi_-))\cdot(U_x(\xi_-,t_0)+b'(\xi_-))\right]\,dt_0\,,
\end{split}
\label{eq:UtxUxx}
\end{equation}
where $\xi_\pm:=x\pm (t-t_0)$.  By Minkowski's inequality and using
$|\cos(x)|\le 1$,
\begin{equation}
\begin{split}
\left\|\frac{\partial}{\partial x}U_{\stackrel{\scriptstyle t}{x}}(\cdot,t)
\right\|_{L^p}&\le \|F''\|_{L^p} + \|G'\|_{L^p} \\
&\quad\quad{}+
\frac{1}{2}\left\|\int_0^tb'''(\cdot+(t-t_0))\,dt_0\right\|_{L^p}
+\frac{1}{2}\left\|\int_0^tb'''(\cdot-(t-t_0))\,dt_0\right\|_{L^p}\\
&\quad\quad{}+\frac{1}{2\epsilon^2}\left\|\int_0^t|b'(\cdot+(t-t_0))|\,dt_0
\right\|_{L^p}+\frac{1}{2\epsilon^2}\left\|
\int_0^t|b'(\cdot-(t-t_0))|\,dt_0
\right\|_{L^p}\\
&\quad\quad{}
+\frac{1}{2\epsilon^2}\left\|\int_0^t|U_x(\cdot+(t-t_0),t_0)|\,dt_0
\right\|_{L^p}+\frac{1}{2\epsilon^2}\left\|
\int_0^t|U_x(\cdot-(t-t_0),t_0)|\,dt_0
\right\|_{L^p}\,.
\end{split}
\end{equation}
Now, by H\"older and Fubini arguments,
\begin{equation}
\begin{split}
\left\|\int_0^t|U_x(\cdot\pm(t-t_0),t_0)|\,dt_0\right\|_{L^p}&=
\left[\int_{-\infty}^{+\infty}\left(\int_0^t|U_x(x\pm (t-t_0),t_0)|\,dt_0
\right)^p\,dx\right]^\frac{1}{p}\\
&\le t^{\tfrac{1}{q}}\left[\int_{-\infty}^{+\infty}\int_0^t|U_x(x\pm (t-t_0),t_0)|^p\,dt_0\,dx\right]^\frac{1}{p}\\
&=t^{\tfrac{1}{q}}\left[\int_0^t\|U_x(\cdot,t_0)\|_{L^p}^p\,dt_0\right]^{1/p}\\
&\le t\sup_{0<t_0<t}\|U_x(\cdot,t_0)\|_{L^p}\,.
\end{split}
\end{equation}
Applying the same argument to the remaining integrals yields the estimate
\begin{equation}
\left\|\frac{\partial}{\partial x}U_{\stackrel{\scriptstyle t}{x}}(\cdot,t)
\right\|_{L^p}\le \|F''\|_{L^p}+\|G'\|_{L^p} + t\|b'''\|_{L^p} +
\frac{t}{\epsilon^2}\|b'\|_{L^p} + \frac{t}{\epsilon^2}
\sup_{0<t<t_0}\|U_x(\cdot,t_0)\|_{L^p}\,.
\end{equation}
Since it has already been established that $U_x$ is in $L^p$ as a
function of $x$ uniformly for $t$ in bounded intervals, the proof is
complete.
\end{proof}

The case most relevant for inverse-scattering theory is $p=1$.  Here
we have the following result.
\begin{cor}
  Suppose the initial data for the Cauchy problem \eqref{eq:Cauchy}
  satisfy $\sin(f),1-\cos(f),$ $f',f'',g,g'\in L^1$.  Then there 
  is a unique global
  weak solution of the problem \eqref{eq:Cauchy} for which $\sin(u), 
  1-\cos(u), u_x,$ $u_{xx}, u_t, u_{tx}\in
  L^1$ for all $t>0$.
\label{cor:L1case}
\end{cor}
\begin{proof}
  We need to translate the given conditions on $f$ and $g$ into
  corresponding conditions on $F$ and $G$ sufficient
  to apply Theorem~\ref{thm:global} and Theorem~\ref{thm:regularity}.
  Since $f'\in L^1$, $f$ is absolutely continuous and uniformly
  bounded, and therefore so is $F$.  Moreover, the limits
  $F(\pm\infty)$ both exist and vanish.  The condition that
  $\sin(f)$ is in $L^1$ therefore guarantees (since $|\sin(x-2\pi
  n)|\ge|x|/2$ for all $n\in\mathbb{Z}$ and for all $x\in\mathbb{R}$
  sufficiently small) that $F\in L^1$.  Next, since
  $F'(x)= f'(x)-b'(x)$ and $F''(x)=f''(x)-b''(x)$, the triangle
  inequality shows that the conditions $f',f''\in L^1$ easily imply that
  $F',F''\in L^1$.  Finally, since $g$ differs from $G$
  only by a factor of $\epsilon$, $g,g'\in L^1$ implies $G,G'\in
  L^1$.  

  From Theorem~\ref{thm:global} and Theorem~\ref{thm:regularity} we
  therefore obtain that $U(\cdot,t)$, $U_x(\cdot,t)$, $U_{xx}(\cdot,t)$, 
  $U_t(\cdot,t)$, and $U_{tx}(\cdot,t)$ all lie in $L^1$ for all
  $t>0$.  Since $u_t=U_t$ and $u_{tx}=U_{tx}$, the fact that $U_t,
  U_{tx}\in L^1$ guarantees that $u_t,u_{tx}\in L^1$.  Since
  $u_x=U_x+b'(x)$ and $u_{xx}=U_{xx}+b''(x)$, the triangle inequality
  shows that $U_x,U_{xx}\in L^1$ guarantees that $u_x, u_{xx}\in L^1$.
  Finally, since $|\sin(u(x,t))|=|\sin(U(x,t)+b(x))|$, the inequality
  \eqref{eq:sinebreakup} shows that $U\in L^1$ implies that
  $\sin(u)\in L^1$, and the inequality (like \eqref{eq:sinebreakup},
  but using instead $1-\cos(x)\le |x|$)
\begin{equation}
\begin{split}
1-\cos(u(x,t))&=1-\cos(U(x,t)+b(x))\\ & = 
[1-\cos(U(x,t)+b(x))]\chi_-(x)+
[1-\cos(U(x,t)-b(-x))]\chi_+(x)\\ &\le |U(x,t)|+c(x)
\end{split}
\end{equation}
shows that $U\in L^1$ implies that $1-\cos(u)\in L^1$, which finishes
the proof.
\end{proof}

In fact, for the $p=1$ case it is possible to show further that an
arbitrary number of $x$-derivatives of $u$ and $u_t$ are in
$L^1(\mathbb{R})$ for all $t>0$ if the same holds true at $t=0$.  The
issue in obtaining higher-order regularity for general $p$ arises from
replacing $\partial/\partial x$ with $\partial^n/\partial x^n$ in
\eqref{eq:UtxUxx}, because while
\begin{equation}
\left|\frac{\partial}{\partial x}\sin(f)\right|=|\cos(f)f'|\le |f'|
\end{equation}
gives an estimate that is linear in already-estimated derivatives, the
corresponding estimate of the $n$th-order partial derivative will
contain, in addition to a term $|f^{(n)}|$, a sum of nonlinear terms
in lower-order derivatives.  In the case of $p=1$, $f^{(k-1)}$ is
controlled in $L^\infty$ by $\|f^{(k)}\|_{L^1}$ by the fundamental
theorem of calculus, so all of the nonlinear terms may be estimated in
$L^1$ by peeling off an appropriate number of uniformly bounded
factors.  For example, to analyze $U_{xxx}$ or
$U_{txx}$, one replaces $\partial/\partial x$ with
$\partial^2/\partial x^2$ in \eqref{eq:UtxUxx} and then it is required to
estimate the $L^p(\mathbb{R})$ norm of a term like
\begin{equation}
\int_0^t\frac{\partial^2}{\partial x^2}\sin(U(x\pm(t-t_0),t_0)+b(x\pm (t-t_0)))\,dt_0\,.
\end{equation}
But, since
\begin{equation}
\left|\frac{\partial^2}{\partial x^2}\sin(f)\right|=\left|\sin(f)(f')^2 +\cos(f)f''\right|\le (f')^2 + |f''|\,,
\end{equation}
the quadratic term would cause some difficulty for general $p$.
However, for $p=1$, the knowledge that $f''\in L^1$ allows one to
further estimate the right-hand side by $K|f'|+|f''|$ for some
constant $K$ that depends on $\|f''\|_{L^1}$.  Then using $f'\in L^1$
as well, the argument proceeds as in the proof of
Theorem~\ref{thm:regularity} and one concludes ultimately that
$U_{xxx}$ and $U_{txx}$ are also in $L^1$ for all $t>0$.  This general
method valid for $p=1$ allows all $x$-derivatives of $U_x$ and $U_t$
to be handled in the same way.

\bibliographystyle{plain}

\end{document}